\documentclass[12pt]{iopart}

\usepackage{verbatim}
\usepackage{graphicx}
\usepackage{subcaption}
\usepackage{dcolumn}
\usepackage{bm}
\usepackage{feynmp-auto}
\usepackage[usenames, dvipsnames]{color}
\usepackage{bbold}
\usepackage{float}
\usepackage{listings}
\usepackage{cancel}
\usepackage{siunitx}
\usepackage{multirow}
\usepackage{cite}
\usepackage{hyperref}

\usepackage{fancyvrb}
\newcommand{\met}{$E_{\text{T}}^{\text{miss}}$}
\newcommand{\mht}{$H_{\text{T}}^{\text{miss}}$}
\newcommand{\htt}{$H_{\text{T}}$}
\newcommand{\ptt}{$p_{\text{T}}$}
\newcommand{\msq}{$M_{\tilde{q}}$}
\newcommand{\mgl}{$M_{\tilde{g}}$}
\newcommand{\mlsp}{$M_{\tilde{\chi}_{1}^{0}}$}
\newcommand{\mnlsp}{$M_{\tilde{\chi}_{2}^{0}}$}

\RecustomVerbatimCommand{\VerbatimInput}{VerbatimInput}%
{fontsize=\tiny,
 frame=lines,  
 framesep=2em, 
 rulecolor=\color{Gray},
 labelposition=topline,
 commandchars=\|\(\), 
 commentchar=*        
}

\expandafter\let\csname equation*\endcsname\relax
\expandafter\let\csname endequation*\endcsname\relax
\usepackage{mathtools}

\begin{document}

\begin{titlepage}
\thispagestyle{empty}

\begin{flushright}
LPT-Orsay-18-73
\end{flushright}

\phantom \\ \phantom \\

\title{Exploring Sensitivity to NMSSM Signatures with Low Missing Transverse Energy at the LHC}
\author{A. Titterton$^{1,2,3}$, U. Ellwanger$^{4}$, H.U. Flaecher$^{1}$, S. Moretti$^{2,3}$,\\ C.H. Shepherd-Themistocleous$^{3}$}\phantom\\
\address{$^1$ H.H. Wills Physics Laboratory, University of Bristol, Bristol, BS8 1TL,\\United Kingdom}
\address{$^2$ School of Physics and Astronomy, University of Southampton, Highfield, Southampton, SO17 1BJ, United Kingdom}
\address{$^3$ Particle Physics Department, Rutherford Appleton Laboratory, Chilton, Didcot, Oxon, OX11 0QX, United Kingdom}
\address{$^4$ Laboratoire de Physique Th\'eorique, UMR 8627, CNRS, Universit\'e de Paris-Sud, Universit\'e Paris-Saclay, 91405 Orsay, France}


\date{\today}
             

\begin{abstract}
We examine scenarios in the Next-to-Minimal Supersymmetric
  Standard Model (NMSSM), where pair-produced squarks and gluinos
  decay via two cascades, each ending in a stable neutralino as
  Lightest Supersymmetric Particle (LSP) and a Standard Model
  (SM)-like Higgs boson, with mass spectra such that the missing
  transverse energy, $E_{\text{T}}^{\text{miss}}$, is very low. Performing two-dimensional
  parameter scans and focusing on the hadronic $H\rightarrow b\bar{b}$
  decay giving a $b\bar{b}b\bar{b}$ + $E_{\text{T}}^{\text{miss}}$ final state we
  explore the sensitivity of a current LHC general-purpose
  jets+$E_{\text{T}}^{\text{miss}}$ analysis to such scenarios.
\end{abstract}

\maketitle

\end{titlepage}

\section{Introduction}

After four years of proton-proton collisions at the LHC with a centre-of-mass energy of $\sqrt{s} = 7 - 8$\,TeV and a further three with $13$\,TeV, searches for physics Beyond the SM (BSM) have so far not observed any significant excesses. Particularly, in searches for Supersymmetry (SUSY)~\cite{SUSY1,SUSY2,SUSY3,SUSY4}, this has allowed lower bounds to be placed on the masses of supersymmetric particles such as squarks, gluinos, gauginos and Higgsinos.

Recent LHC analyses such as~\cite{CMS-SUS-16}, utilising the $\alpha_{\text{T}}$ kinematic variable~\cite{alphaT, alphaT2}, have pushed the lower bounds on the squark mass $M_{\tilde{q}}$ and gluino mass $M_{\tilde{g}}$ well in excess of 1\,TeV and the mass of a neutralino LSP $M_{\tilde{X}^{0}_{1}}$ also as high as $1$\,TeV for certain regions of parameter space of simplified SUSY models~\cite{ATLAS_SUS,ATLAS_SUS2,ATLAS_SUS3,CMS-SUS-16-036,CMS-SUS-16-033,CMS-SUS-16-049}. These experimental limits are of course dependent upon various properties of the decay cascade such as the masses of other sparticles, the decay branching fractions and the kinematic distributions of the decay products, but have still ruled out a large area of parameter space in the MSSM.

The majority of the SUSY search effort so far has relied upon the notion that an $R$-parity-conserving supersymmetric model is expected to generate events featuring large missing transverse energy, \met. In addition, long SUSY decay cascades often imply many hadronic jets with large transverse momentum, \ptt, meaning events whose jet \ptt\ scalar sum, \htt, is also very high.

Starting from the NMSSM~\cite{NMSSM} scenarios proposed in~\cite{UlrichAna}, we consider the case where the Next-to-LSP (NLSP) decays into a LSP plus a SM-like Higgs boson, $H$. In a scenario where the LSP is very soft in this decay the \met\ is reduced considerably. An example of such a scenario would be for a very light LSP where the NLSP mass, \mnlsp, is just slightly above \mlsp\ +$M_H$,
since the much heavier Higgs boson would inherit most of the momentum from the NLSP\@. However, in the MSSM, most of the heavier sparticles would prefer to skip this NLSP decay step and decay straight into the LSP instead, thus still generating high \met.

\looseness=1
If, in contrast, the LSP were the singlino of the NMSSM~\cite{Ellwanger}, having only weak couplings to sparticles, then the heavier sparticles will decay into a bino-like NLSP, which plays the role of the MSSM-like LSP\@. This NLSP then decays into a soft, ``true'' LSP and a Higgs boson, thus allowing for these low-\met\ scenarios. In this SUSY scenario, the singlino is the fermionic counterpart of a singlet superfield $\hat{S}$, where $S$ is the gauge singlet field to which the coupling of the two Higgs doublets of the MSSM solves the $\mu$-problem~\cite{KimNilles}.

Furthermore, in the case where we have \mnlsp\ not much smaller than \msq\ or \mgl\ but still close to \mlsp\ + $125$\,GeV/$c^{2}$, the now heavy LSP will gain fairly little extra momentum compared to the initially produced sparticles, even though it will inherit more of this than the now (in comparison) lighter Higgs boson. Additionally, the small mass gaps in the decay cascade will mean low-\ptt\ jets, implying events with low \htt\ as well as low \met, in contrast to the typical signature of a more minimal SUSY model.

As discussed in~\cite{UlrichAna}, these NMSSM scenarios with low \met\ generally have weaker constraints imposed by experimental searches, however, the focus therein was mainly on Run I of the LHC, with all sparticle masses in the $1 - 1.5$\,TeV/$c^{2}$ range. In order to gain understanding of the sensitivity to these models which might be attained with current search efforts in Run II, the CMS $\alpha_{\text{T}}$-based general-purpose BSM analysis in~\cite{CMS-SUS-16} is examined, which focuses on an all-hadronic final state and uses $35.9$\,fb$^{-1}$ of CMS data from Run II of the LHC at a centre-of-mass energy of $\sqrt{s} = 13$\,TeV.

A similar quantity to \met\ which is frequently used in searches involving hadronic final states is missing-$H_{\text{T}}$, denoted \mht. \mht\ is defined as the norm of the two-dimensional vector sum of the transverse momenta of the \emph{hadronic jets} in an event within various acceptance regions, such as \ptt, $\eta$ and various isolation requirements. On the other hand \met\ includes all objects within these acceptance regions, such as leptons. Following the approach of \mht\ throughout~\cite{CMS-SUS-16}, this will be used in lieu of \met\  throughout this paper, noting that these quantities should be essentially equivalent in the presence of a sufficiently tight veto on events containing final state leptons.

Starting from the eight Benchmark Points (BPs), denoted BP1, \ldots, BP8 and presented in table~\ref{ulrich_ana_table}, taken from~\cite{UlrichAna}, parameter scans are developed which envelop these points in order to determine the current sensitivity of this CMS analysis. Monte Carlo (MC) simulated events are then generated at parton-level, with the decays, hadronisation and detector response calculated for each point in these scans.

\begin{table}
\center
	\begin{tabular}{ |c|c|c|c| }
	\hline
	Point & $M_{\tilde{q}}$ [GeV/$c^{2}$] & $M_{\tilde{g}}$ [GeV/$c^{2}$] & $M_{\tilde{t},\tilde{b}}$ [GeV/$c^{2}$]\\
 	\hline
 	BP1 & $1000$ & $1010$ & decoupled\\
 	BP2 & $1400$ & $1410$ & decoupled\\
 	BP3 & $1100$ & $900$ & decoupled\\
 	BP4 & $1500$ & $1300$ & decoupled\\
 	BP5 & $1400$ & $1410$ & $M_{\tilde{t}} = 750$\\
 	BP6 & $1100$ & $1110$ & $M_{\tilde{b}} = 750$\\
 	BP7 & $1500$ & $1300$ & $M_{\tilde{t}} = 750$\\
 	BP8 & $1400$ & $1200$ & $M_{\tilde{b}} = 750$\\
 	\hline
 	\end{tabular}
 	\caption{Original BPs in~\cite{UlrichAna}.}
 	\label{ulrich_ana_table}
\end{table}

\section{Testing scenarios with low {\boldmath $E_{\text{T}}^{\text{miss}}$}{E(T)**(miss)}\label{testing}}

Current searches for SUSY have yet to find substantial evidence, increasing the lower bounds on sparticle masses for simplified models. However, for certain regions of parameter space, it is possible that a supersymmetric decay cascade may in fact give a very low \met\ signal whilst still having a LSP mass of only a few GeV/$c^{2}$, weakening the lower bounds on the masses of the squarks and  gluino.

\begin{figure}
\centering
  \includegraphics[keepaspectratio=true,width=90mm]{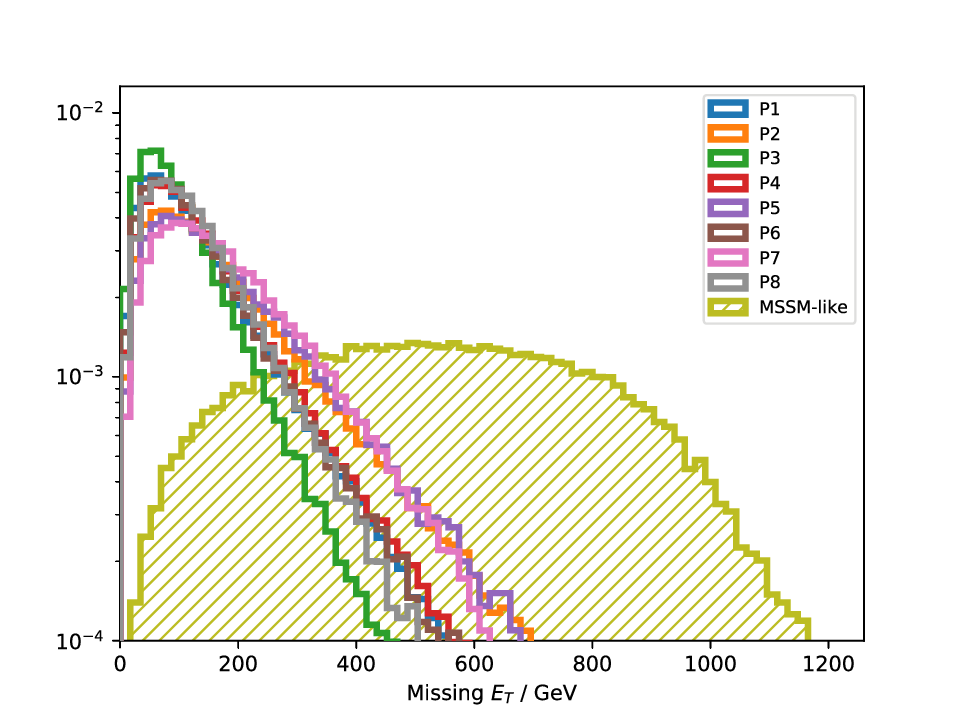}
\caption{\met\ distribution for the eight benchmark points defined in~\cite{UlrichAna} along with an MSSM-like simplified scenario. \label{P1_MET}}
\end{figure}

For example, the \met\ distributions are shown in figure~\ref{P1_MET} for each of the eight BPs in table~\ref{ulrich_ana_table}, featuring a LSP mass of $3$\,GeV/$c^{2}$ with the squark and gluino masses around 1\,TeV/$c^{2}$, along with a simplified MSSM-like scenario which exhibits typically larger values of \met. This MSSM-like scenario features the same sparticle masses as BP1, such as $1$\,TeV/$c^{2}$ squarks and a $3$\,GeV/$c^{2}$ LSP.

Here, it is clear that the mean \met\ is quite low for these NMSSM scenarios, more akin to that from SM  processes such as top quark pair production. In turn, this suggests current experimental searches concentrating on a hadronic jets plus \met\ final state will likely not be optimally tuned to this type of SUSY signature.

In~\cite{UlrichAna} the Higgs boson in question is SM-like with a mass of around $125$\,GeV/$c^{2}$ and the final state under consideration contains two bottom quark jets and two $\tau$ leptons, produced when one of the two  Higgs bosons decays to a bottom quark-antiquark pair and the other to a $\tau$ lepton-antilepton pair. This was chosen in order to suppress SM background processes compared with an all-hadronic final state such as that where both Higgs bosons decay into a bottom quark-antiquark pair.

However, since the analysis considered in~\cite{CMS-SUS-16} concentrates specifically on all-hadronic final states, this paper will focus on the scenario where each Higgs boson decays into a bottom quark-antiquark pair, with no $\tau$ leptons produced, thus seeking a $4b$+\met\ final state. This retains the maximum proportion of signal events, since the largest branching fraction for the Higgs boson decay is that to a bottom quark-antiquark pair. Likewise, the contribution towards the background from SM events involving large numbers of hadronic jets formed through Quantum Chromo-Dynamics (QCD) interactions, i.e., multijet events, will also increase.

Thus denoting this SM-like Higgs boson as $H$, our final state decays will appear as
\begin{align}
	\text{NLSP} &\rightarrow H + \text{LSP}\notag \\
	H &\rightarrow b\bar{b} \text{ (jets)}. \label{finaldecay}
\end{align}

\subsection{Other factors affecting signal processes}

Whilst the masses of the neutralinos and  Higgs boson will dictate the behaviour of some of the kinematical variables, these have no real effect on the initial sparticle production cross-section, which is to a large extent dependent only upon the masses of the squarks and  gluino.

As the squark and gluino masses increase, so does the mass gap between these and the NLSP. This in turn will lead to high-\ptt \hspace{1mm}jets from the SUSY decay cascade as well as increased NLSP momentum, resulting in higher \htt \hspace{1mm}and larger Higgs boson \ptt.

\section{Simulation techniques}

Firstly, the mass spectra are chosen in order to exploit these light-LSP, low-\met\ scenarios. In order to generate simulated events for processes calculated at matrix element level, MadGraph~\cite{madgraph}, which has built-in support for NMSSM processes at Leading Order (LO), is used. However MadGraph cannot generate NMSSM events at Next-to-LO (NLO), therefore, the cross-section value stated by MadGraph is not used. Instead, squark and gluino production cross-sections are calculated (inclusively) using Prospino~\cite{prospino} at~NLO.

The initial squarks, antisquarks and gluinos, plus up to two additional hard jets, are generated by MadGraph at LO, after which Pythia8~\cite{Pythia} decays the particles according to their respective Branching Ratios (BRs). A fast simulation of the CMS detector is then performed using Delphes~\cite{DELPHES}.

Taking, for example the, first two BPs in~\cite{UlrichAna}, denoted BP1 and BP2 in table~\ref{ulrich_ana_table}, we see that gluinos are in both cases around $10$\,GeV/$c^{2}$ heavier than squarks. The gluinos thus decay first into squarks, with each squark decaying into a NLSP and a correspondingly flavoured quark: $\tilde{g}\rightarrow\tilde{q}+q$; $\tilde{q}\rightarrow\tilde{\chi}_{2}^{0}+q$.

BP3 and BP4 differ in that the gluino is $200$\,GeV/$c^{2}$ lighter than the squarks. In these scenarios the left-handed squarks always decay into the gluino and a correspondingly flavoured quark, whilst the right-handed squarks decay either into a gluino-quark combination or skip this step entirely and decay directly into a NLSP and a quark, with corresponding BRs of 70\% and 30\%, respectively, with these fractions close to those estimated by NMSDECAY~\cite{NMSDCAY1, NMSDECAY2}.

For BP5 and BP6 the squarks are lighter than the gluino, as in points BP1 and BP2. However, the respective stop/sbottom-type squark is now lighter than the gluino, sufficiently so that gluino two-body decays are possible. In these two BPs the gluino is assumed to always decay into a stop (BP5) or sbottom (BP6) squark and corresponding top/bottom quark, which in turn decays into a NLSP and correspondingly flavoured quark. The first- and
second-generation squarks decay with 100\% BR into the NLSP and corresponding quark as in  BP1 and BP2.

BP7 and BP8 also involve the stop (BP7) or sbottom (BP8) squark, but with the gluino lighter than the first two generation squarks, as in BP3 and BP4. Much like in BP3 and BP4, the left-handed squarks always decay into a gluino and a correspondingly flavoured quark, whilst the right-handed squark decays with 70\% and 30\% BR into either a gluino and a quark or directly into a NLSP and quark, respectively. In both of these points the gluino always decays into the respective stop/sbottom squark and top/bottom quark, with the third generation squark decaying furthermore into a NLSP and corresponding top/bottom quark.

\looseness=-1
Figure~\ref{feynman} shows example decay cascades for each of these BPs. BP2 and BP4 have been omitted, since the possible diagrams do not differ from those for BP1 and BP3, however, extra diagrams are included for BP3 and BP7 to illustrate the possible routes by which the right-handed squarks may decay. Additionally, the alternative BP7 decay chain may apply to  BP8  by simply switching each of the stop/top for the corresponding sbottom/bottom squark/quark.

\begin{figure}[p]
\centering
	\begin{subfigure}[t]{0.45\textwidth}
		\centering
  		\includegraphics[keepaspectratio=true,width=\columnwidth]{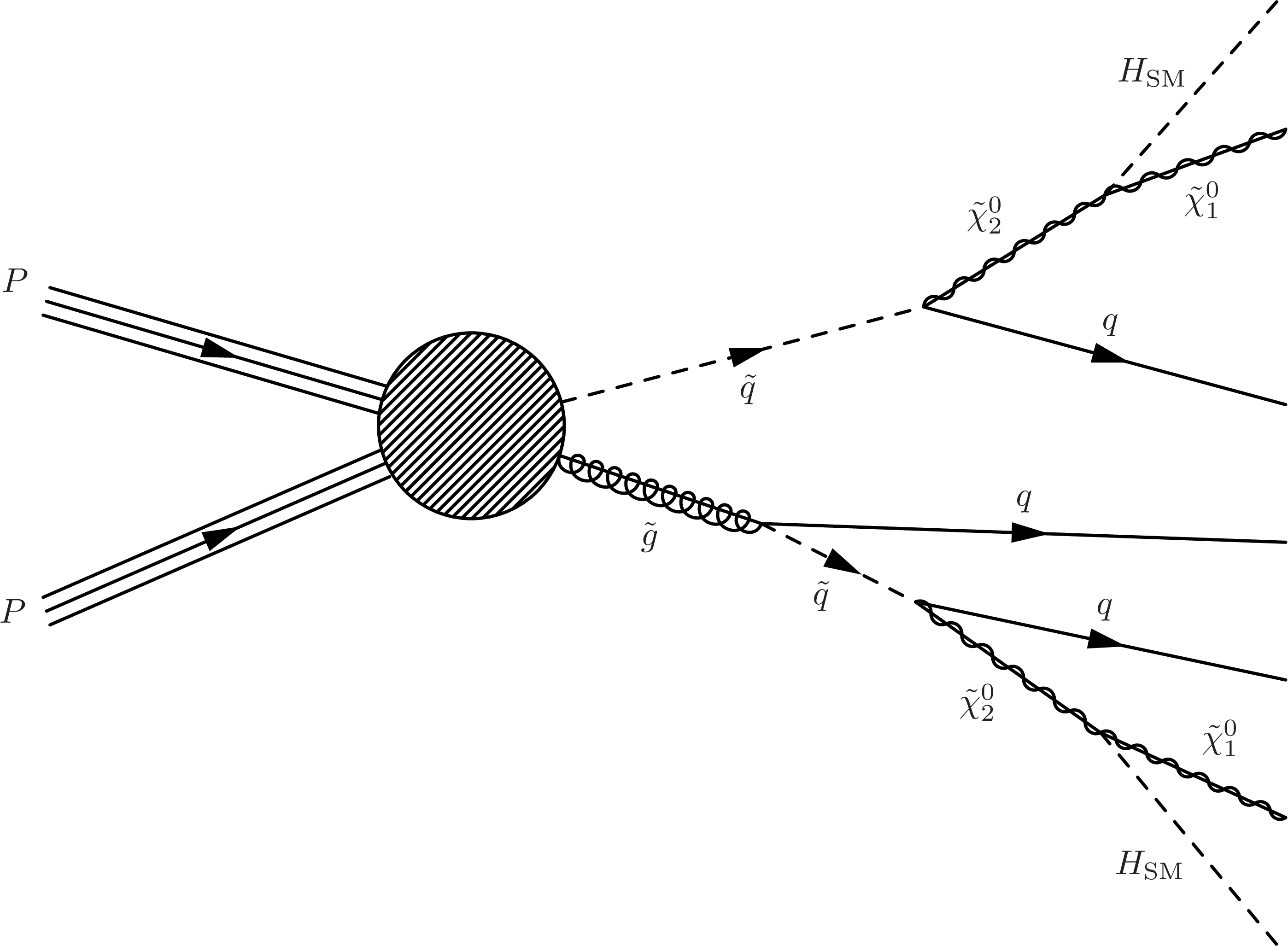}
  		\caption{BP1 possible decay cascade.\label{P1_diagram}}
	\end{subfigure}\quad%
	\begin{subfigure}[t]{0.45\textwidth}
		\centering
  		\includegraphics[keepaspectratio=true,width=\columnwidth]{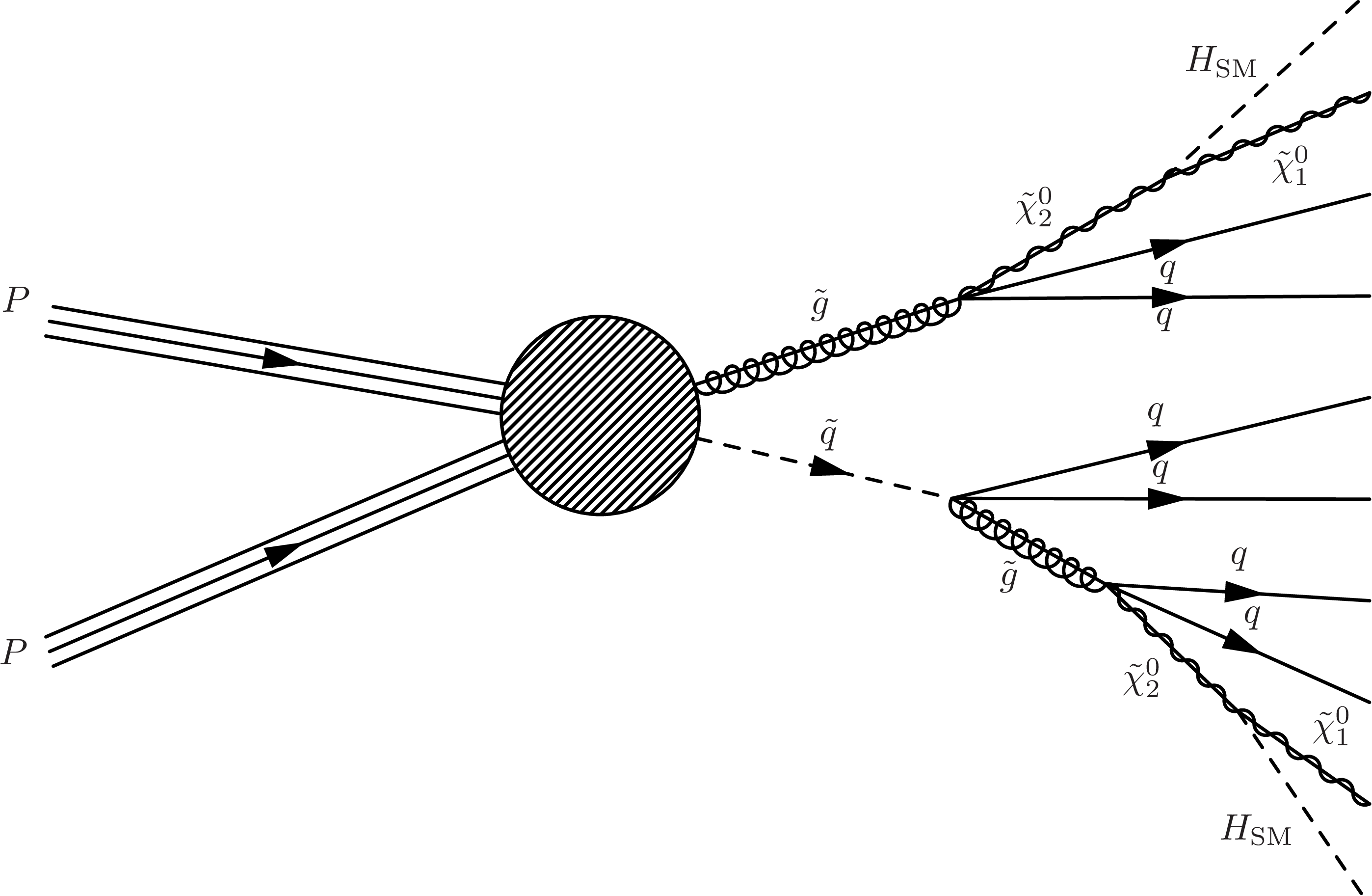}
  		\caption{BP3 possible decay cascade.\label{P3_diagram}}
	\end{subfigure}

\bigskip

\bigskip

	\begin{subfigure}[t]{0.45\textwidth}
		\centering
  		\includegraphics[keepaspectratio=true,width=\columnwidth]{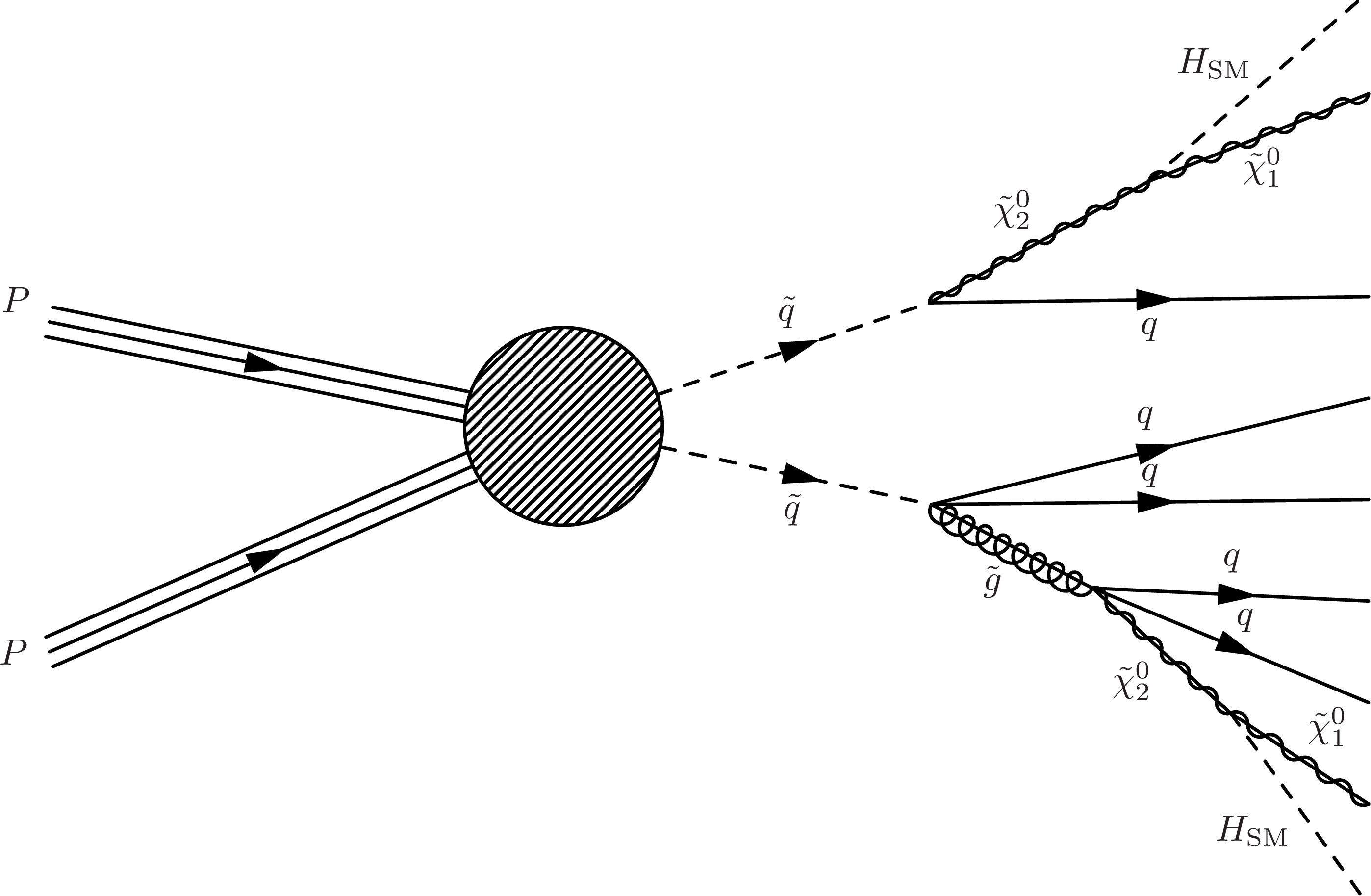}
  		\caption{BP3 alternative decay cascade.\label{P3_diagram_alt}}
	\end{subfigure}\quad%
	\begin{subfigure}[t]{0.45\textwidth}
		\centering
  		\includegraphics[keepaspectratio=true,width=\columnwidth]{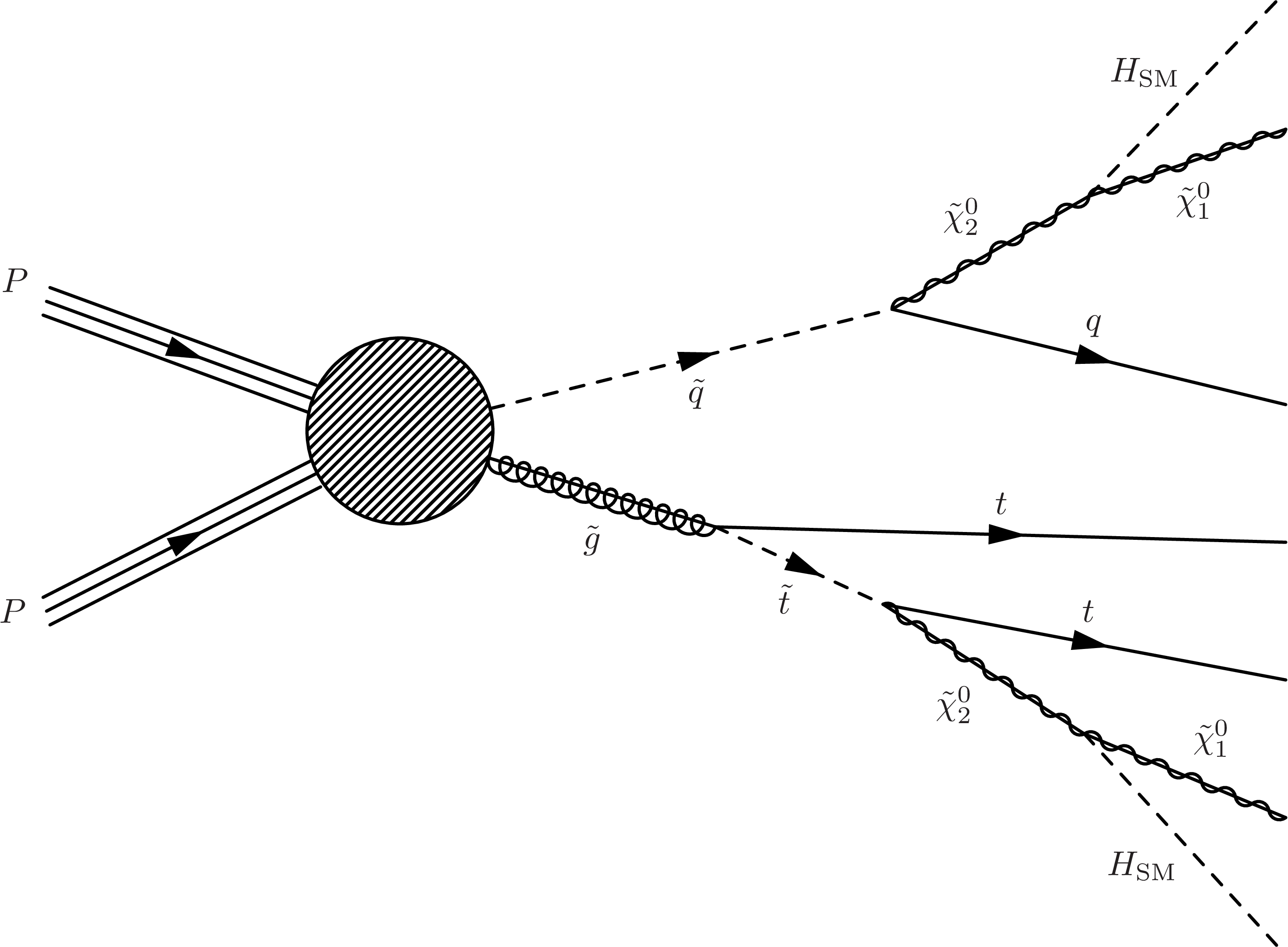}
  		\caption{BP5 possible decay cascade.\label{P5_diagram}}
	\end{subfigure}

\bigskip

\bigskip

	\begin{subfigure}[t]{0.45\textwidth}
		\centering
  		\includegraphics[keepaspectratio=true,width=\columnwidth]{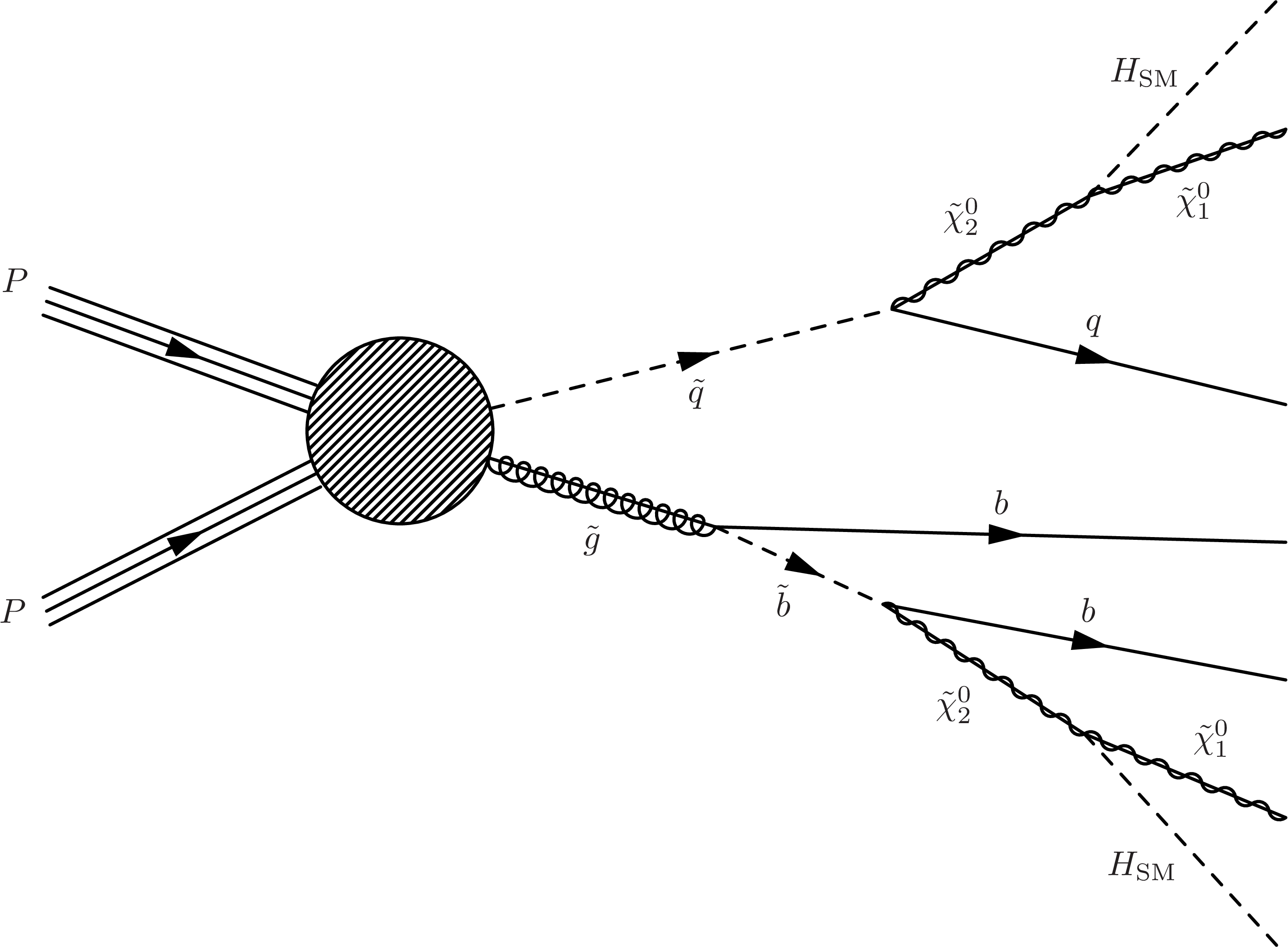}
  		\caption{BP6 possible decay cascade.\label{P6_diagram}}
	\end{subfigure}\quad%
	\begin{subfigure}[t]{0.45\textwidth}
		\centering
  		\includegraphics[keepaspectratio=true,width=\columnwidth]{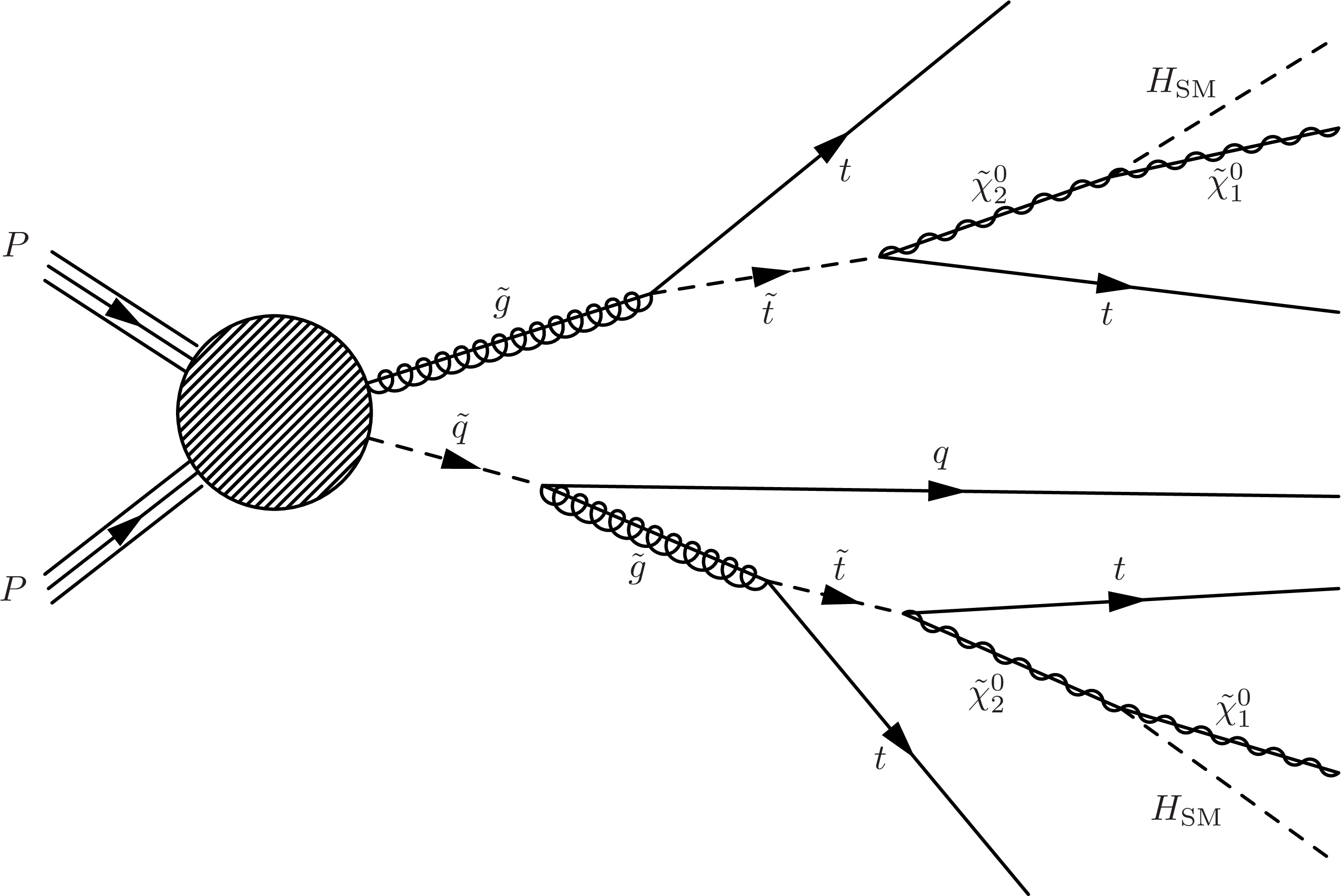}
  		\caption{BP7 possible decay cascade.\label{P7_diagram}}
	\end{subfigure}
\end{figure}

The Feynman diagrams in figure~\ref{feynman} show examples of the processes by which we may produce a final state with two LSPs and two Higgs bosons. In each of these diagrams we produce one squark and one gluino directly, with each decaying via an example cascade, however, both squark-squark and gluino-gluino pair production processes are included additionally in the event generation. These processes may also involve extra hadronic jets produced along with the initial sparticles or typically softer radiated jets.

\begin{figure}[ht]\ContinuedFloat
\centering
	\begin{subfigure}[t]{0.45\textwidth}
		\centering
  		\includegraphics[keepaspectratio=true,width=\columnwidth]{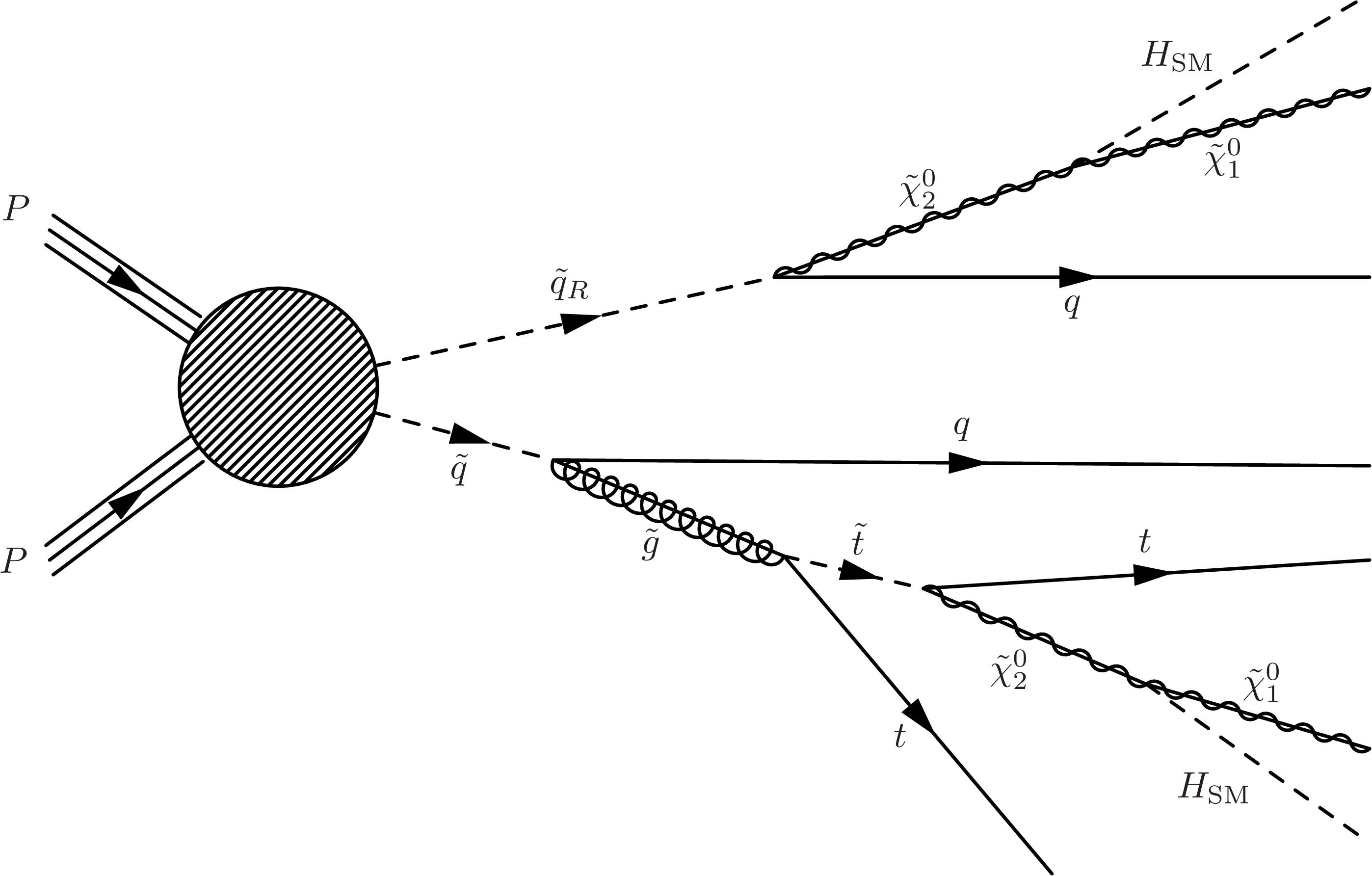}
  		\caption{BP7 alternative decay cascade.\label{P7_diagram_alt}}
	\end{subfigure}\quad%
	\begin{subfigure}[t]{0.45\textwidth}
		\centering
  		\includegraphics[keepaspectratio=true,width=\columnwidth]{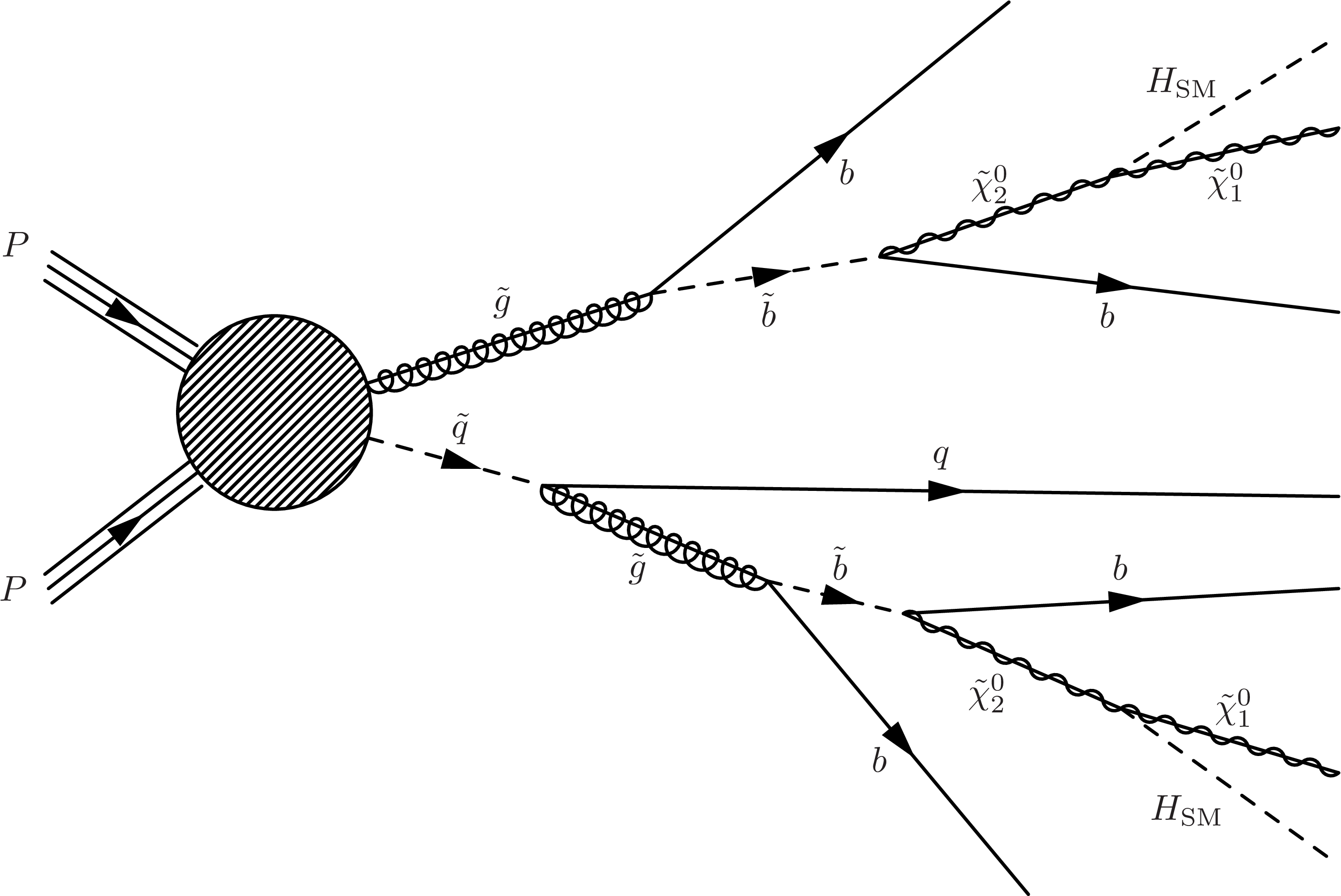}
  		\caption{BP8 possible decay cascade.\label{P8_diagram}}
	\end{subfigure}
\caption{Feynman diagrams showing example processes by which we may produce two singlino LSP along with two Standard Model-like Higgs bosons. An example diagram is given for each of the BPs in~\cite{UlrichAna} which contain unique mass hierarchies. Here, BP1, BP3, BP5, BP6, BP7 and BP8 are shown in (a), (b,c), (d),  (e), (f,g) and (h),   respectively.\label{feynman}}
\end{figure}

\begin{figure}[t]
\centering
	\begin{subfigure}[t]{0.45\textwidth}
		\centering
  		\includegraphics[keepaspectratio=true,width=60mm]{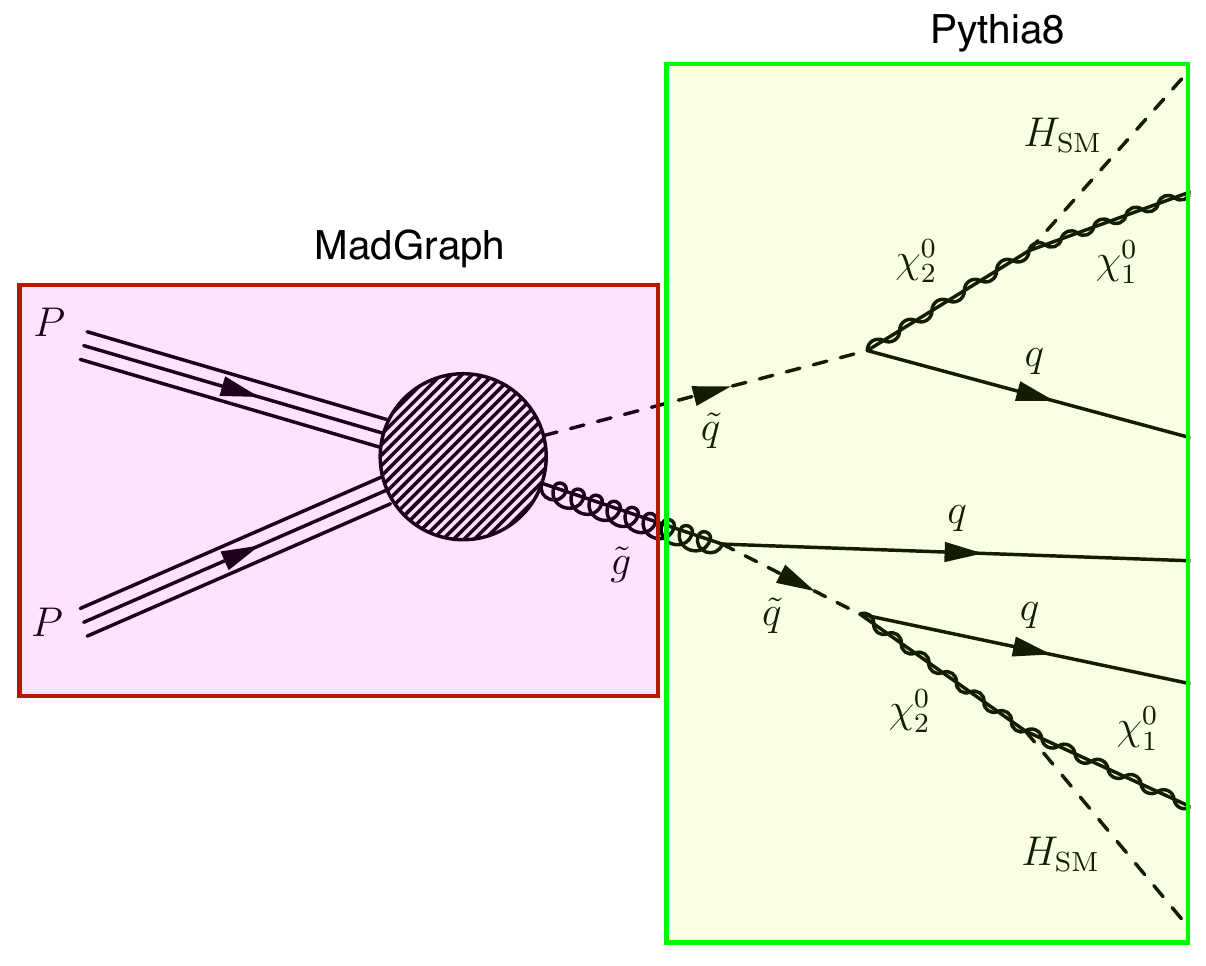}
  		\caption{Gluino decays in Pythia. \label{feynmandiagram2}}
	\end{subfigure}\quad%
	\begin{subfigure}[t]{0.45\textwidth}
		\centering
  		\includegraphics[keepaspectratio=true,width=62mm]{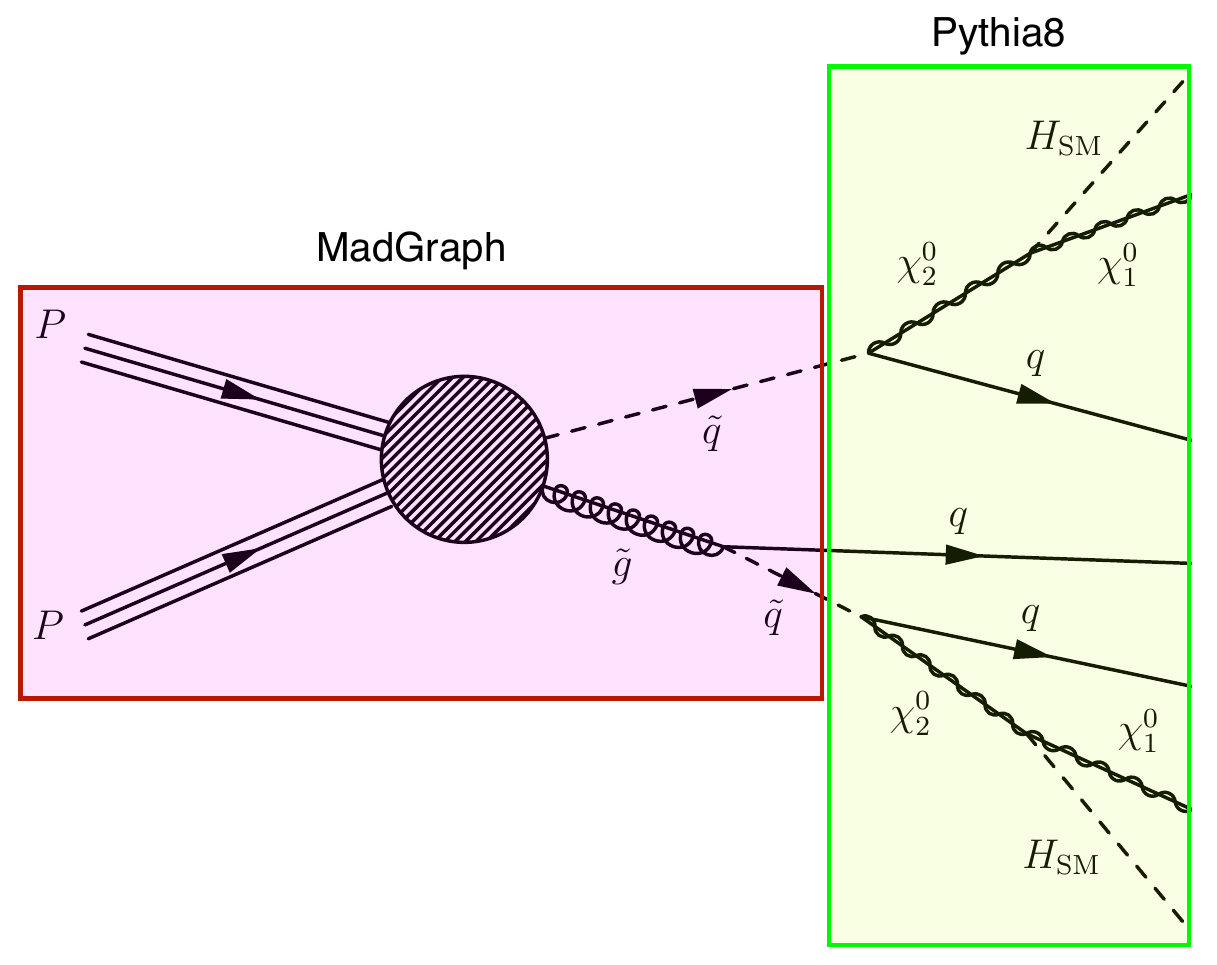}
  		\caption{Gluino decays in MadGraph. \label{feynmandiagram}}
	\end{subfigure}
\caption{MadGraph event generation of two squarks and one jet (left) and one squark and one gluino (right), sharing the same Feynman diagram after Pythia has performed the gluino decay.\label{doublecounting}}
\end{figure}

Considering  the diagram in figure~\ref{P1_diagram}, the initial gluino and squark may be produced at parton-level in MadGraph, with the second squark produced by Pythia from the decay of the gluino, shown in figure~\ref{feynmandiagram2}. However, if one were to direct MadGraph to produce two squarks, allowing the production of extra jets, some of the subprocesses considered would involve a squark stemming from the decay of a gluino, all of which would be calculated at parton-level, as in figure~\ref{feynmandiagram}. Thus there is an overlap between the events generated by MadGraph when asked for two squarks and those generated when producing gluinos, which are in turn decayed by Pythia into squarks.

In order to remove the possibility of subprocesses being counted more than once in the overall calculation, it is required that any squark or gluino whose decay is performed at parton-level must be off-shell. Since any squarks or gluinos whose decay is performed by Pythia will have been treated as \emph{final state} particles by MadGraph, they must be on-shell. Thus, by summing the complementary on- and off-shell terms, the entire momentum space over which the squarks and gluinos may decay is obtained.

\section{Mass scans and event selections}

\subsection{Two-dimensional mass scans}

Considering each of the existing BPs in~\cite{UlrichAna}, where \mnlsp$=130$\,GeV/$c^{2}$ and \mlsp$=3$\,GeV/$c^{2}$, a two-dimensional mass scan is constructed. The mass gaps $M_{\tilde{q}} - M_{\tilde{g}}$ and $M_{\tilde{\chi}_{2}^{0}} - M_{\tilde{\chi}_{1}^{0}}$ are kept constant, with $M_{\tilde{q},\tilde{g}}$ and $M_{\tilde{\chi}_{1}^{0}, \tilde{\chi}_{2}^{0}}$ now treated as two independent parameters.

However, it may be noted that BP1 and BP2 are essentially the same, but for $M_{\tilde{q},\tilde{g}}$ being $400$\,GeV/$c^{2}$ heavier in the latter. Therefore, one would find that a mass scan about BP1 would encapsulate BP2 anyway. This is also the case for BP3 with respect to BP4, however, BP5 to BP8 transform into four independent scans, since for BP5 and BP7 the stop squark is involved in the decay cascades but not the sbottom squark. For BP6 and BP8 the converse~applies.

Whilst it is expected that the sensitivity to these NMSSM scenarios be lowest for the lightest LSP, due to the suppressed \mht\ distribution, the masses of the NLSP and LSP in each scan are increased as high as is possible whilst remaining below the mass of the lighter of the squark and gluino. This is done in order to cover the entirety of the available sparticle mass range, such that it is possible to compare the same NMSSM scenario with a range of LSP masses. Thus, the six independent mass scans are defined as in table~\ref{scan_table}.

\begin{table}
\centering\setlength{\tabcolsep}{4pt}

	\resizebox*{1\textwidth}{!}{\begin{tabular}{ |c|c|c|c|c|c| }
	\hline
	& $M_{\tilde{q}}$ [GeV/$c^{2}$] & $M_{\tilde{g}}$ [GeV/$c^{2}$] & $M_{\tilde{\chi}_{1}^{0}}$ [GeV/$c^{2}$] & $M_{\tilde{\chi}_{2}^{0}}$ [GeV/$c^{2}$] & $M_{\tilde{t},\tilde{b}}$ [GeV/$c^{2}$]\\
 	\hline
 	BP1/BP2 & $1200\rightarrow 3000$ & $M_{\tilde{q}} + 10$ & $3\rightarrow \{M_{\tilde{q}} - 20\}$ & $M_{\tilde{\chi}_{1}^{0}} + 127$ & decoupled\\
 	BP3/BP4 & $1200\rightarrow 3000$ & $M_{\tilde{q}} - 200$ & $3\rightarrow \{M_{\tilde{g}} - 20\}$ & $M_{\tilde{\chi}_{1}^{0}} + 127$ & decoupled\\
 	BP5 & $1200\rightarrow 3000$ & $M_{\tilde{q}} + 10$ & $3\rightarrow \{M_{\tilde{t}} - 200\}$ & $M_{\tilde{\chi}_{1}^{0}} + 127$ & $M_{\tilde{t}} = M_{\tilde{q}} - 250$\\
 	BP6 & $1200\rightarrow 3000$ & $M_{\tilde{q}} + 10$ & $3\rightarrow \{M_{\tilde{b}} - 20\}$ & $M_{\tilde{\chi}_{1}^{0}} + 127$ & $M_{\tilde{b}} = M_{\tilde{q}} - 250$\\
 	BP7 & $1200\rightarrow 3000$ & $M_{\tilde{q}} - 200$ & $3\rightarrow \{M_{\tilde{t}} - 200\}$ & $M_{\tilde{\chi}_{1}^{0}} + 127$ & $M_{\tilde{t}} = M_{\tilde{g}} - 250$\\
 	BP8 & $1200\rightarrow 3000$ & $M_{\tilde{q}} - 200$ & $3\rightarrow \{M_{\tilde{b}} - 20\}$ & $M_{\tilde{\chi}_{1}^{0}} + 127$ & $M_{\tilde{b}} = M_{\tilde{g}} - 250$\\
 	\hline
 	\end{tabular}}\relax
 	\caption{Table showing various mass ranges in the scans.}
 	\label{scan_table}
\end{table}

For the first two scans the NLSP mass is increased up to just below the lighter of the squark and gluino masses whilst still allowing for on-shell decay. For the remaining four scans, where the respective stop or sbottom squark is non-decoupled, its mass is set to be $250$\,GeV/$c^{2}$ lower than the lightest of the squark and gluino, such that the gluino may still decay into the relevant third generation squark, along with an appropriate quark. In these cases, the NLSP mass may still be increased, so long that the involved third generation squark may still decay in an on-shell fashion into its respectively flavoured quark and a~NLSP.

\subsection{Event selections}

The considered experimental analysis~\cite{CMS-SUS-16} contains many measurement bins for various observables in the data, background and signal channels, in particular for the number of hadronic jets ($N_{\text{jets}}$), the number of $b$-tagged hadronic jets ($N_{b\text{-jets}}$), \htt\ and \mht. Focusing on a $b\bar{b}b\bar{b} + $\met\ final state, with plenty of jets from both cascades, we consider the bin with the highest number of jets, i.e.,  the one for which $N_{\text{jets}} \geq 6$. Anticipating a high chance of bottom quarks being mis-identified, especially if the Higgs bosons are boosted, we consider both the $N_{b\text{-jets}} = 2$, $N_{b\text{-jets}} = 3$ and $N_{b\text{-jets}} \geq 4$ bins. Additionally, since we are primarily interested in topologies generating high \htt, we focus on the upper-most \htt\ $> 1200$\,GeV/$c$ bin, with the exception of the case where we have four or more $b$-tagged hadronic jets, where the only bin in~\cite{CMS-SUS-16} is \htt\ $> 400$\,GeV/$c$.

Furthermore, in this analysis, a cut on `biased Delta-phi' ($\Delta\phi^{\star}$) is applied to reduce the QCD background. Additional cuts in~\cite{CMS-SUS-16} regarding vetoing events containing isolated leptons and photons are also performed, as well as failing events which contain forward/backward-oriented hadronic jets. The event selection is therefore detailed as~\mbox{follows}:
\begin{itemize}
	\item At least 6 hadronic jets, where any jet must have \ptt $\geq 40$\,GeV/$c$.
	\item $N_{b\text{-jets}}=2$, $N_{b\text{-jets}}= 3$, $N_{b\text{-jets}}\geq 4$, i.e.,  separate bins.
	\item $N_{b\text{-jets}}$, $H_{\text{T}}$\ and \mht \hspace{1.5mm}binning defined in table~\ref{yield_table}.
	\item $H_{\text{T}} \geq 1200$\,GeV/$c$ for events where $N_{b\text{-jets}} \leq 3$, or $H_{\text{T}} \geq 400$\,GeV/$c$ where $N_{b\text{-jets}} \geq 4$.
	\item $\Delta\phi^{\star} \geq 0.5$.
\end{itemize}

In addition, event vetoes are defined such that events will fail should they contain any of the following isolated objects:
	\begin{itemize}
		\item photons with \ptt\hspace{1mm}$> 20$\,GeV/$c$ and $|\eta| < 2.5$,
		\item electrons and muons with \ptt\hspace{1mm}$> 10$\,GeV/$c$ and $|\eta| < 2.5$,
		\item jets with \ptt\hspace{1mm}$> 40$\,GeV/$c$ and $|\eta| > 2.4$,
	\end{itemize}
	where isolation is defined as being separated from other objects by angular distance $\Delta R > 0.3$ for photons or $\Delta R > 0.4$ for electrons, muons and jets.

Finally an \htt-dependent cut on $\alpha_{\text{T}}$ is implemented as described in table~\ref{alpha_T_table}, noting that for all but the $N_{b\text{-jets}} \geq 4$ bin there is no $\alpha_{\text{T}}$ cut due to the \htt\ requirement.

As defined in~\cite{alphaT}, in the case where an event contains just two hadronic jets, $\alpha_{\text{T}}$ is calculated as follows:
\begin{align*}
	\alpha_{T} = \frac{E^{j2}_{T}}{M_{T}},
\end{align*}
where $E^{j2}_{T}$ is the transverse energy of the second-leading hadronic jet and
\begin{align*}
M_{T} = \sqrt{\left(\sum_{i=1}^{2}E_{T}^{j_{i}}\right)^{2} - \left(\sum_{i=1}^{2}p_{x}^{j_{i}}\right)^{2} - \left(\sum_{i=1}^{2}p_{y}^{j_{i}}\right)^{2}}.
\end{align*}
For events with more than two hadronic jets, the latter are combined to create \emph{pseudo-jets} in such a way that the difference in \ptt\ between these two pseudo-jets is minimised~\cite{alphaT2}.

\subsection{Signal, background and data event yields}

Table~\ref{alpha_T_table} contains the data and background yields from~\cite{CMS-SUS-16} for each of the $N_{b\text{-jets}}$ and \mht\ bins, satisfying the remaining event selection criteria.
\sisetup{
table-number-alignment=center,
separate-uncertainty=true,
table-figures-integer = 1,
table-figures-decimal = 5}
\begin{table}
\center
	\begin{tabular}{ |c|cr@{--}l|cr@{--}l|c|c S[separate-uncertainty,table-figures-uncertainty=1]| }
	\hline
	\hspace{3mm}Bin\hspace{3mm} & \hspace{2mm} & \multicolumn{2}{c}{$H_T$ [GeV/$c$]}\hspace{1mm} \vline & \hspace{2mm} & \multicolumn{2}{c}{{$H_T^{\rm miss}$ [GeV/$c$]}}\hspace{1mm} \vline & Data Yield &\  & {Background Yield}\hspace{2mm}\\
 	\hline
 	$=$ 2b & & 1200 & $\infty$ & & 200 & 400 & 0 & & 2.51 \pm 1.02\\
 	$=$ 2b & & 1200 & $\infty$ & & 400 & 600 & 0 & & 1.65 \pm 0.44\\
 	$=$ 2b & & 1200 & $\infty$ & & 600 & 900 & 2 & & 0.62 \pm 0.32\\
 	$=$ 2b & & 1200 & $\infty$ & & 900 & $\infty$ & 0 & & 0.19 \pm 0.18\\
 	$=$ 3b & & 1200 & $\infty$ & & 200 & 400 & 1 & & 0.40 \pm 0.16\\
 	$=$ 3b & & 1200 & $\infty$ & & 400 & 600 & 0 & & 0.25 \pm 0.08\\
 	$=$ 3b & & 1200 & $\infty$ & & 600 & 900 & 1 & & 0.09 \pm 0.04\\
 	$=$ 3b & & 1200 & $\infty$ & & 900 & $\infty$ & 0 & & 0.02 \pm 0.02\\
 	$\geq$ 4b & & 400 & $\infty$ & & 200 & $\infty$ & 4 & & 2.46 \pm 0.70\\
	\hline
 	\end{tabular}
 	\caption{Data and background yields for the bins used in this analysis, taken from~\cite{CMS-SUS-16}.}
 	\label{yield_table}
\end{table}

\begin{table}
\center
	\begin{tabular}{ |c|c|c|c|c|c|c| }
	\hline
	\htt\ [GeV/$c$] & $200-250$ & $250-300$ & $300-350$ & $350-400$ & $400-900$ & $900-\infty$ \\
 	\hline
 	$\alpha_{\text{T}}$ & $> 0.65$ & $> 0.6$ & $> 0.55$ & $> 0.53$ & $> 0.52$ & $> 0$\\
 	\hline
 	\end{tabular}
 	\caption{Table detailing the \htt-dependent $\alpha_{\text{T}}$ cuts.}
 	\label{alpha_T_table}
\end{table}

These data yields and background estimations are then used to calculate lower bounds on the sparticle masses given the signal yields for each mass point in each mass scan, with the systematic uncertainty on the signal yields assumed to be $25$\%.

\subsection{Validation of cut and count analysis tools}

In order to check that one may rely on the signal event yields calculated by the software used to implement the event selection and indeed that the estimation of the systematic uncertainty is appropriate, it is important to compare these yields with those in~\cite{CMS-SUS-16}. However of course the experimental analysis in question does not feature an NMSSM low-\met\ scenario such as those under consideration in this work, so a reference benchmark model is chosen, $T1bbbb$~\cite{T1bbbb}.

\begin{figure}\centering
  \includegraphics[keepaspectratio=true,width=90mm]{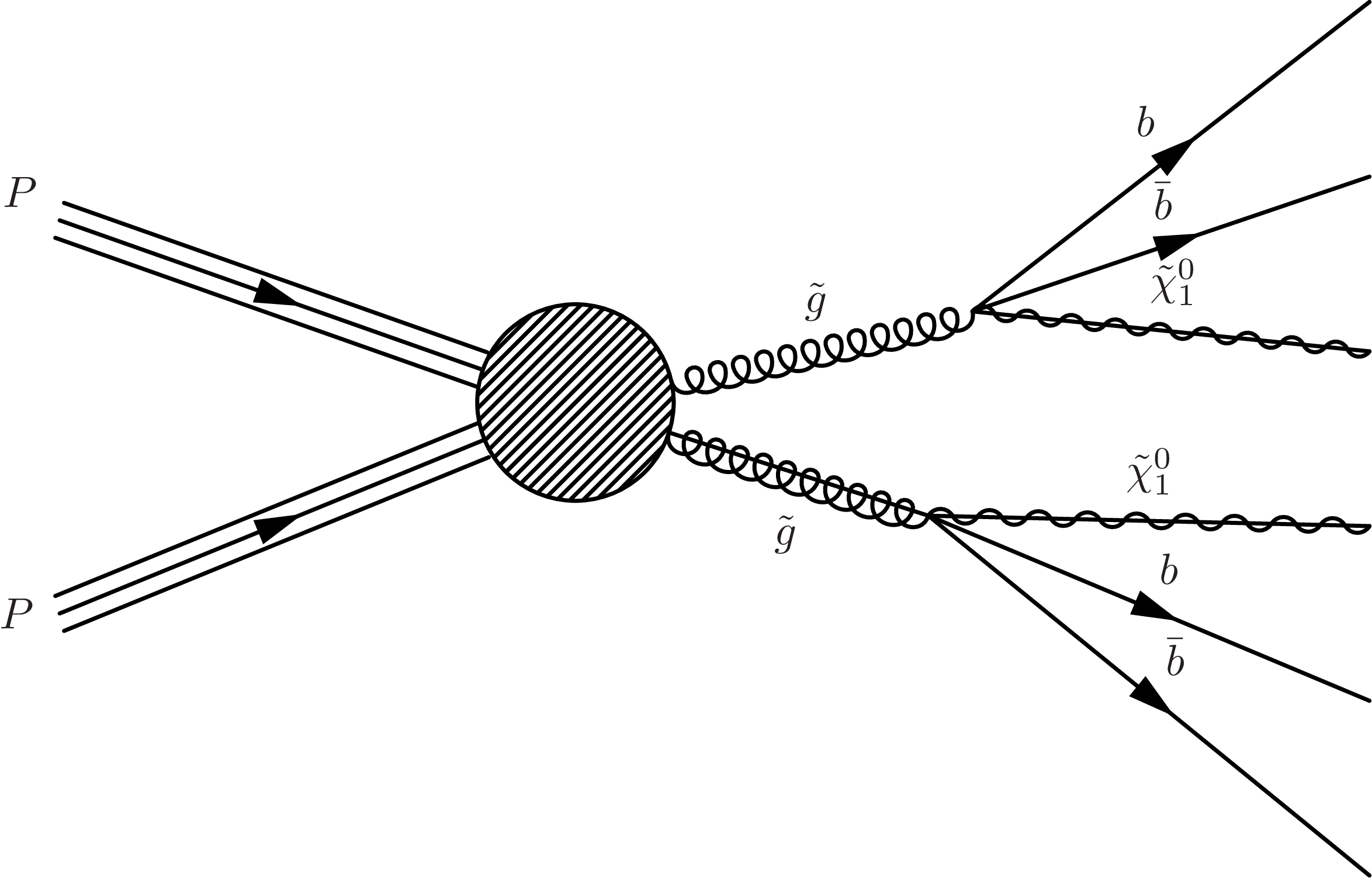}
\caption{Feynman diagram showing gluino pair production and decay in
  the $T1bbbb$ benchmark~model. \label{T1bbbb}}
\end{figure}

$T1bbbb$ is a simplified Supersymmetric model whereby pair produced gluinos each undergo a three-body decay into a bottom quark-antiquark pair and an LSP neutralino, shown in figure~\ref{T1bbbb}. In this example the gluino has a mass of $1900$\,GeV/$c^{2}$ and the LSP $100$\,GeV/$c^{2}$. In addition to the pair produced gluinos, up to two hard jets are considered at parton level in the event generation, as was done with the NMSSM signal mass points.

\begin{table}
\center
	\begin{tabular}{ |l|c|c| }
	\hline
    \multirow{3}{*}{Event Selection} & \multicolumn{2}{|c|}{Benchmark Model}\\
    \multirow{3}{*}{} & \multicolumn{2}{|c|}{{T1$bbbb$: $M_{\tilde{q}} = 1900$\,GeV, $M_{\chi^{0}_{1}} = 100$\,GeV}}\\
    \cline{2-3}
	\multirow{3}{*}{} & Efficiencies from~\cite{CMS-SUS-16} & Delphes \& own software\\
 	\hline
    Before selection & 100.0 & 100.0\\
    Isolated muon, electron, photon vetos & 99.4 & 98.2\\
    $p_{T}^{j_{1}} > 100$\, & 98.7 & 98.1\\
    $0.1 < f_{h^{\pm}}^{j_{1}} < 0.95$ & 93.9 & 98.1\\
    $H_{\text{T}} > 200$\,GeV & 93.9 & 98.1\\
    $H_{\text{T}}^{\text{miss}} > 200$\,GeV & 88.5 & 92.2\\
    Event veto for forward jets ($|\eta| > 2.4$) & 69.9 & 74.4\\
    $H_{\text{T}}^{\text{miss}}$/$E_{\text{T}}^{\text{miss}} < 1.25$ & 69.3 & 73.7\\
    $n_{\text{jet}}$- and $H_{\text{T}}$-dependent $\alpha_{\text{T}}$ thresholds & 69.2 & 73.7\\
    $\Delta\phi^{*}_{\text{min}} > 0.5$ & 25.1 & 23.7\\
  	\hline
 	\end{tabular}
 	\caption{Cumulative percentages of events passing the event selections compared with those from~\cite{CMS-SUS-16} for a standard reference benchmark model, $T1bbbb$~\cite{T1bbbb}.}
 	\label{cutflow_table}
\end{table}

As shown in table~\ref{cutflow_table} it is clear that the respective efficiencies of each of the event selections are all within a few percent of those taken directly from~\cite{CMS-SUS-16} for the same example benchmark model. Therefore, in order to take into account other sources of uncertainty such as Initial State Radiation (ISR), the choice of $25$\% appears to be appropriate.

\section{Signal properties}

We examine the observable properties of these BPs, in order to explore which event selections have the greatest impact on the experimental sensitivity. Additionally we also consider the MC \emph{truth} values, i.e., the generated values without detector simulation being applied, for quantities such as $b$ jet angular separation, since this quantity taking a value below the resolution of the detector will have a large effect on the efficiency of bottom quark tagging and the ability to resolve both $b$ jets stemming from the decay of each Higgs~boson.

First consider the BP1-BP8-type mass scans, taking from each two mass points where one has the lightest $3$\,GeV/$c^{2}$ LSP and the other a mid-range $953$\,GeV/$c^{2}$ LSP more typical of simplified SUSY models, choosing an example squark mass of $2$\,TeV/$c^{2}$. Plotting quantities of interest for all events, before any event selection is applied, and normalising to unity for comparison, it can be shown how many of the cuts in~\cite{CMS-SUS-16} will likely thwart most of the signal events generated in the low-\mlsp\ regions. These are then overlayed with the corresponding normalised distributions for background processes containing QCD multijet events and background contributions from top quark pair production.

Additionally, a simplified MSSM-like model is presented for comparison. This scenario is derived from the BP1-type scan, with the difference being that the NLSP is dropped, with the squarks and gluinos decaying instead directly to the LSP\@. Thus no Higgs bosons are produced and the LSP momentum is no longer suppressed, removing the possibility of a low \met\ and Higgs boson enriched scenario. In this model the squark mass is set at $2$\,TeV/$c^{2}$ and the effective ``LSP'' has mass $3$\,GeV/$c^{2}$.

\boldmath
\subsection{Total scalar $H_{\text{T}}$}
\unboldmath

A dominant feature of many SUSY cascades is high $H_{\text{T}}$, due to the large number of jets produced in the many decays. We note that these low-LSP mass scenarios deliver an \htt\ distribution with mean well over $2$\,TeV/$c$.

\begin{figure}
\centering
	\begin{subfigure}[t]{0.3\textwidth}
		\centering
  		\includegraphics[keepaspectratio=true,width=\columnwidth]{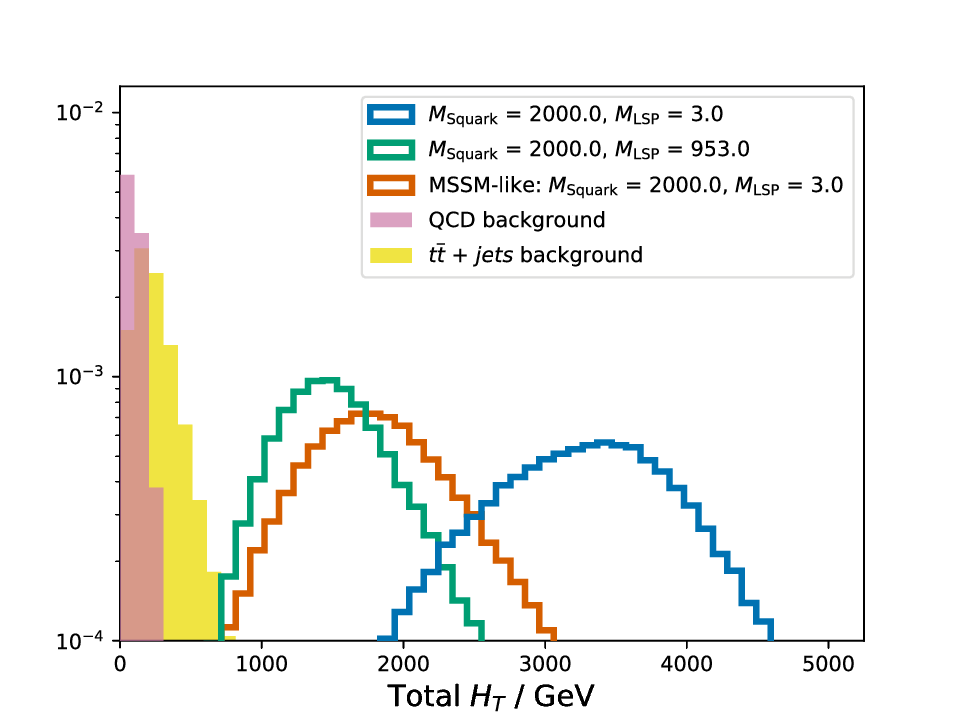}
  		\caption{BP1\label{P1_HT}}
	\end{subfigure}\quad%
	\begin{subfigure}[t]{0.3\textwidth}
		\centering
  		\includegraphics[keepaspectratio=true,width=\columnwidth]{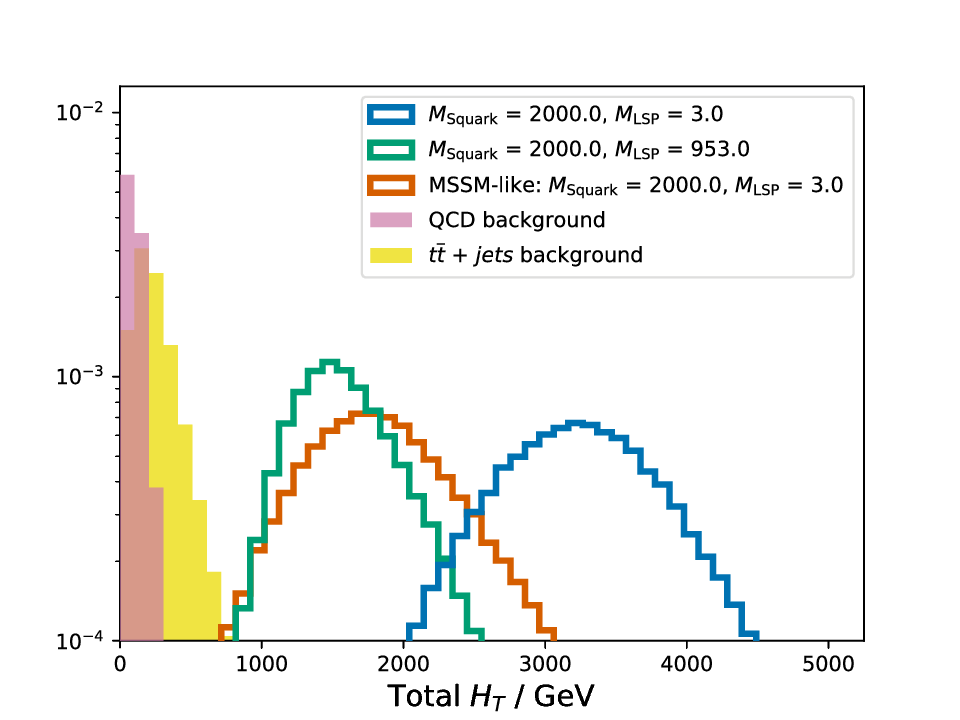}
  		\caption{BP3\label{P3_HT}}
	\end{subfigure}\quad%
	\begin{subfigure}[t]{0.3\textwidth}
		\centering
  		\includegraphics[keepaspectratio=true,width=\columnwidth]{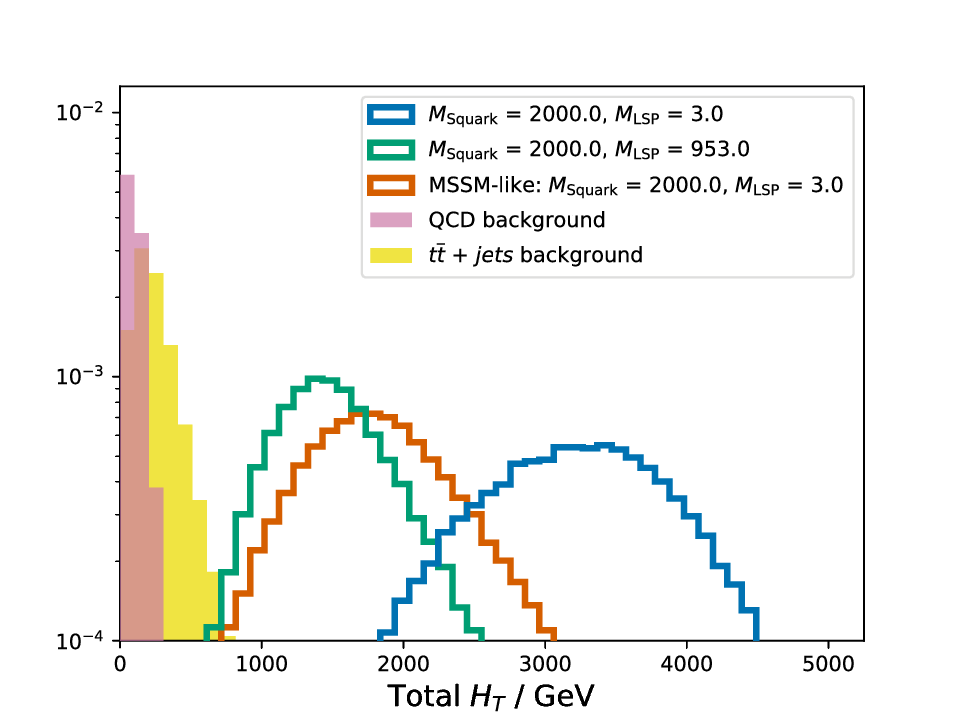}
  		\caption{BP5\label{P5_HT}}
	\end{subfigure}\\[5pt]

	\begin{subfigure}[t]{0.3\textwidth}
		\centering
  		\includegraphics[keepaspectratio=true,width=\columnwidth]{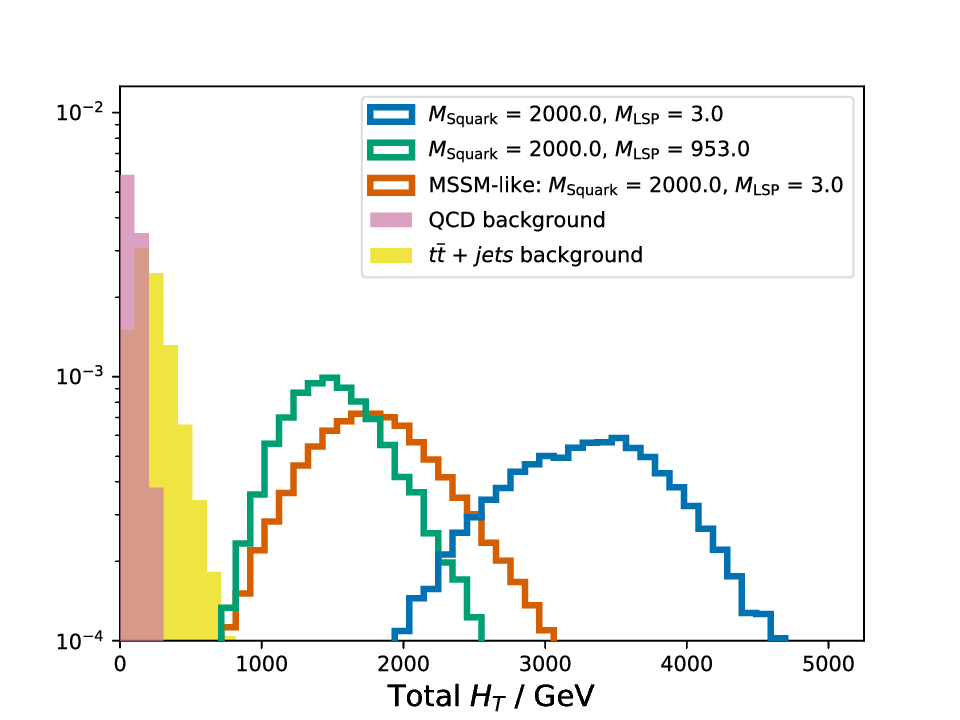}
  		\caption{BP6\label{P6_HT}}
	\end{subfigure}\quad%
	\begin{subfigure}[t]{0.3\textwidth}
		\centering
  		\includegraphics[keepaspectratio=true,width=\columnwidth]{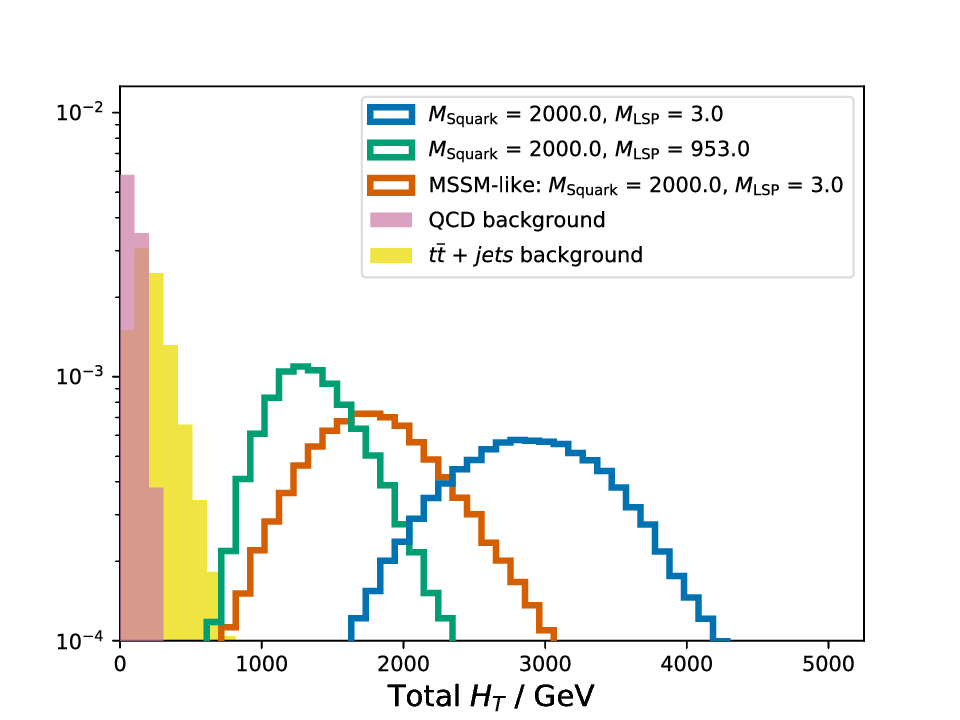}
  		\caption{BP7\label{P7_HT}}
	\end{subfigure}\quad%
	\begin{subfigure}[t]{0.3\textwidth}
		\centering
  		\includegraphics[keepaspectratio=true,width=\columnwidth]{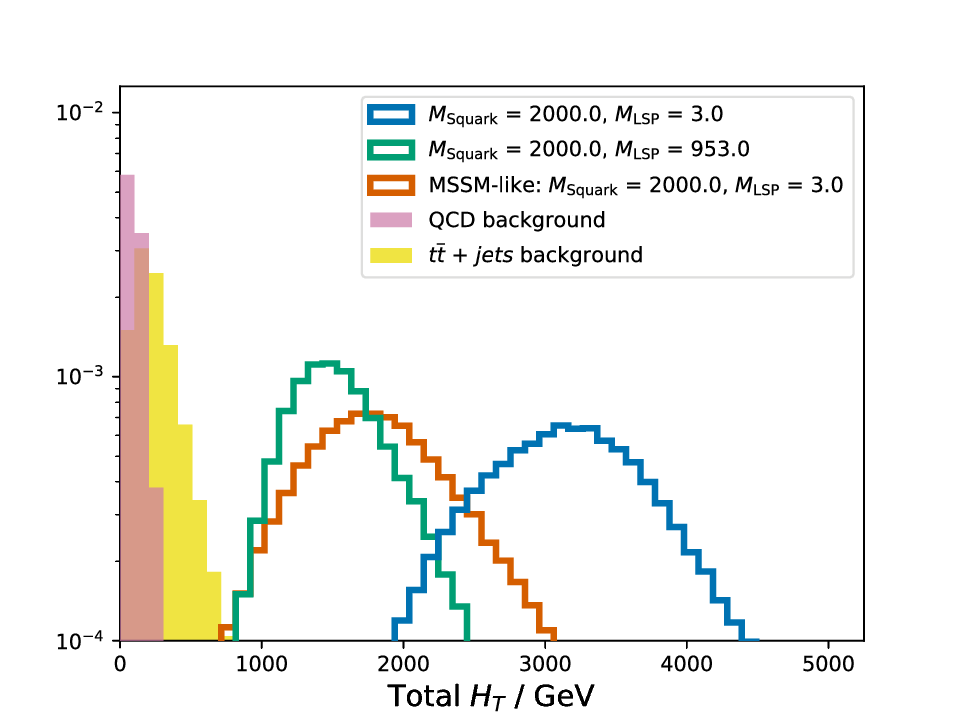}
  		\caption{BP8\label{P8_HT}}
	\end{subfigure}
\caption{$H_{\text{T}}$ distributions for low and mid-range $M_{\text{LSP}}$ near the observed limit in the BP1-type scan, compared with QCD and $t\bar{t}$ background processes and an MSSM-like scenario with a light~LSP.\label{HT_compare}}
\end{figure}

Figure~\ref{HT_compare} shows the \htt\  distributions for two example points from each of the six mass scans. Whilst it is clear that the mean \htt\  for both QCD multijet and top pair production background processes is far lower than that for the signal processes, their respective cross-sections are much higher. Therefore, whilst the peaks of these distributions are well separated, it would be expected for the tails of both background processes to still be significant compared with the signal processes when considering the event yields in an analysis.

\begin{figure}
\centering
	\begin{subfigure}[t]{0.3\textwidth}
		\centering
  		\includegraphics[keepaspectratio=true,width=\columnwidth]{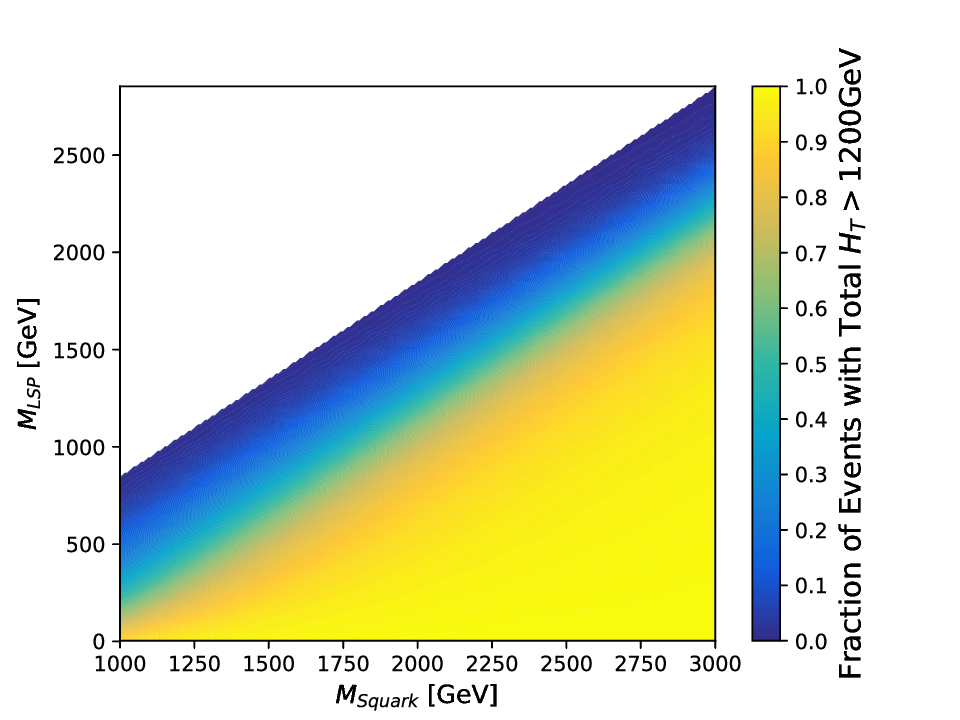}
  		\caption{BP1-type Mass Scan\label{P1_HTeff}}
	\end{subfigure}\quad%
	\begin{subfigure}[t]{0.3\textwidth}
		\centering
  		\includegraphics[keepaspectratio=true,width=\columnwidth]{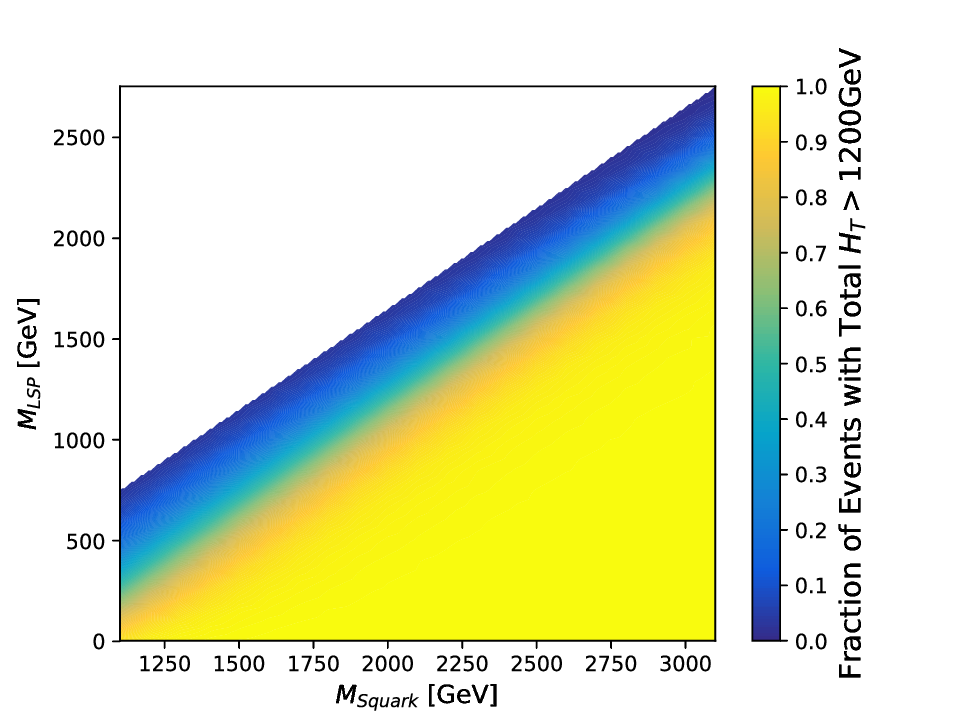}
  		\caption{BP3-type Mass Scan\label{P3_HTeff}}
	\end{subfigure}\quad%
	\begin{subfigure}[t]{0.3\textwidth}
		\centering
  		\includegraphics[keepaspectratio=true,width=\columnwidth]{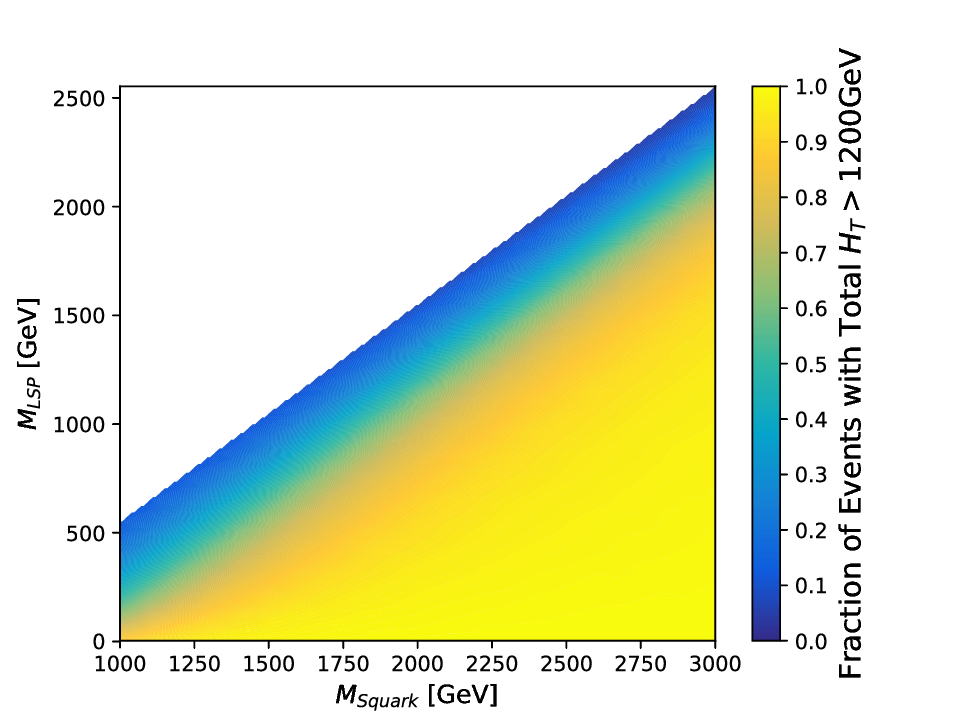}
  		\caption{BP5-type Mass Scan\label{P5_HTeff}}
	\end{subfigure}\\[5pt]

	\begin{subfigure}[t]{0.3\textwidth}
		\centering
  		\includegraphics[keepaspectratio=true,width=\columnwidth]{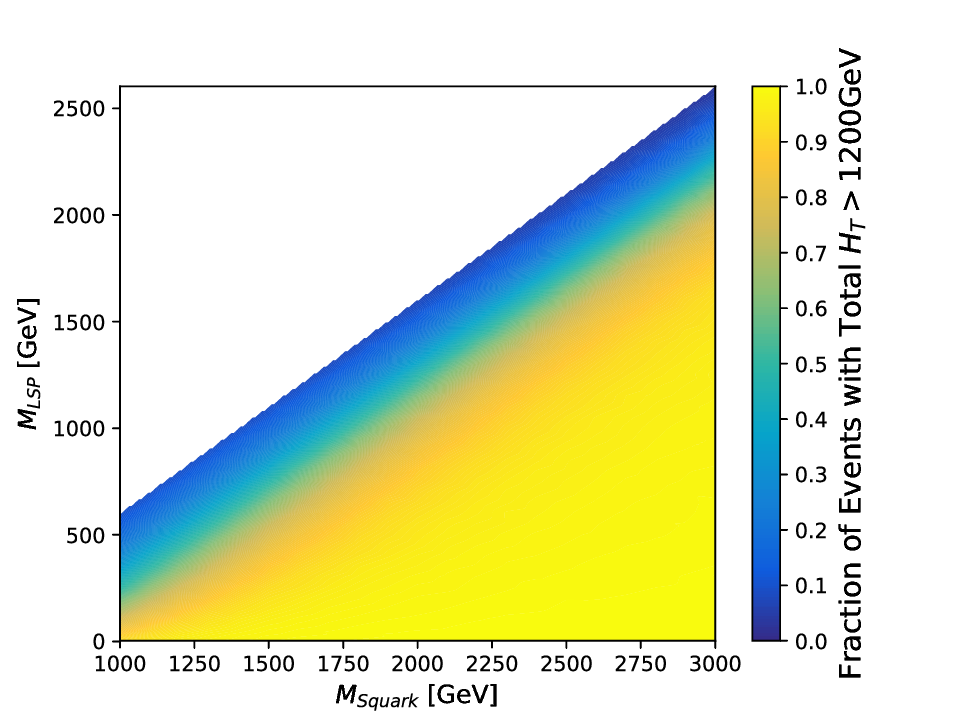}
  		\caption{BP6-type mass scan\label{P6_HTeff}}
	\end{subfigure}\quad%
	\begin{subfigure}[t]{0.3\textwidth}
		\centering
  		\includegraphics[keepaspectratio=true,width=\columnwidth]{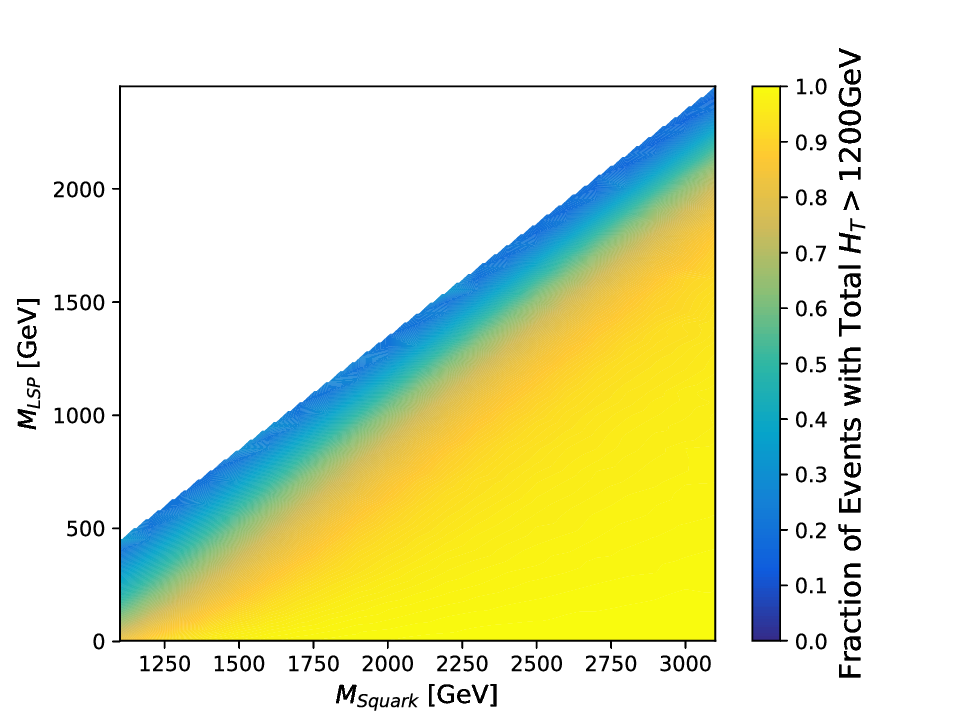}
  		\caption{BP7-type mass scan\label{P7_HTeff}}
	\end{subfigure}\quad%
	\begin{subfigure}[t]{0.3\textwidth}
		\centering
  		\includegraphics[keepaspectratio=true,width=\columnwidth]{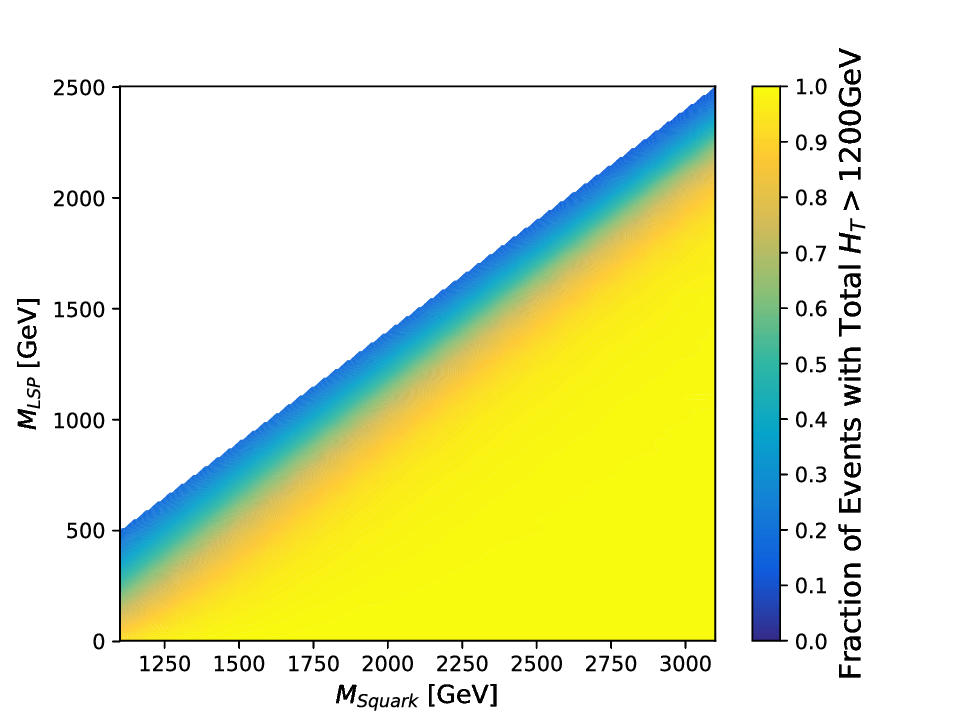}
  		\caption{BP8-type mass scan\label{P8_HTeff}}
	\end{subfigure}
\caption{Fraction of events with total $H_{\text{T}} > 1200$\,GeV for the BP1-BP8-type mass scans.\label{HTeff}}
\end{figure}

In fact, considering the fraction of events passing the $1200$\,GeV/$c$ minimum \htt\ requirement across the BP1-type mass scan range, shown in figure~\ref{HTeff}, it becomes clear that for many of the mass points essentially all events pass this cut. The regions where this is not the case are generally limited to those where the LSP mass is close to the masses of the squarks and gluinos, thus reducing the \ptt\ of the jet(s) emitted as the squark or gluino decays.

However, it is also apparent, in figure~\ref{HTeff}, that the mean \htt\ is much lower where the LSP mass approaches that of the squarks and gluinos. This is the case since the mass gaps in the decay cascade are reduced and so the \ptt\ of any emitted quark is suppressed.

\boldmath
\subsection{$H_{\text{T}}^{\text{miss}}$}
\unboldmath

Whilst of course the existence of these LSP with non-zero momenta dictates there must also be some missing net transverse energy, so long that the two LSPs are exactly not back-to-back.

\begin{figure}
\centering
	\begin{subfigure}[t]{0.3\textwidth}
		\centering
  		\includegraphics[keepaspectratio=true,width=\columnwidth]{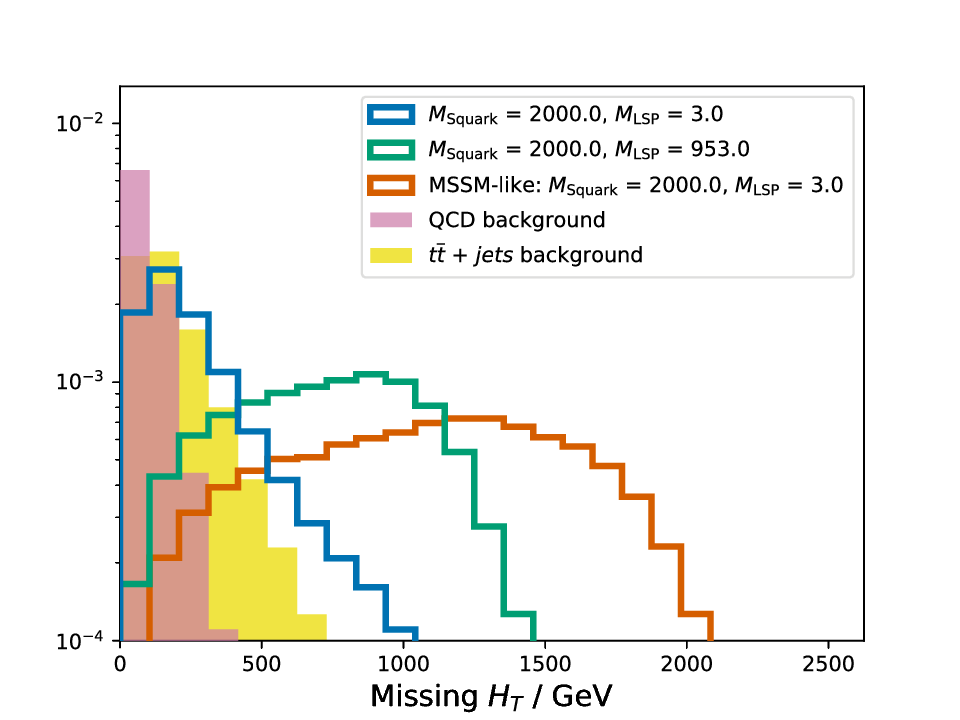}
  		\caption{BP1\label{P1_MHT}}
	\end{subfigure}\quad%
	\begin{subfigure}[t]{0.3\textwidth}
		\centering
  		\includegraphics[keepaspectratio=true,width=\columnwidth]{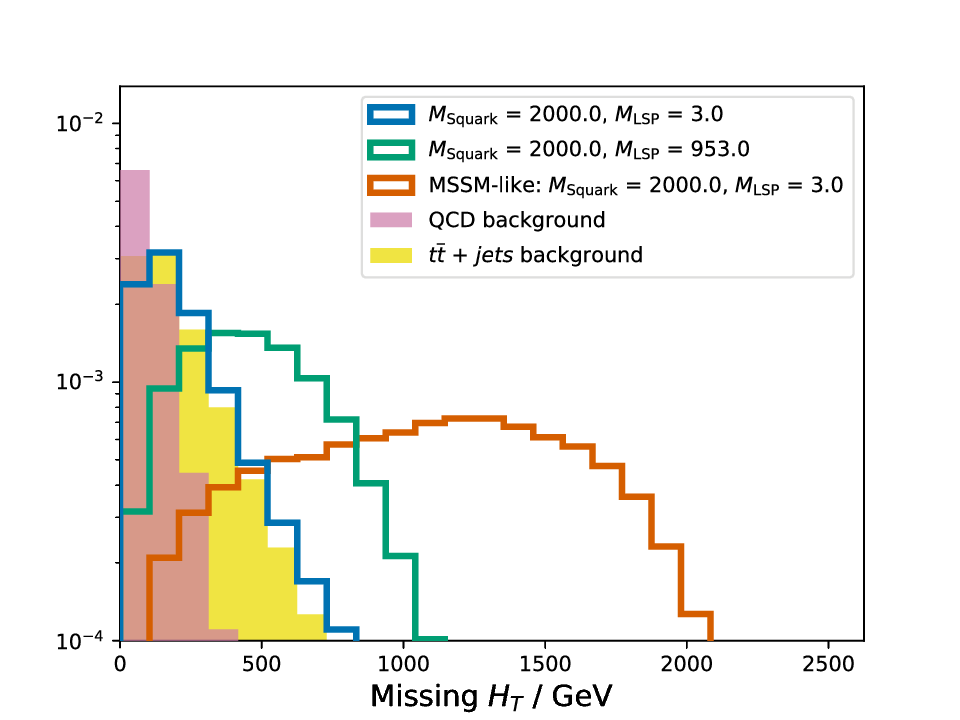}
  		\caption{BP3\label{P3_MHT}}
	\end{subfigure}\quad%
	\begin{subfigure}[t]{0.3\textwidth}
		\centering
  		\includegraphics[keepaspectratio=true,width=\columnwidth]{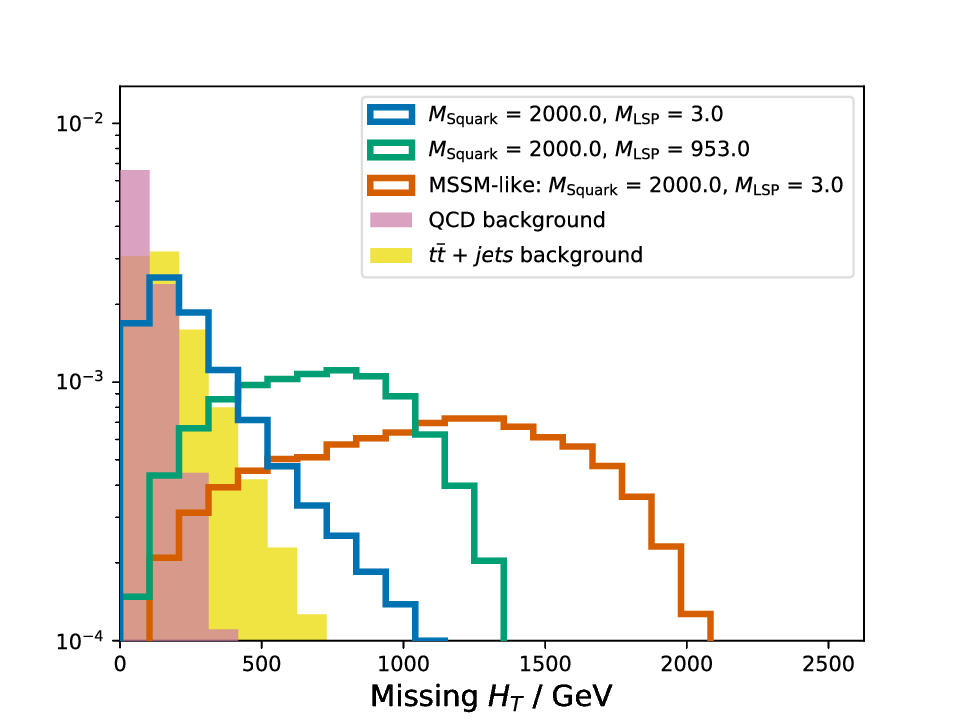}
  		\caption{BP5\label{P5_MHT}}
	\end{subfigure}\\[5pt]

	\begin{subfigure}[t]{0.3\textwidth}
		\centering
  		\includegraphics[keepaspectratio=true,width=\columnwidth]{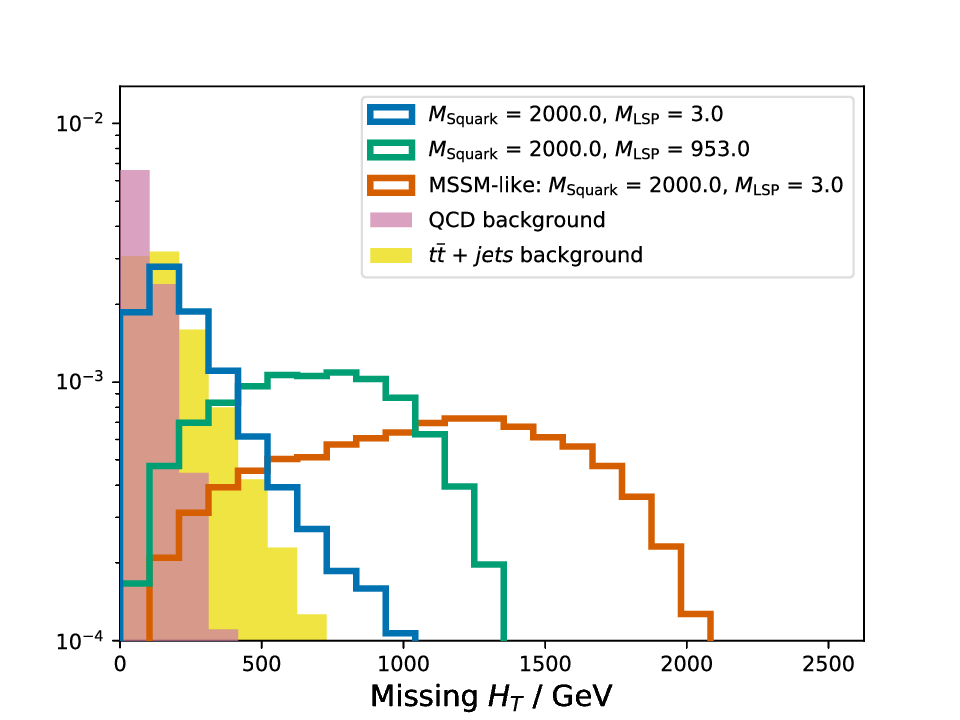}
  		\caption{BP6\label{P6_MHT}}
	\end{subfigure}\quad%
	\begin{subfigure}[t]{0.3\textwidth}
		\centering
  		\includegraphics[keepaspectratio=true,width=\columnwidth]{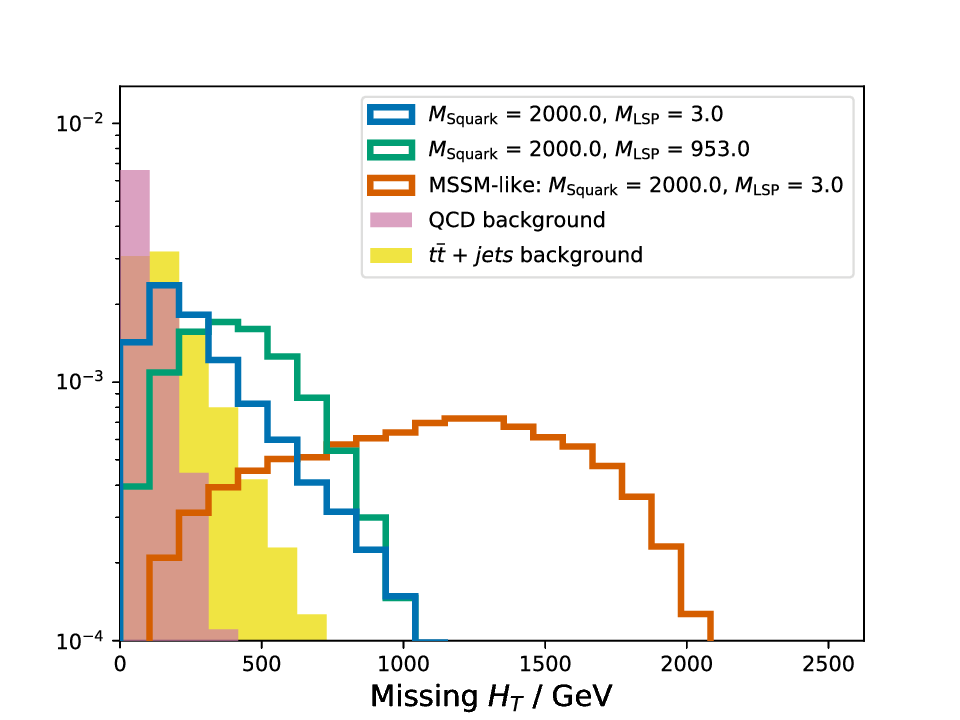}
  		\caption{BP7\label{P7_MHT}}
	\end{subfigure}\quad%
	\begin{subfigure}[t]{0.3\textwidth}
		\centering
  		\includegraphics[keepaspectratio=true,width=\columnwidth]{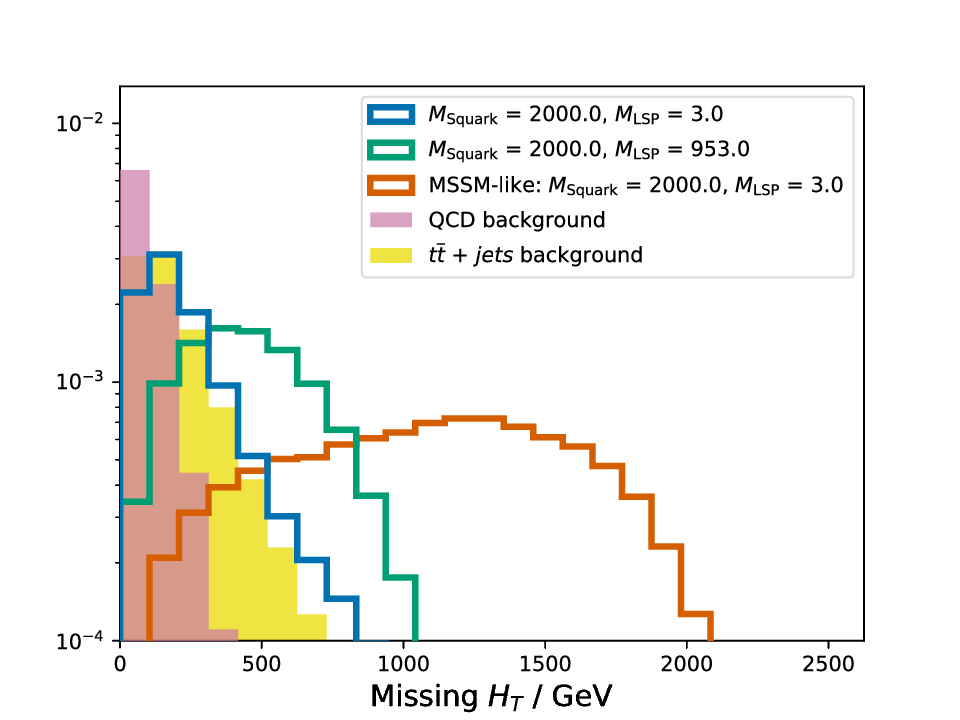}
  		\caption{BP8\label{P8_MHT}}
	\end{subfigure}
\caption{\mht\ distributions for low and mid-range $M_{\text{LSP}}$ near the observed limit in the BP1-type scan, compared with QCD and $t\bar{t}$ background processes.\label{MHT_compare}}
\end{figure}

In figure~\ref{MHT_compare} rather distinct distributions between the two mass points may be observed. The heavy LSP scenarios give a fairly wide spread of \mht\ values due to the presence of such a heavy, boosted and invisible LSP\@. The light-LSP scenarios however suppress this, since the considerably heavier Higgs boson will inherit most of the momentum from the NLSP decay, leaving a soft LSP.

Even in these light-LSP scenarios, there are still events whose \mht\ is quite high. This larger \mht\ can arise via Higgs bosons decaying to final states other than a bottom quark-antiquark pair, though the branching fractions for decays such as $H \rightarrow ZZ^{*} \rightarrow \nu \bar{\nu} \nu \bar{\nu}$ are very small. A more likely phenomenon is for one or more of the bottom quarks stemming from the decays of the Higgs bosons to decay into a charm quark and an electron or muon, along with the appropriately-flavoured neutrino, with the neutrino momentum contributing to the overall \mht. Additionally, hadronic Higgs boson decay final states such as $b\bar{b}$ are difficult to reconstruct owing to resolution and, given the higher number of jets compared with, say, a cleaner $H \rightarrow \gamma\gamma$ decay channel, meaning a larger contribution to the uncertainty in \mht\ and \htt.

Here the \mht\ decreases drastically in the limit of a light-LSP in the NMSSM, with the peak well below $200$\,GeV/$c$, the minimum \mht\ requirement used from~\cite{CMS-SUS-16}. It is clear for these areas of mass space many of the events are lost due to \mht\ cuts, thus decreasing experimental sensitivity to this type of model.

\begin{figure}
\centering
	\begin{subfigure}[t]{0.3\textwidth}
		\centering
  		\includegraphics[keepaspectratio=true,width=\columnwidth]{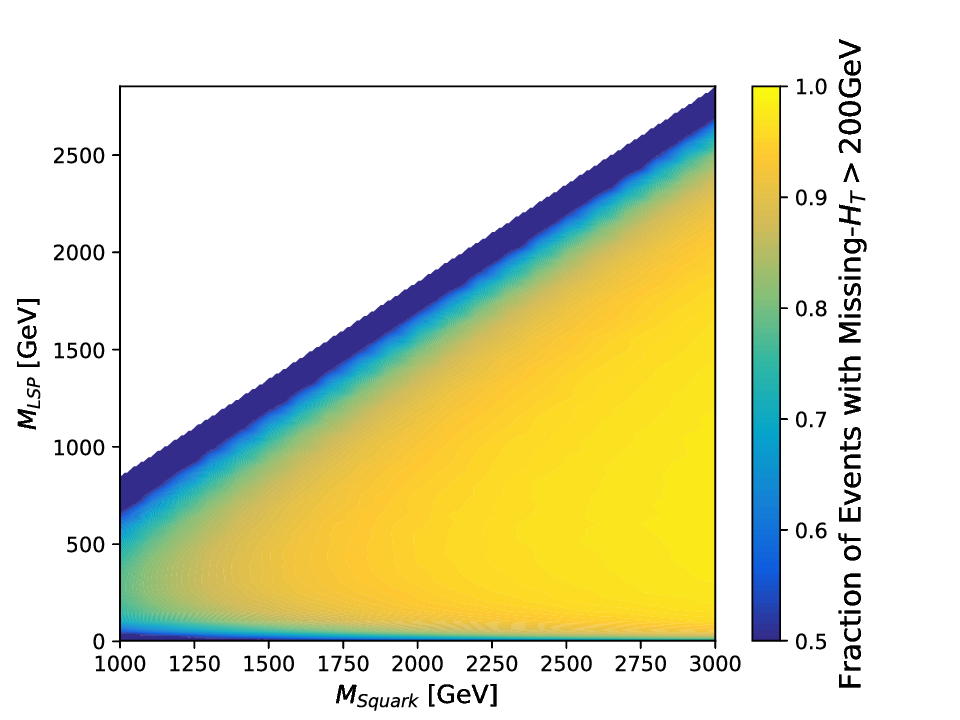}
  		\caption{BP1-type Mass Scan\label{P1_MHTeff}}
	\end{subfigure}\quad%
	\begin{subfigure}[t]{0.3\textwidth}
		\centering
  		\includegraphics[keepaspectratio=true,width=\columnwidth]{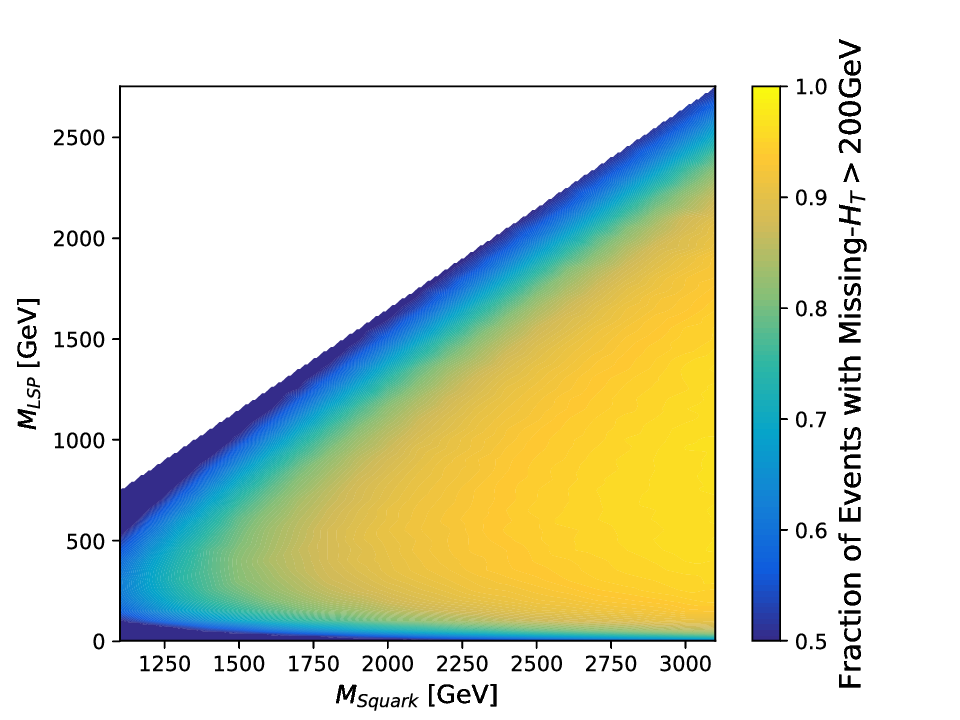}
  		\caption{BP3-type Mass Scan\label{P3_MHTeff}}
	\end{subfigure}\quad%
	\begin{subfigure}[t]{0.3\textwidth}
		\centering
  		\includegraphics[keepaspectratio=true,width=\columnwidth]{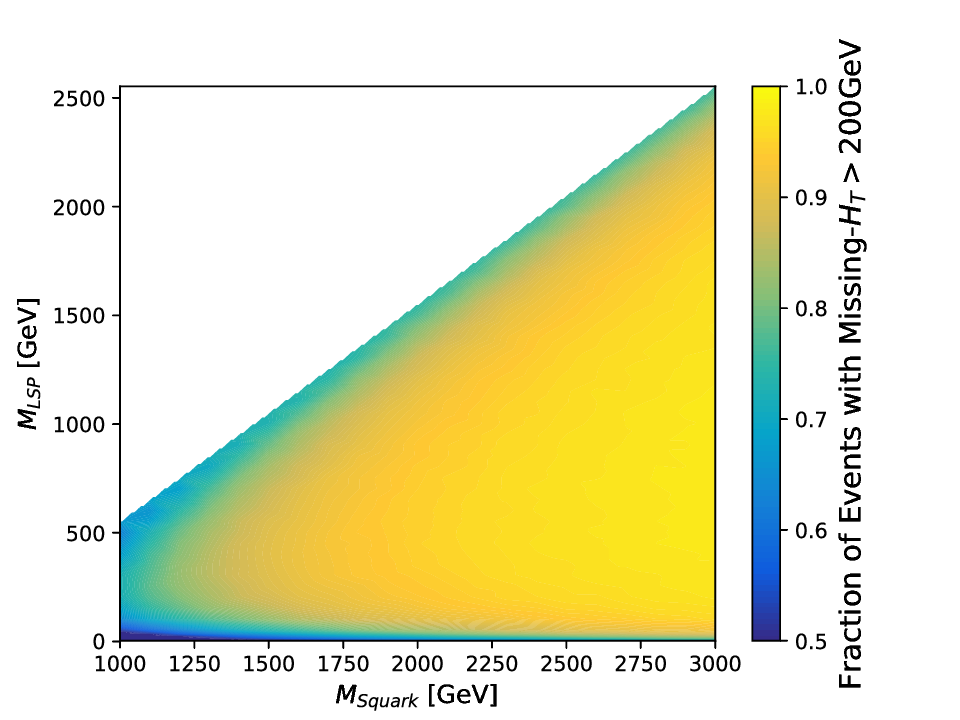}
  		\caption{BP5-type Mass Scan\label{P5_MHTeff}}
	\end{subfigure}\\[5pt]

	\begin{subfigure}[t]{0.3\textwidth}
		\centering
  		\includegraphics[keepaspectratio=true,width=\columnwidth]{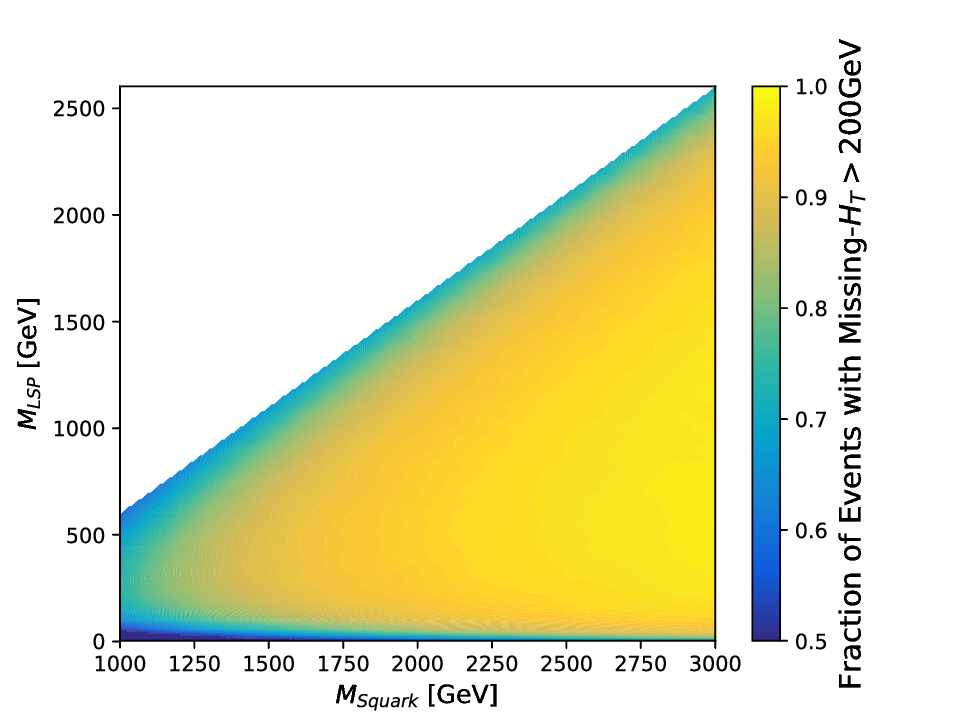}
  		\caption{BP6-type mass scan\label{P6_MHTeff}}
	\end{subfigure}\quad%
	\begin{subfigure}[t]{0.3\textwidth}
		\centering
  		\includegraphics[keepaspectratio=true,width=\columnwidth]{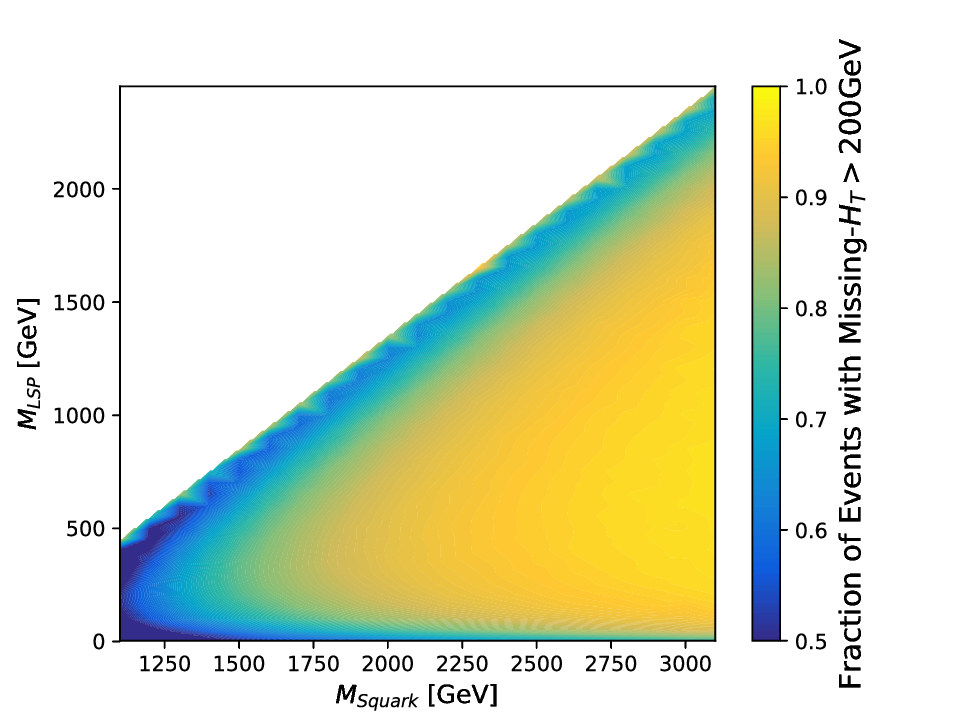}
  		\caption{BP7-type mass scan\label{P7_MHTeff}}
	\end{subfigure}\quad%
	\begin{subfigure}[t]{0.3\textwidth}
		\centering
  		\includegraphics[keepaspectratio=true,width=\columnwidth]{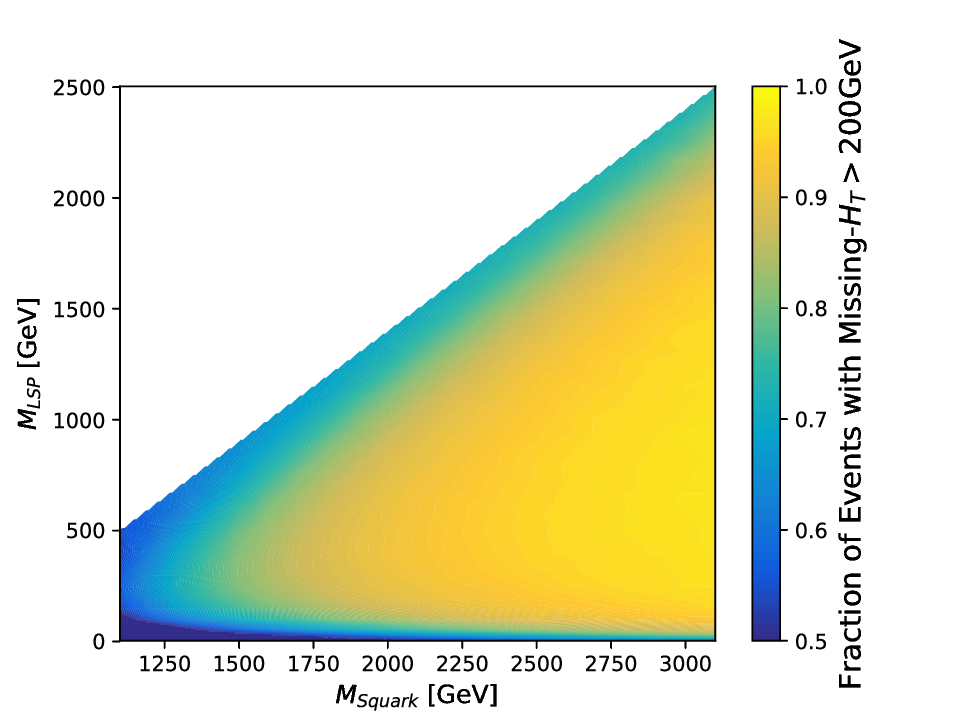}
  		\caption{BP8-type mass scan\label{P8_MHTeff}}
	\end{subfigure}
\caption{Fraction of events with \mht\ $> 200$\,GeV/$c$ for the BP1-BP8-type mass scans.\label{MHTeff}}
\end{figure}

Additionally, as seen in the colour map plots in figure~\ref{MHTeff}, the fraction of events with \mht\ greater than the lower edge of the minimum \mht\ bin, $200$\,GeV/$c$, rises considerably as the LSP mass increases. Above a certain threshold, however, this fraction begins to drop since, as the LSP mass approaches that of the squarks and gluinos, the mass gaps in the decay cascade shrinks, meaning that we are left with heavy LSPs with very little kinetic energy, and thus low \mht.

\subsection{Number of hadronic jets}

Typical SUSY decay cascades often involve a large number of hadronic jets being produced. The model considered in this paper features two such cascades ending in Higgs boson decays, and some including stop and sbottom-type squarks whose decays produce even more squarks.

\begin{figure}
\centering
	\begin{subfigure}[t]{0.3\textwidth}
		\centering
  		\includegraphics[keepaspectratio=true,width=\columnwidth]{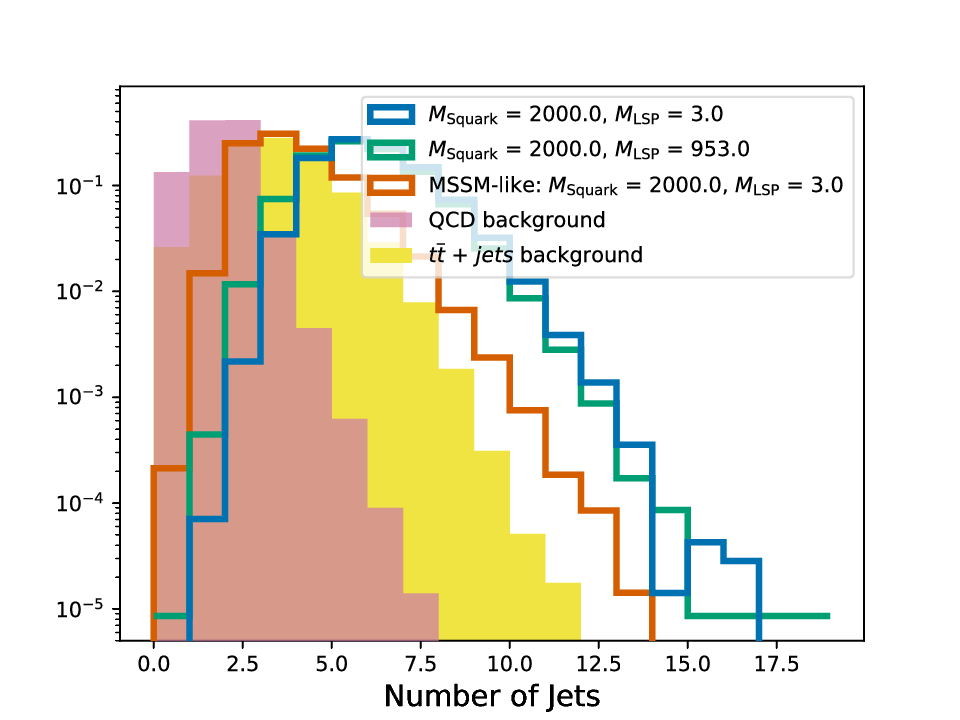}
  		\caption{BP1\label{P1_NJet}}
	\end{subfigure}\quad%
	\begin{subfigure}[t]{0.3\textwidth}
		\centering
  		\includegraphics[keepaspectratio=true,width=\columnwidth]{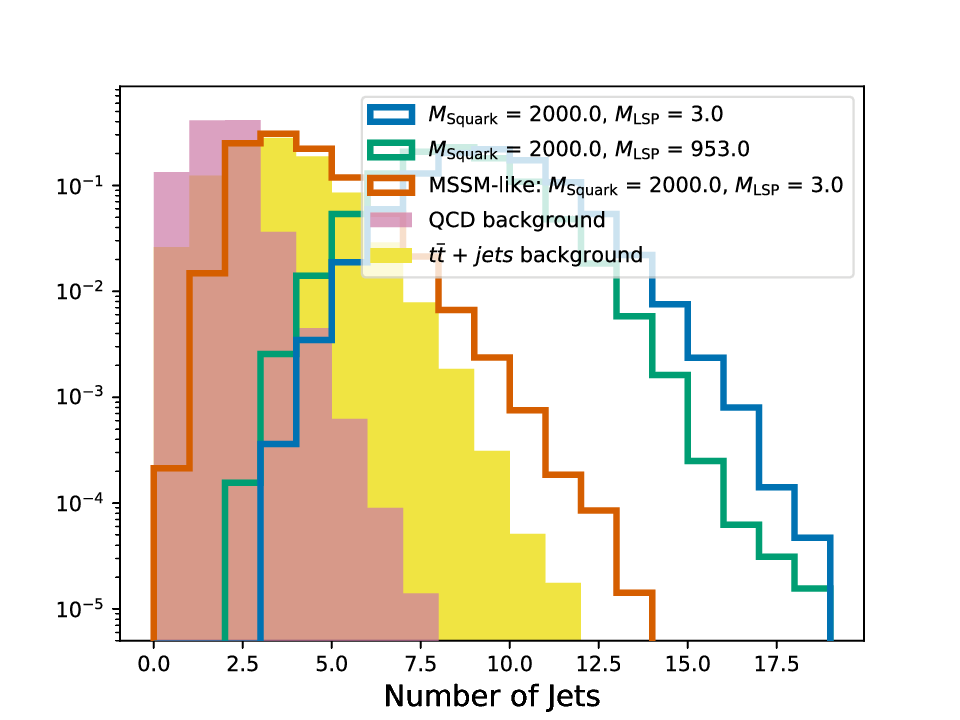}
  		\caption{BP3\label{P3_NJet}}
	\end{subfigure}\quad%
	\begin{subfigure}[t]{0.3\textwidth}
		\centering
  		\includegraphics[keepaspectratio=true,width=\columnwidth]{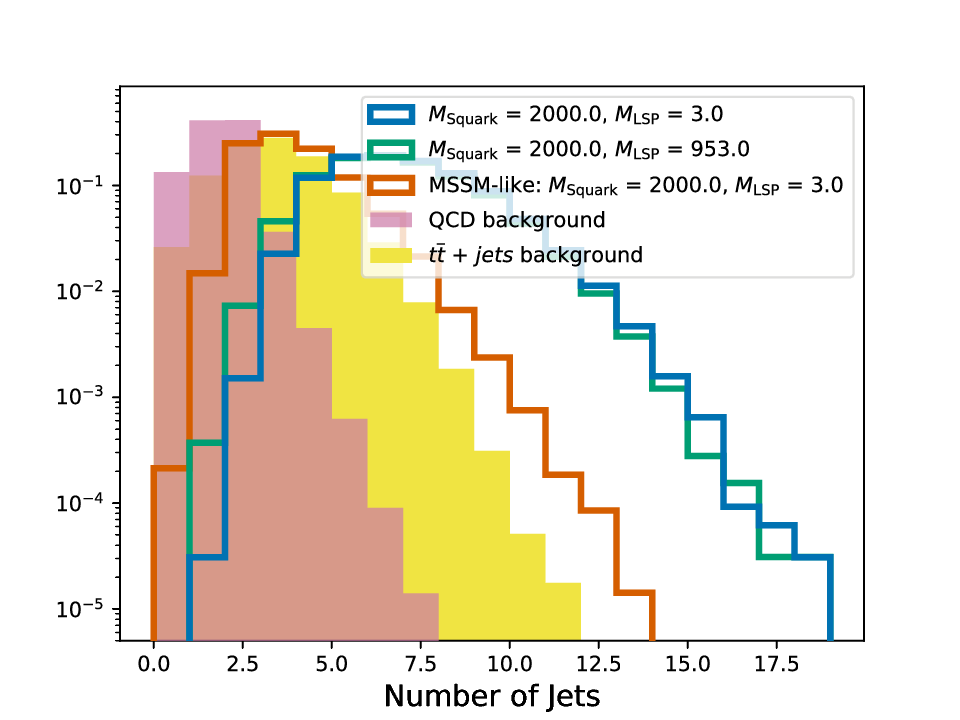}
  		\caption{BP5\label{P5_NJet}}
	\end{subfigure}\\[5pt]

	\begin{subfigure}[t]{0.3\textwidth}
		\centering
  		\includegraphics[keepaspectratio=true,width=\columnwidth]{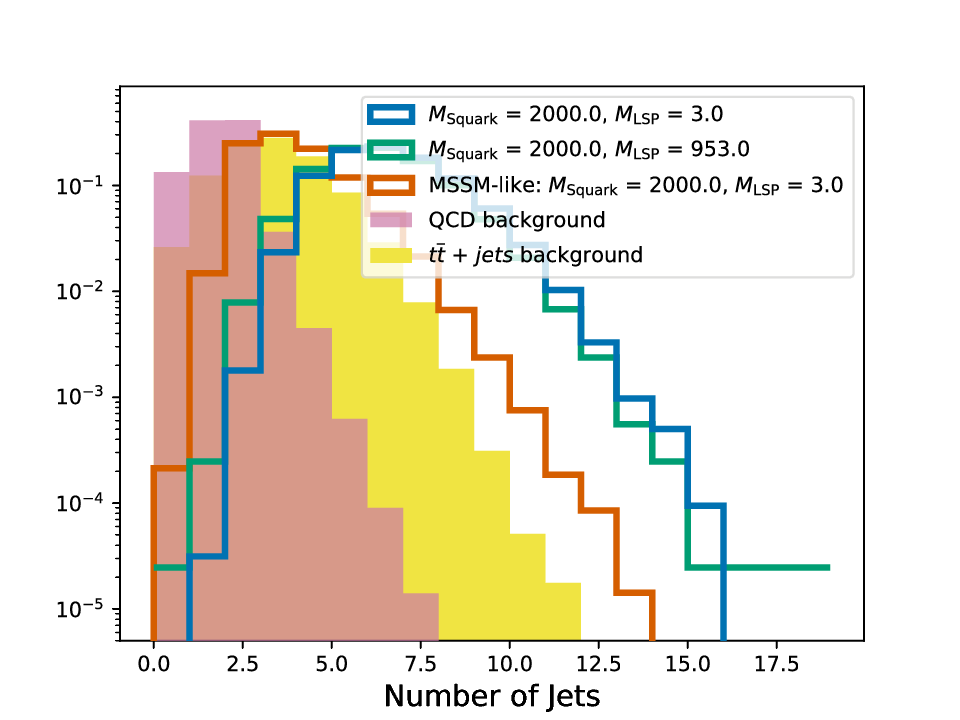}
  		\caption{BP6\label{P6_NJet}}
	\end{subfigure}\quad%
	\begin{subfigure}[t]{0.3\textwidth}
		\centering
  		\includegraphics[keepaspectratio=true,width=\columnwidth]{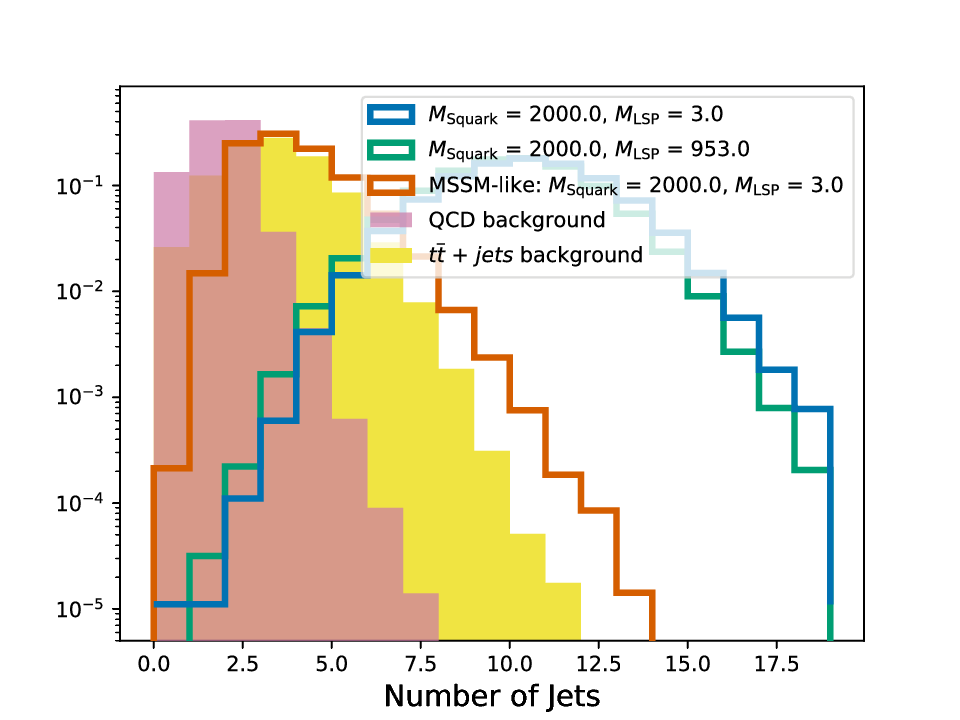}
  		\caption{BP7\label{P7_NJet}}
	\end{subfigure}\quad%
	\begin{subfigure}[t]{0.3\textwidth}
		\centering
  		\includegraphics[keepaspectratio=true,width=\columnwidth]{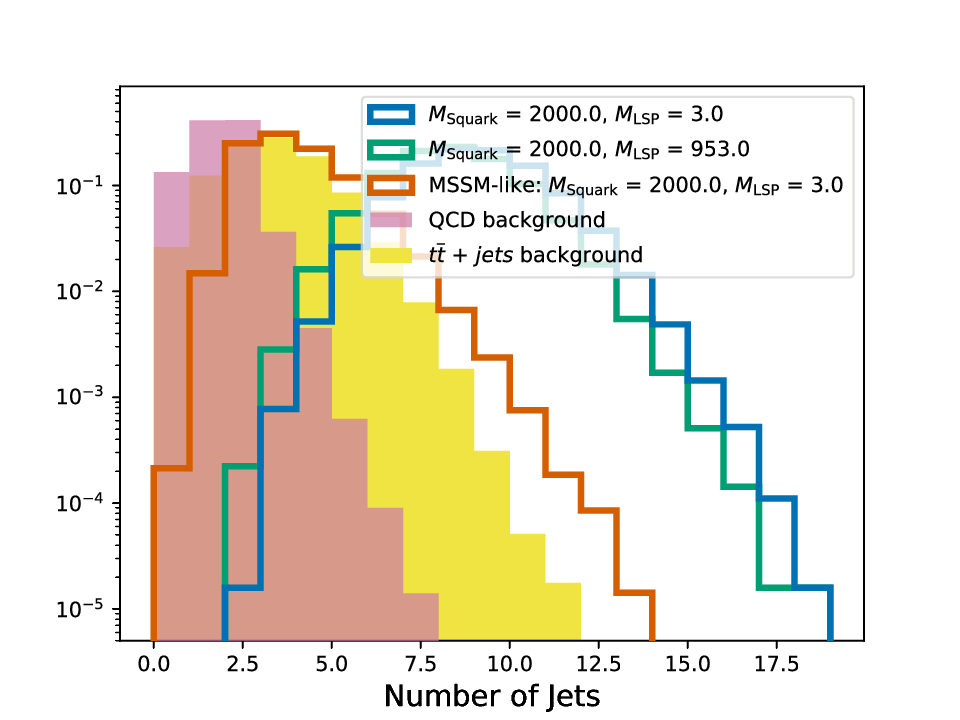}
  		\caption{BP8\label{P8_NJet}}
	\end{subfigure}
\caption{Number of hadronic jets for low and mid-range $M_{\text{LSP}}$ near the observed limit in the BP1-type scan, compared with QCD and $t\bar{t}$ background processes and an MSSM-like scenario with a light LSP.\label{NJet_compare}}
\end{figure}

Figure~\ref{NJet_compare} shows the number of hadronic jets as defined in~\cite{CMS-SUS-16}, with \ptt $> 40$\,GeV/$c$ and $|\eta| < 2.4$, for the usual signal and background  processes.

\begin{figure}
\centering
	\begin{subfigure}[t]{0.3\textwidth}
		\centering
  		\includegraphics[keepaspectratio=true,width=\columnwidth]{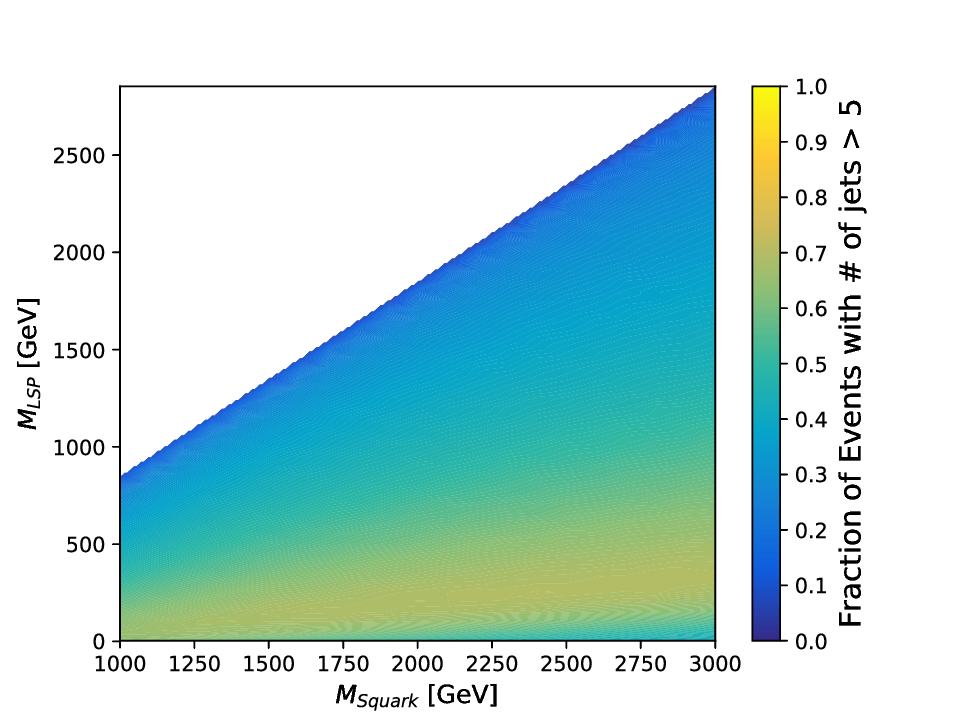}
  		\caption{BP1-type Mass Scan\label{P1_NJeteff}}
	\end{subfigure}\quad%
	\begin{subfigure}[t]{0.3\textwidth}
		\centering
  		\includegraphics[keepaspectratio=true,width=\columnwidth]{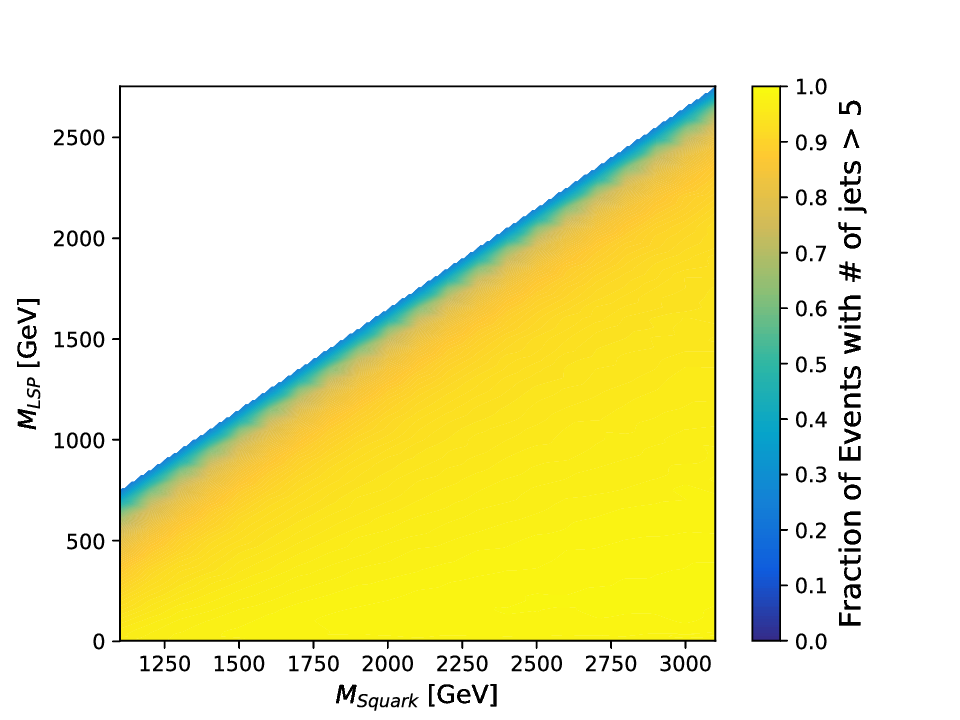}
  		\caption{BP3-type Mass Scan\label{P3_NJeteff}}
	\end{subfigure}\quad%
	\begin{subfigure}[t]{0.3\textwidth}
		\centering
  		\includegraphics[keepaspectratio=true,width=\columnwidth]{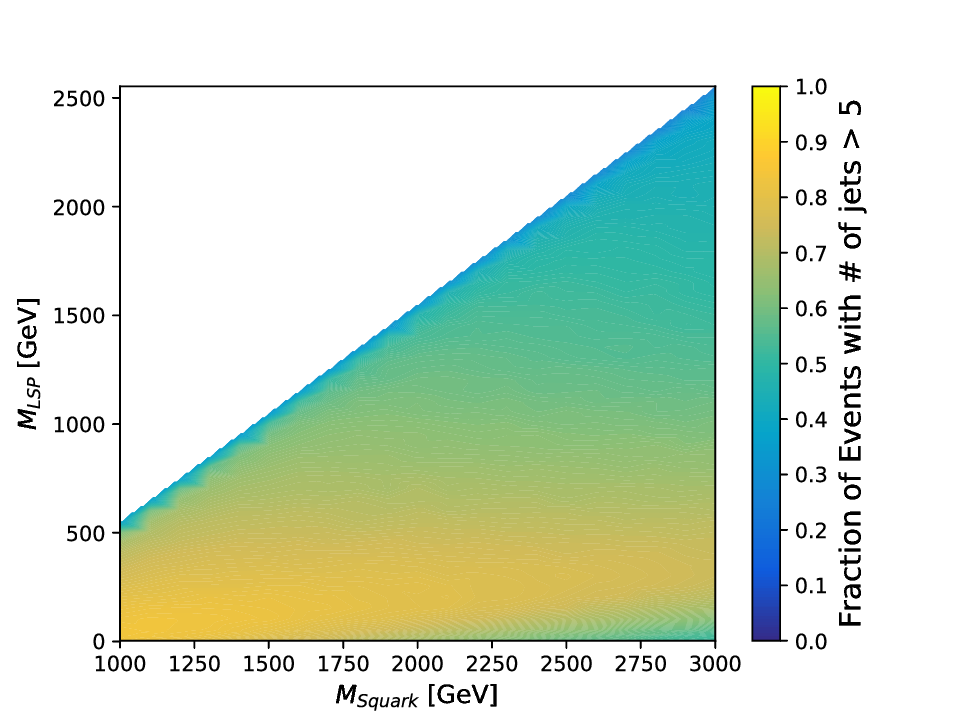}
  		\caption{BP5-type Mass Scan\label{P5_NJeteff}}
	\end{subfigure}\\[5pt]

	\begin{subfigure}[t]{0.3\textwidth}
		\centering
  		\includegraphics[keepaspectratio=true,width=\columnwidth]{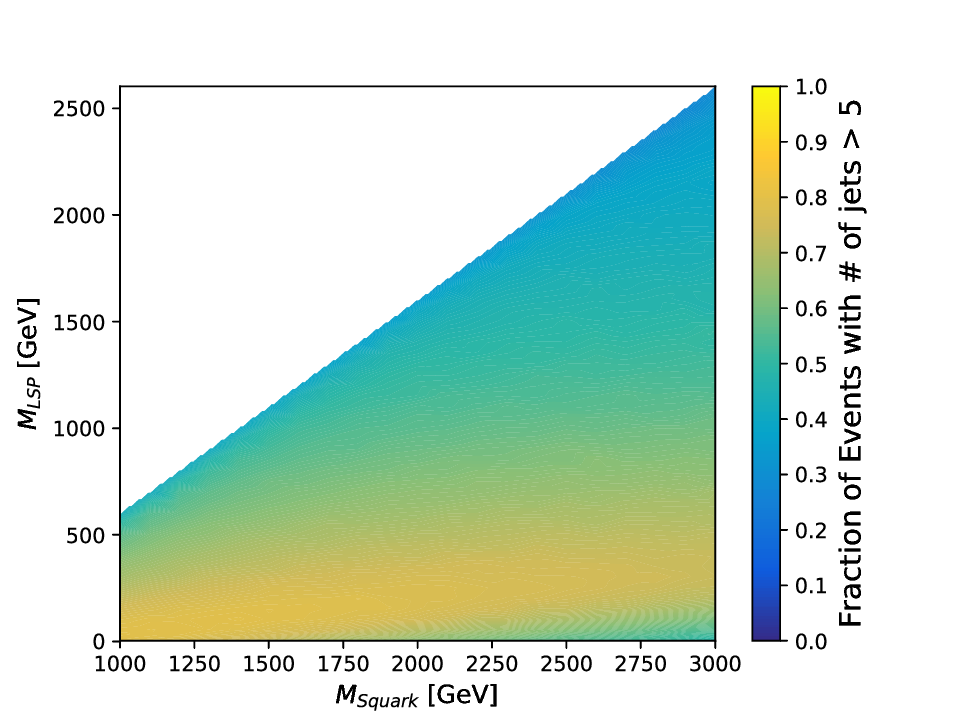}
  		\caption{BP6-type mass scan\label{P6_NJeteff}}
	\end{subfigure}\quad%
	\begin{subfigure}[t]{0.3\textwidth}
		\centering
  		\includegraphics[keepaspectratio=true,width=\columnwidth]{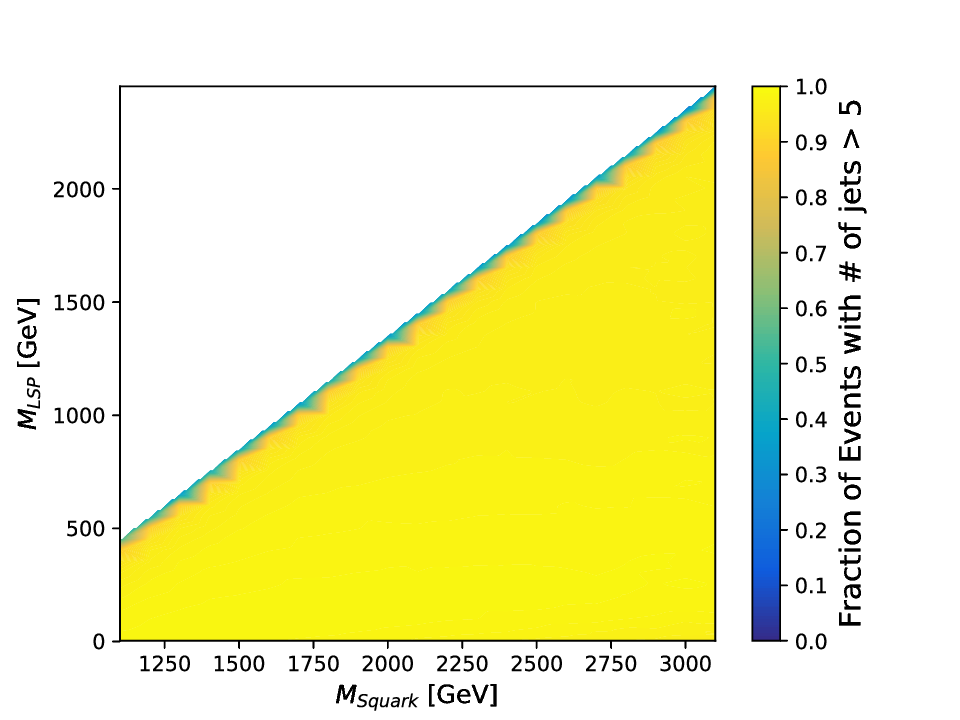}
  		\caption{BP7-type mass scan\label{P7_NJeteff}}
	\end{subfigure}\quad%
	\begin{subfigure}[t]{0.3\textwidth}
		\centering
  		\includegraphics[keepaspectratio=true,width=\columnwidth]{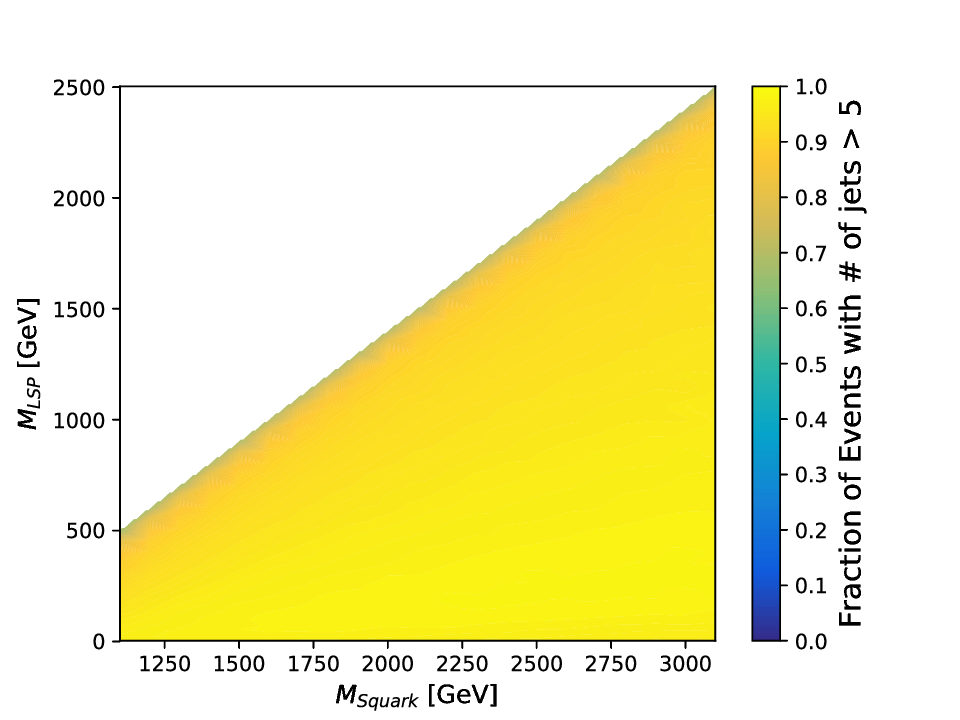}
  		\caption{BP8-type mass scan\label{P8_NJeteff}}
	\end{subfigure}
\caption{Fraction of events with total number of hadronic jets $> 5$ for the BP1-BP8-type mass~scans.\label{NJeteff}}
\end{figure}

Considering the fraction of events containing at least six hadronic jets, it can be seen in figure~\ref{NJeteff} that whilst in general most signal events pass this selection, the three scans where the gluino is lighter than the squark have a much higher efficiency with respect to this cut.

The primary reason for this behaviour lies in the decay products of the sparticles in the various mass hierarchies. The decay cascade for the BP3, BP6 and BP8-type scans involve squarks decaying to a gluino and a quark, where the gluino decays into two quarks and a neutralino. Conversely, scenarios where the squark is lighter than the gluino involve the gluino decaying into a squark and a quark, with the squark decaying into a neutralino and only one quark.

This three-body gluino decay means that each cascade, of which these scenarios include two, produces an extra quark, thus increasing the expected number of hadronic jets per event in the detector.

Additionally in each mass scan the fraction of events passing this selection is generally highest for a lighter LSP, dropping considerably as the LSP mass approaches the squark/gluino mass. This drop in efficiency for heavy LSP is due to the small mass gaps in the decay cascades meaning softer hadronic jets, such that the \ptt\ of some of these jets will fall below the $40~$GeV/$c$ minimum threshold and so will not be considered.

\boldmath
\subsection{Number of $b$-tagged hadronic jets}
\unboldmath

In this NMSSM scenario where the LSP is a singlino, two SM-like Higgs bosons will be produced. The BRs for SM-like Higgs boson decay dictate that the most likely decay is that to a bottom quark-antiquark pair, therefore it is expected that a large number of the hadronic jets in each event will be \emph{tagged} as being a bottom quark, or $b$-\emph{tagged}.

\begin{figure}
\centering
	\begin{subfigure}[t]{0.3\textwidth}
		\centering
  		\includegraphics[keepaspectratio=true,width=\columnwidth]{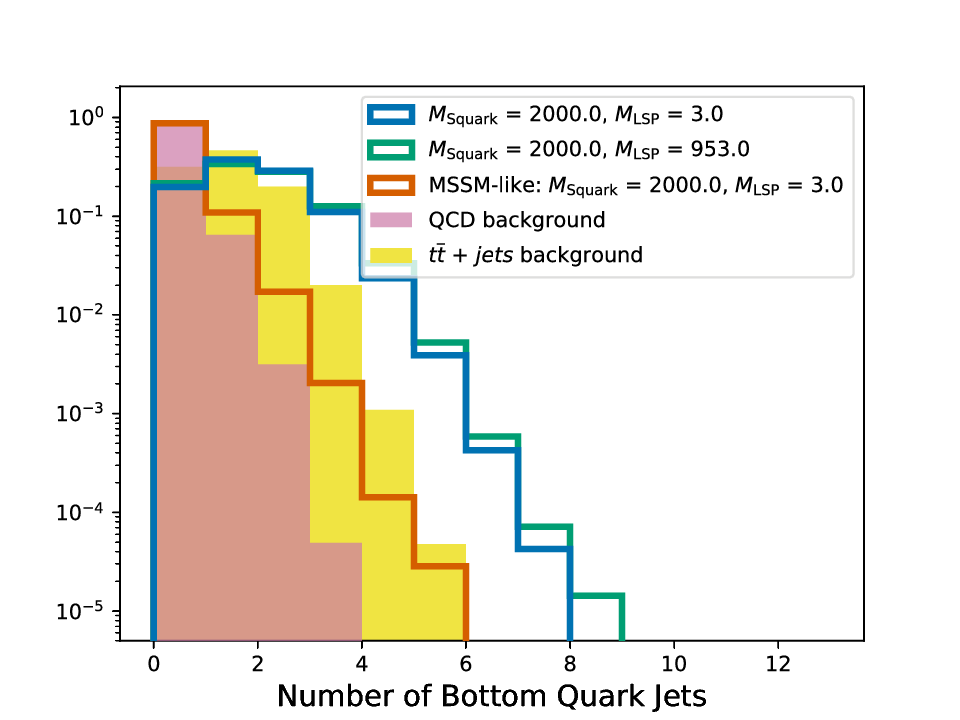}
  		\caption{BP1\label{P1_NBJet}}
	\end{subfigure}\quad%
	\begin{subfigure}[t]{0.3\textwidth}
		\centering
  		\includegraphics[keepaspectratio=true,width=\columnwidth]{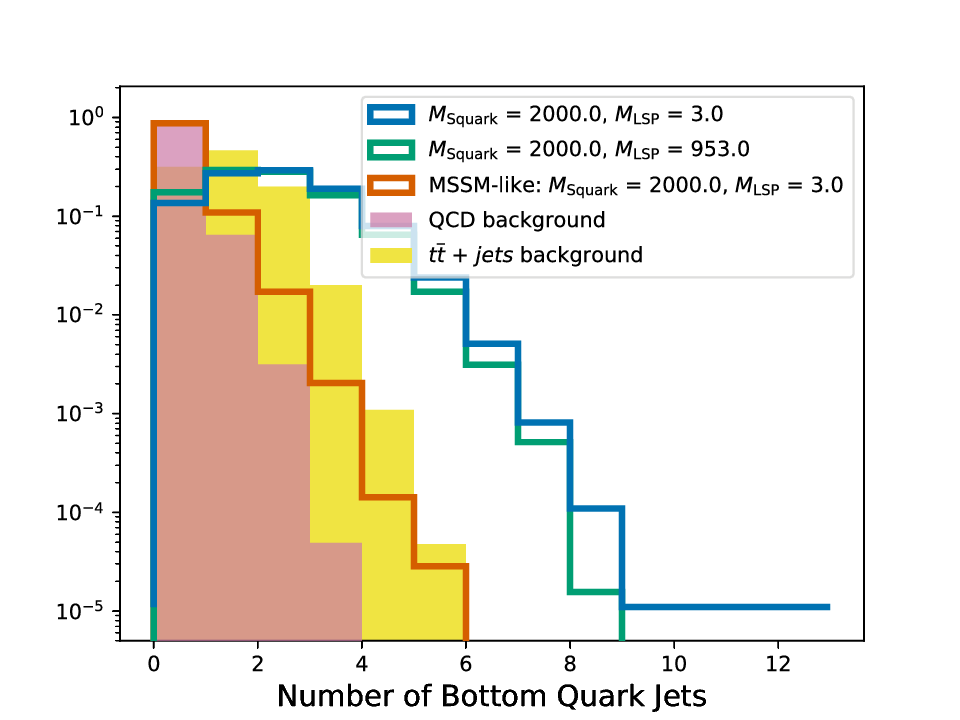}
  		\caption{BP3\label{P3_NBJet}}
	\end{subfigure}\quad%
	\begin{subfigure}[t]{0.3\textwidth}
		\centering
  		\includegraphics[keepaspectratio=true,width=\columnwidth]{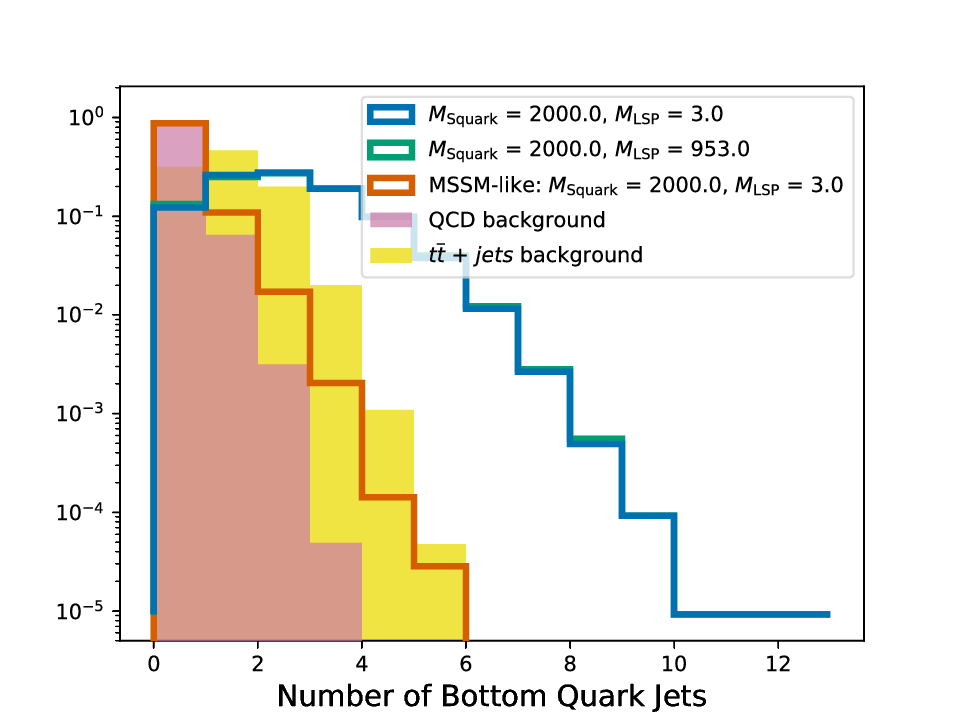}
  		\caption{BP5\label{P5_NBJet}}
	\end{subfigure}\\[5pt]

	\begin{subfigure}[t]{0.3\textwidth}
		\centering
  		\includegraphics[keepaspectratio=true,width=\columnwidth]{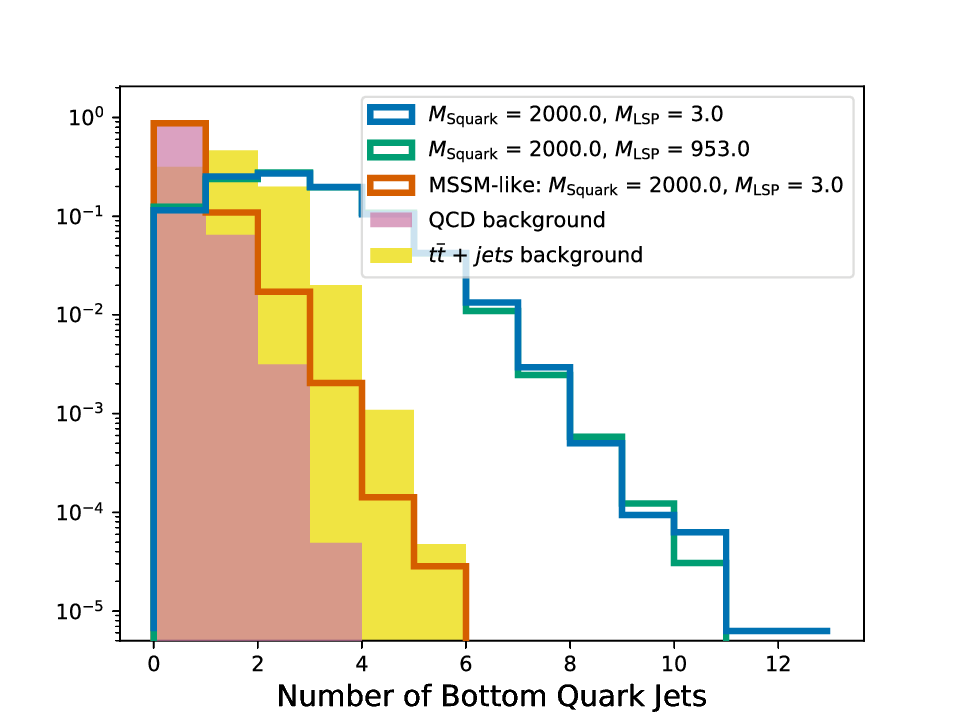}
  		\caption{BP6\label{P6_NBJet}}
	\end{subfigure}\quad%
	\begin{subfigure}[t]{0.3\textwidth}
		\centering
  		\includegraphics[keepaspectratio=true,width=\columnwidth]{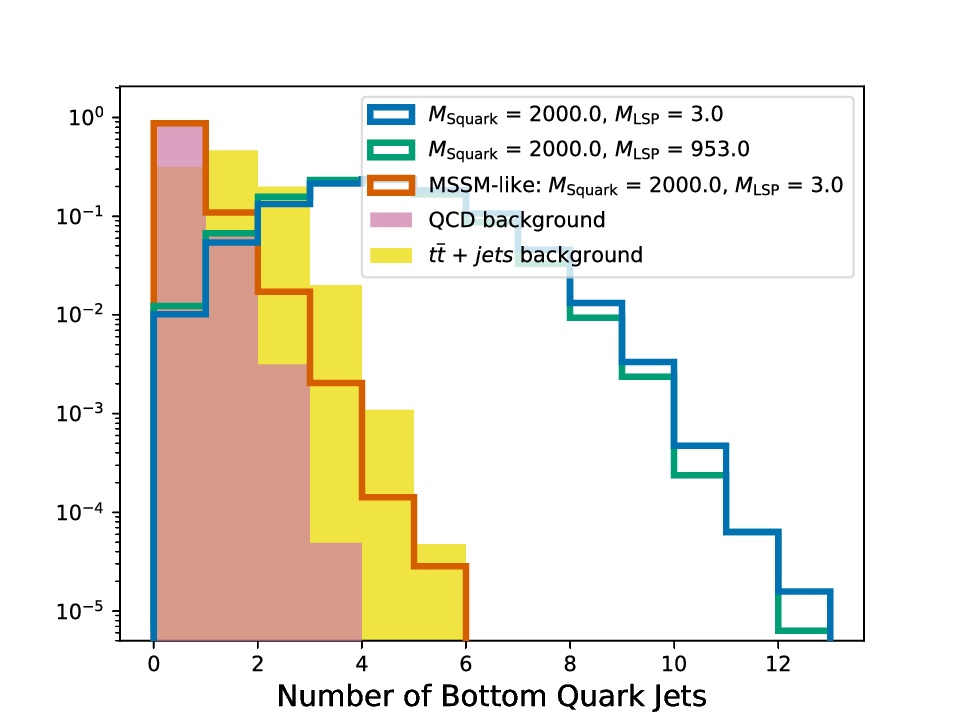}
  		\caption{BP7\label{P7_NBJet}}
	\end{subfigure}\quad%
	\begin{subfigure}[t]{0.3\textwidth}
		\centering
  		\includegraphics[keepaspectratio=true,width=\columnwidth]{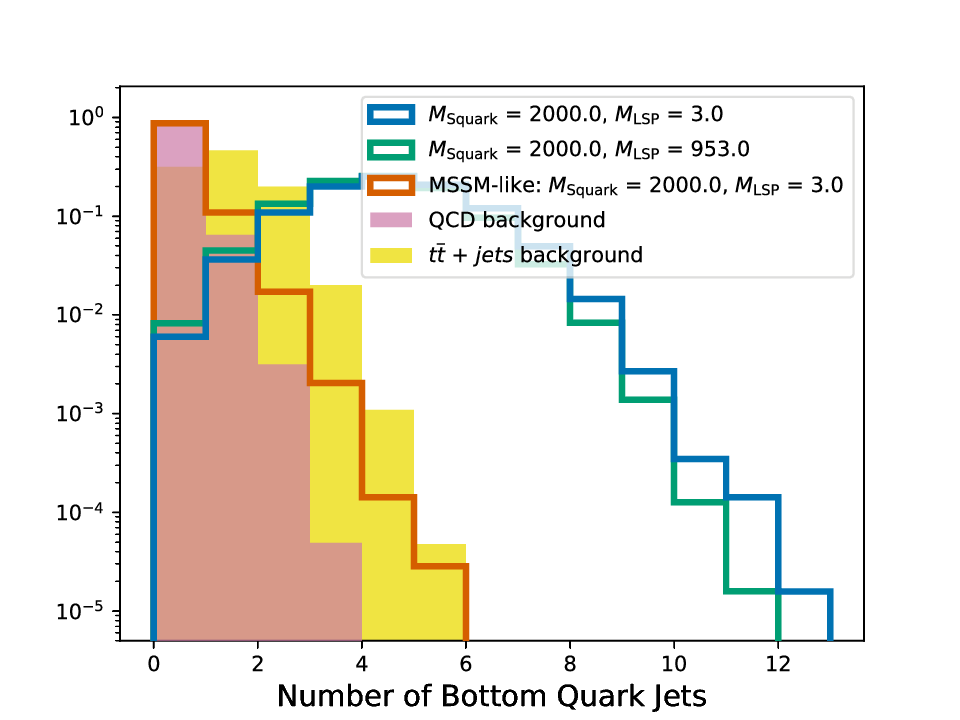}
  		\caption{BP8\label{P8_NBJet}}
	\end{subfigure}
\caption{Number of $b$-tagged hadronic jets for low and mid-range $M_{\text{LSP}}$ near the observed limit in the BP1-type scan, compared with QCD and $t\bar{t}$ background processes and an MSSM-like scenario with a light LSP.\label{NBJet_compare}}
\end{figure}

Due to uncertainty, of course, not all $b$-tagged jets are necessarily hadronic jets containing bottom quarks, nor will all bottom quarks form $b$-tagged jets. However, the efficiency of correctly $b$-tagging a bottom quark is around $70\%$, whereas the likelihood of $b$-tagging a lighter flavour quark is only $1\%$ or so. As such the average number of $b$-tagged jets for each of the example mass points in each mass scan is quite large, as shown in Figure~\ref{NBJet_compare}.

\begin{figure}
\centering
	\begin{subfigure}[t]{0.3\textwidth}
		\centering
  		\includegraphics[keepaspectratio=true,width=\columnwidth]{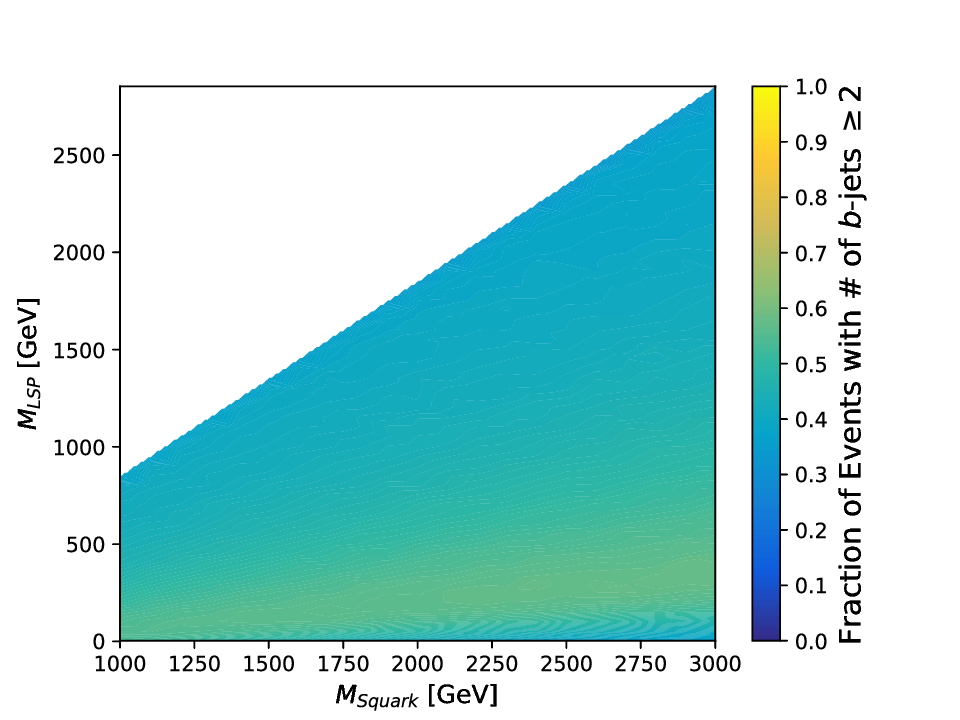}
  		\caption{BP1-type Mass Scan\label{P1_NBJeteff}}
	\end{subfigure}\quad%
	\begin{subfigure}[t]{0.3\textwidth}
		\centering
  		\includegraphics[keepaspectratio=true,width=\columnwidth]{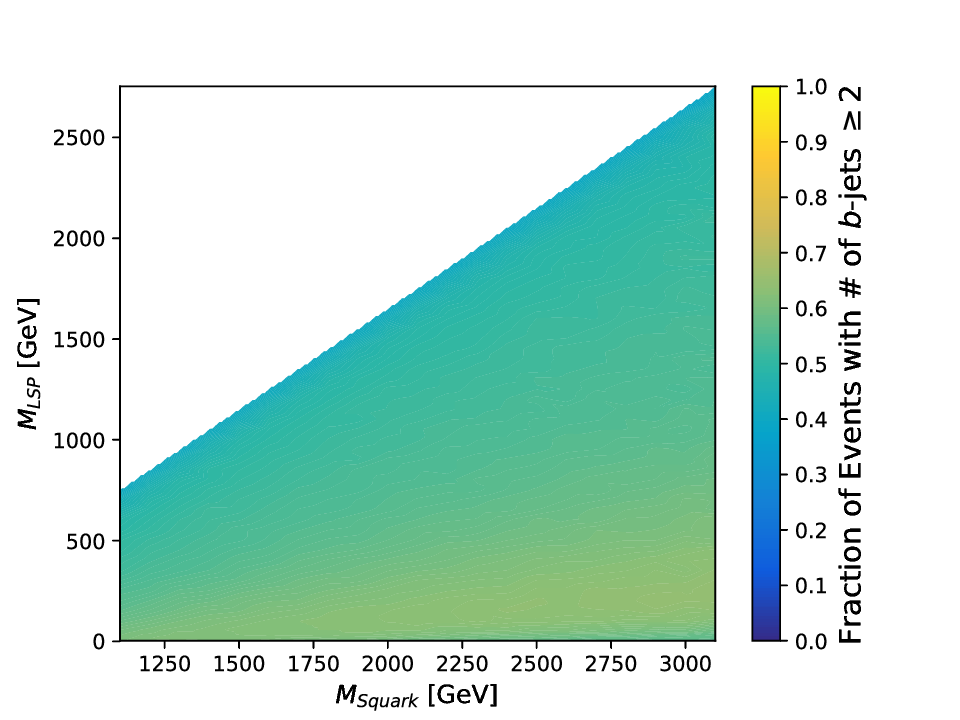}
  		\caption{BP3-type Mass Scan\label{P3_NBJeteff}}
	\end{subfigure}\quad%
	\begin{subfigure}[t]{0.3\textwidth}
		\centering
  		\includegraphics[keepaspectratio=true,width=\columnwidth]{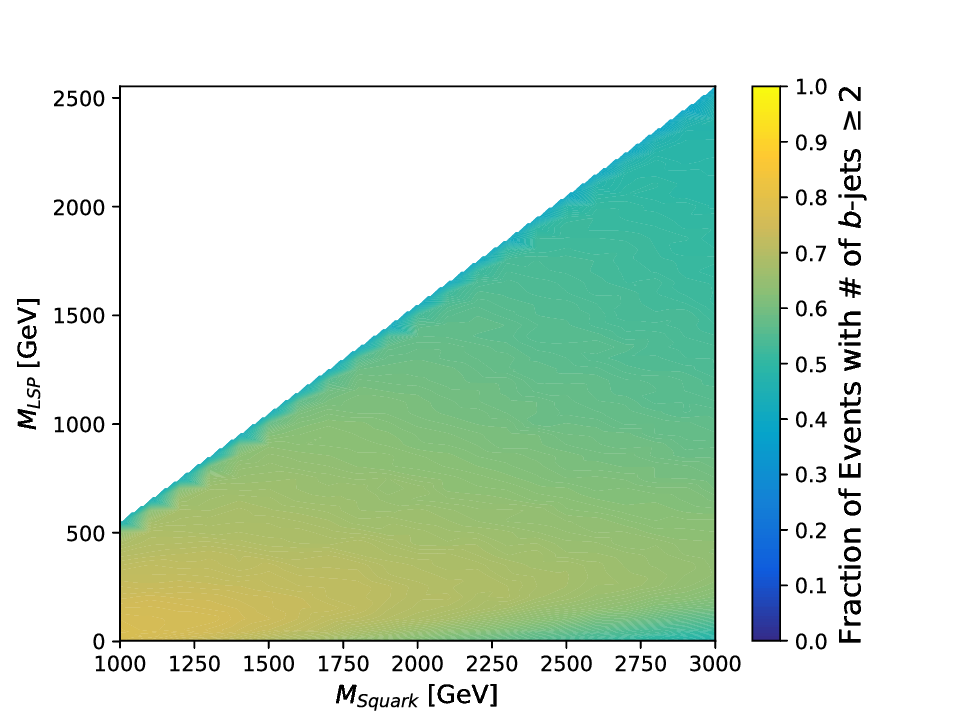}
  		\caption{BP5-type Mass Scan\label{P5_NBJeteff}}
	\end{subfigure}\\[5pt]

	\begin{subfigure}[t]{0.3\textwidth}
		\centering
  		\includegraphics[keepaspectratio=true,width=\columnwidth]{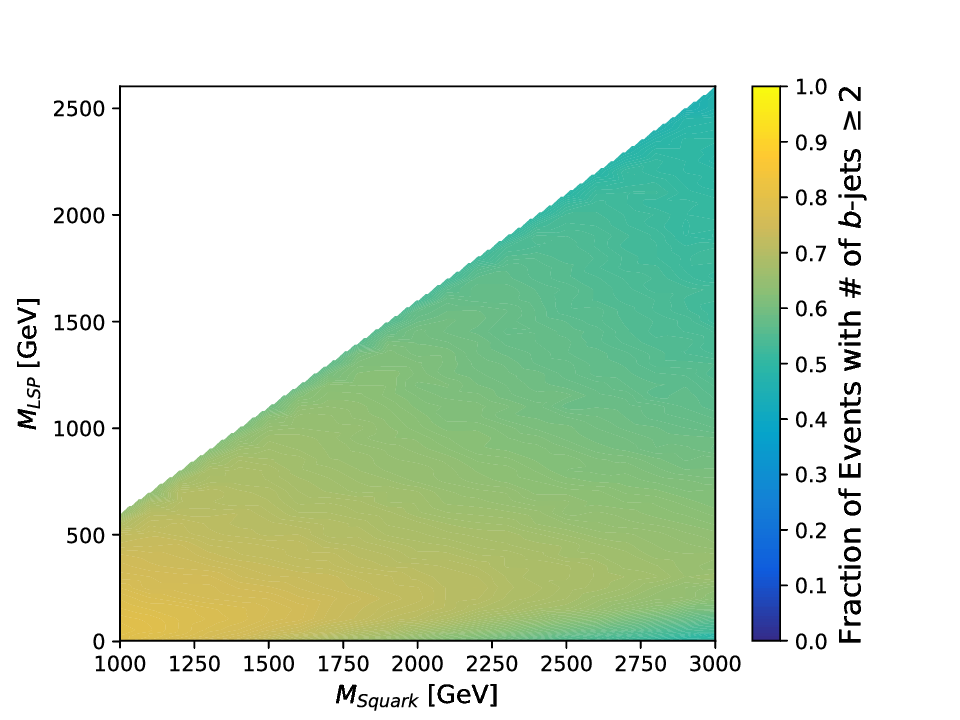}
  		\caption{BP6-type mass scan\label{P6_NBJeteff}}
	\end{subfigure}\quad%
	\begin{subfigure}[t]{0.3\textwidth}
		\centering
  		\includegraphics[keepaspectratio=true,width=\columnwidth]{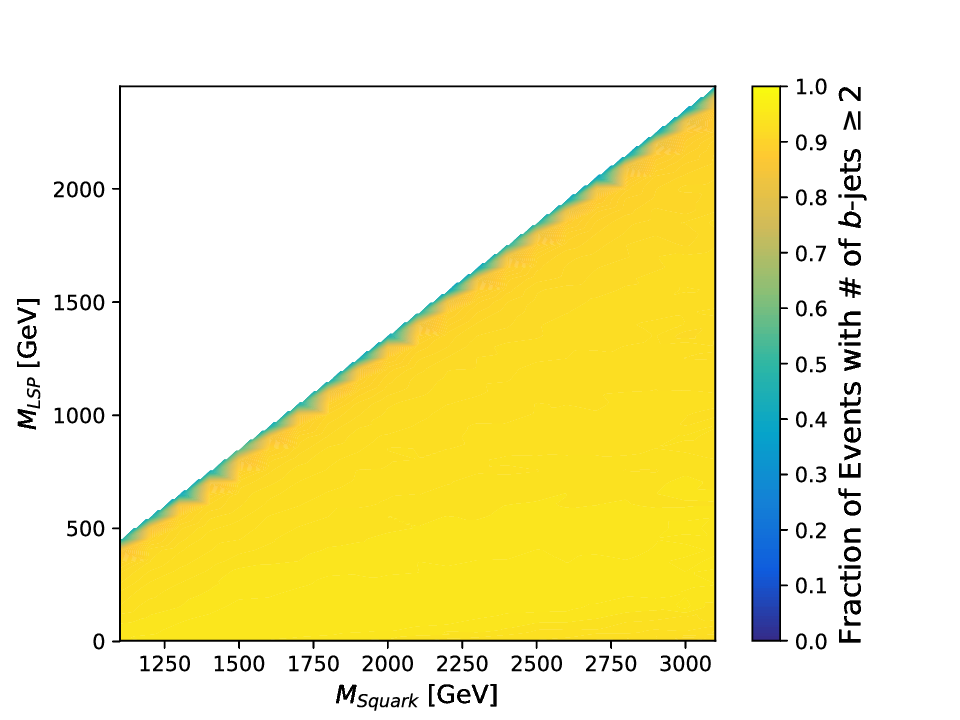}
  		\caption{BP7-type mass scan\label{P7_NBJeteff}}
	\end{subfigure}\quad%
	\begin{subfigure}[t]{0.3\textwidth}
		\centering
  		\includegraphics[keepaspectratio=true,width=\columnwidth]{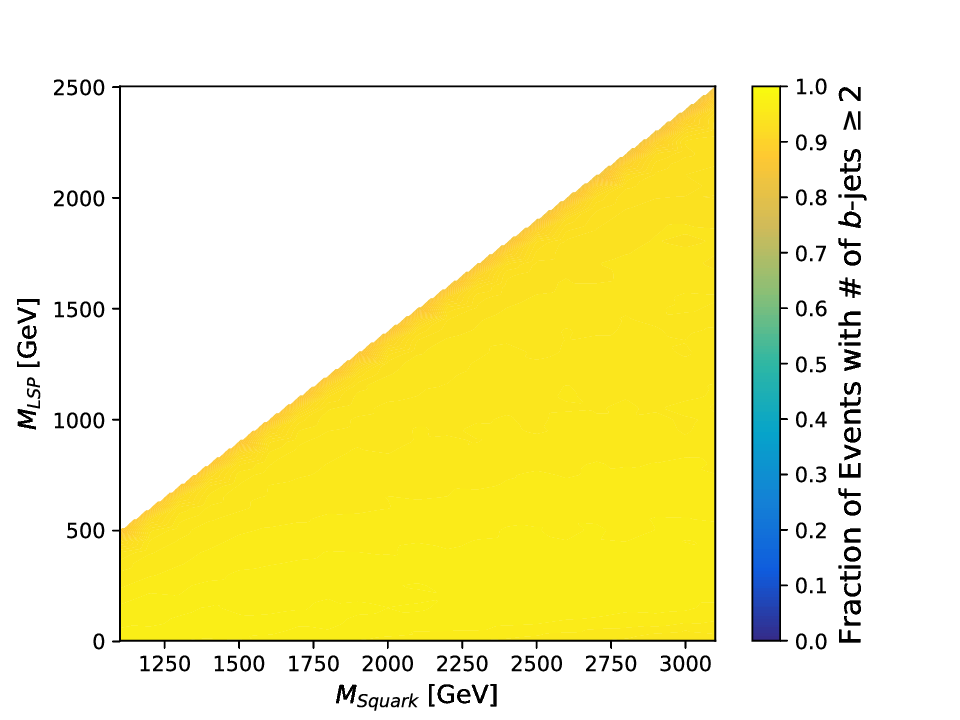}
  		\caption{BP8-type mass scan\label{P8_NBJeteff}}
	\end{subfigure}
\caption{Fraction of events with total number of $b$-tagged hadronic jets greater than or equal to 2 for the BP1-BP8-type mass scans.\label{NBJeteff}}
\end{figure}

Considering the efficiency of the binning imposed, taking the fraction of events with at least two $b$-tagged hadronic jets, it may be noted that in general at least around half of the events in these signal points contain at least two $b$-tagged hadronic jets, as shown in figure~\ref{NBJeteff}. However it is clear that in the BP7 and BP8-type scans this efficiency increases to almost 100\%, shown also by the large number of such jets in the example mass points shown for these scans in figures~\ref{P7_NBJet} and~\ref{P8_NBJet}.

The high $b$-tag multiplicities in these scans stem from the decay cascades, shown in figure~\ref{feynman}. In the case of the BP7 scan, up to four top quarks are produced, whose decays may lead to the production of bottom quarks. More simply, in the case of the BP8-type scan, it is possible to obtain up to four bottom quarks without even considering the decay of each Higgs boson.

Additionally, it may be noted that whilst the correlations between sparticle masses and the fraction of events containing at least two $b$-tagged hadronic jets is not as clear as was seen in figure~\ref{MHTeff} for \mht, say, there is generally a decrease in efficiency as the LSP mass approaches the masses of the squark and gluino. Similarly to the case for the number of hadronic jets, this is likely since the jet \ptt\ distribution is softer as the mass gaps in the SUSY cascade decrease, meaning some of the bottom quarks will not have enough transverse momentum required to pass the $40~$GeV/$c$ threshold necessary to be considered.

\boldmath
\subsection{$\Delta\phi^{*}$}
\unboldmath

Minimum {biased Delta-phi}, $\Delta\phi^{*}$, is a variable used in~\cite{CMS-SUS-16} in order to reduce the background contribution from QCD multijet events, designed in such a way that events with genuine \met\ or \mht\ would be expected to have large $\Delta\phi^{*}$ values whereas SM processes should generate small values, typically less than 0.5.

This quantity is defined as follows. For each event, first consider the hadronic jets with \ptt\ $> 40$\,GeV/$c$ and $|\eta| < 2.4$. For each of these jets the difference in azimuthal angle $\phi$ is calculated between the jet and the \mht\ calculated \emph{without that jet}.

The minimum of these $\Delta\phi$ values is then taken as minimum $\Delta\phi^{*}$, since this would be expected to be most robust against sources of spurious \mht\ such as erroneous measurement of the momentum of hadronic jets.

\begin{figure}
\centering
	\begin{subfigure}[t]{0.3\textwidth}
		\centering
  		\includegraphics[keepaspectratio=true,width=\columnwidth]{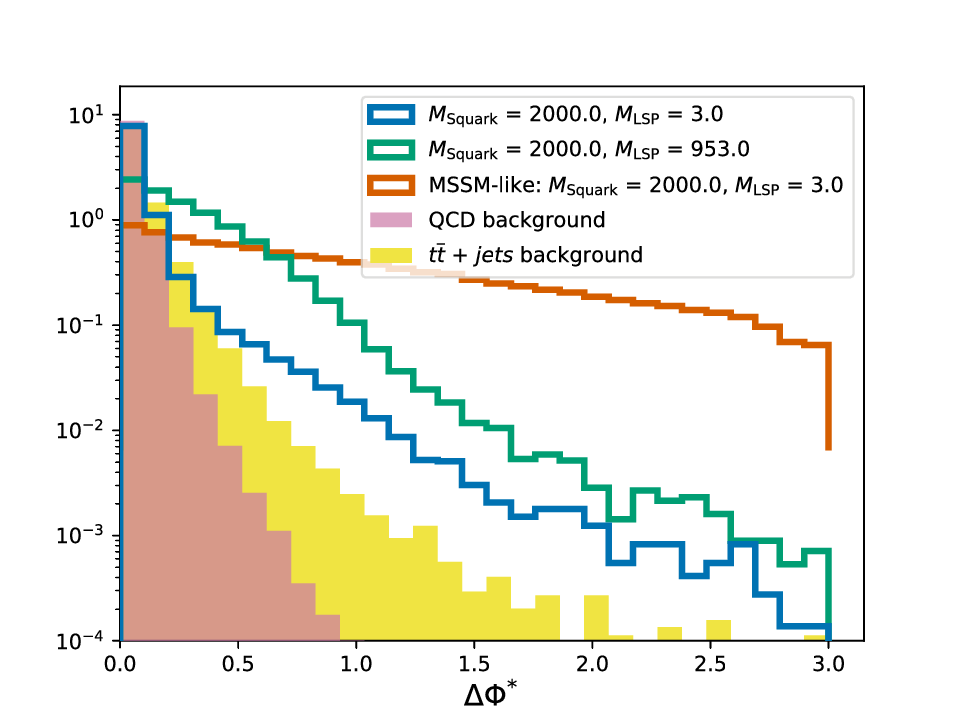}
  		\caption{BP1\label{P1_BDP}}
	\end{subfigure}\quad%
	\begin{subfigure}[t]{0.3\textwidth}
		\centering
  		\includegraphics[keepaspectratio=true,width=\columnwidth]{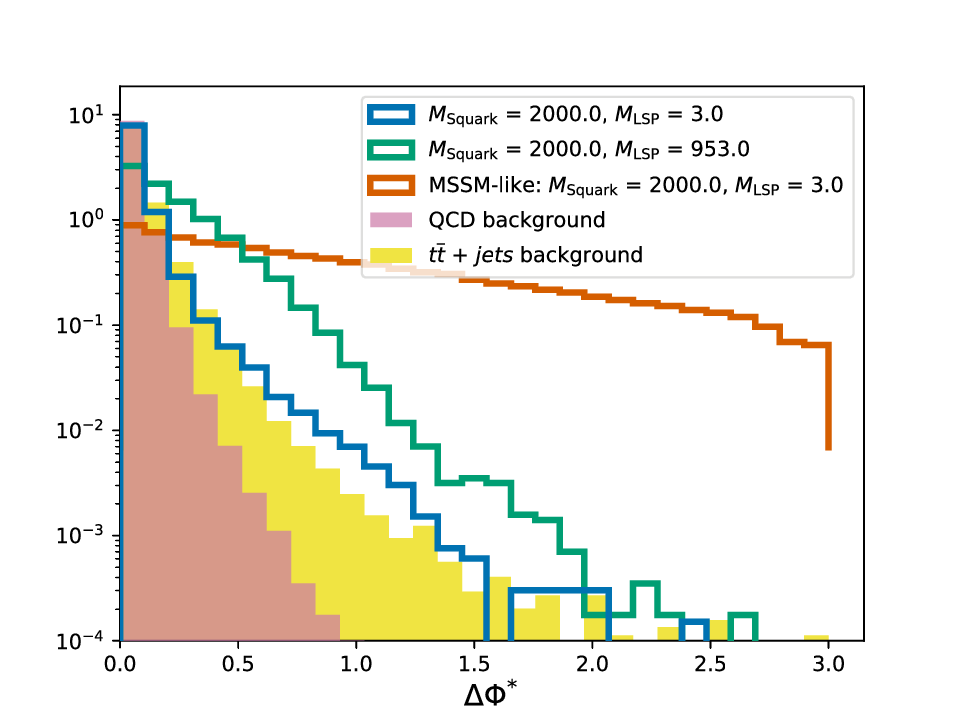}
  		\caption{BP3\label{P3_BDP}}
	\end{subfigure}\quad%
	\begin{subfigure}[t]{0.3\textwidth}
		\centering
  		\includegraphics[keepaspectratio=true,width=\columnwidth]{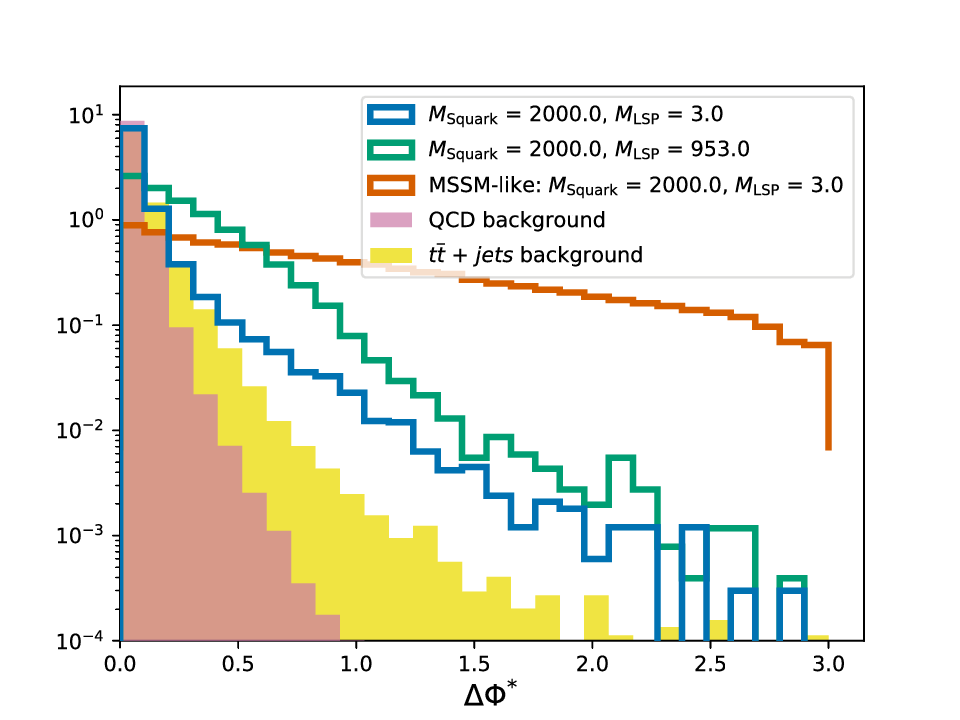}
  		\caption{BP5\label{P5_BDP}}
	\end{subfigure}\\[5pt]

	\begin{subfigure}[t]{0.3\textwidth}
		\centering
  		\includegraphics[keepaspectratio=true,width=\columnwidth]{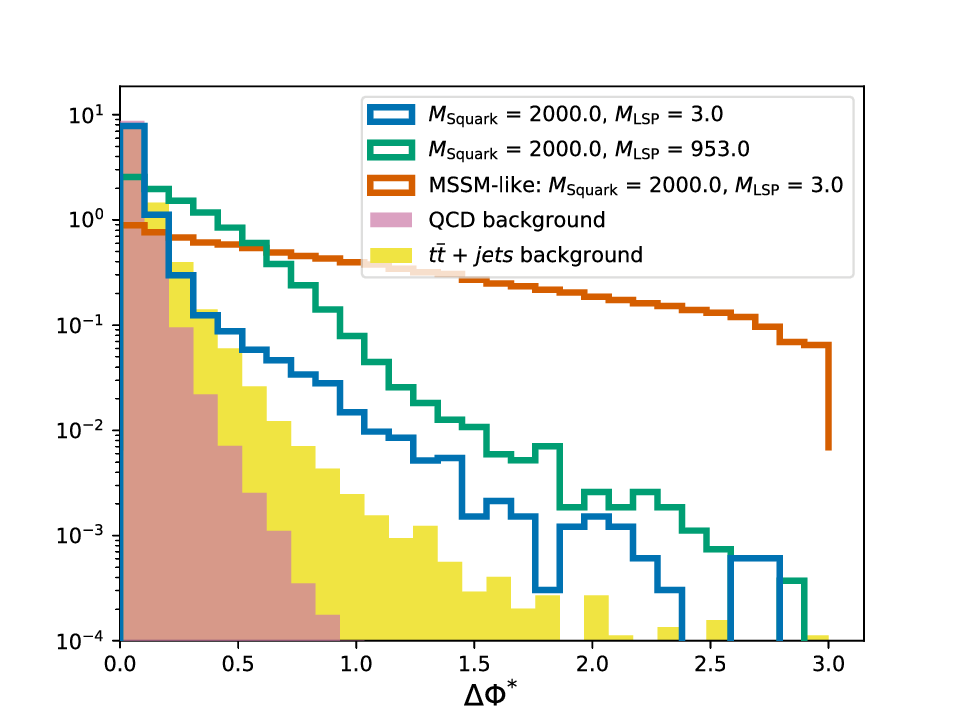}
  		\caption{BP6\label{P6_BDP}}
	\end{subfigure}\quad%
	\begin{subfigure}[t]{0.3\textwidth}
		\centering
  		\includegraphics[keepaspectratio=true,width=\columnwidth]{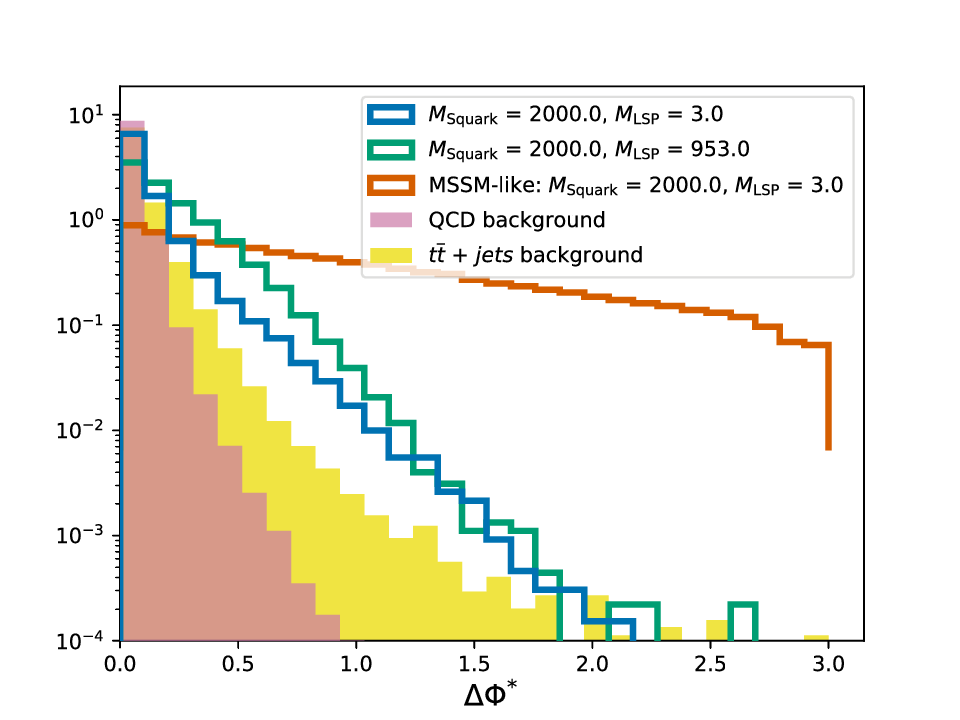}
  		\caption{BP7\label{P7_BDP}}
	\end{subfigure}\quad%
	\begin{subfigure}[t]{0.3\textwidth}
		\centering
  		\includegraphics[keepaspectratio=true,width=\columnwidth]{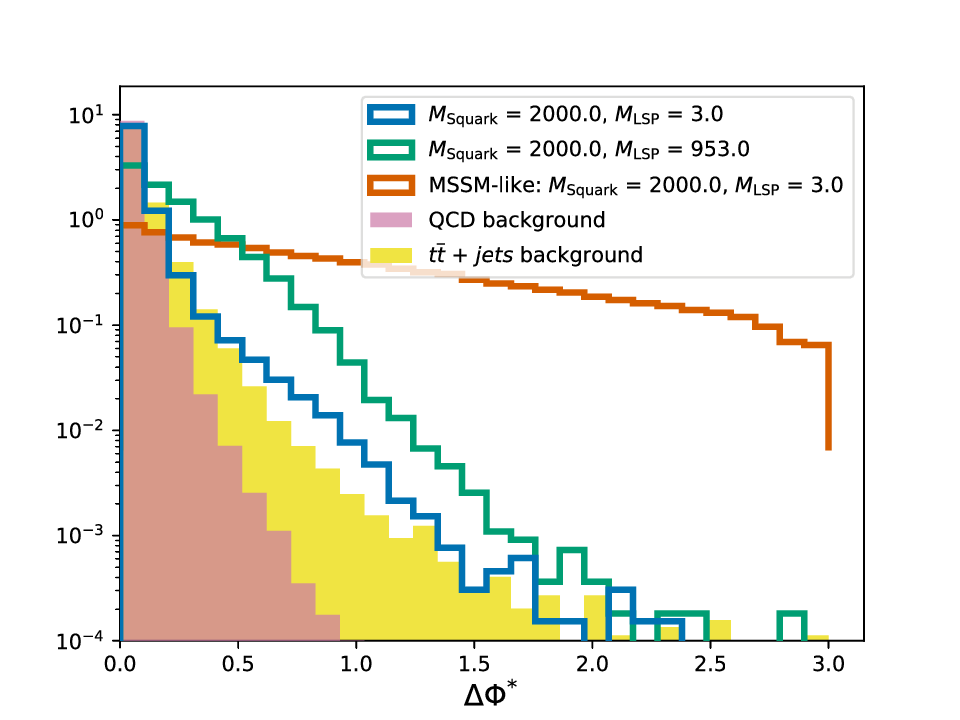}
  		\caption{BP8\label{P8_BDP}}
	\end{subfigure}
\caption{$\Delta\phi^{*}$ distributions for low and mid-range $M_{\text{LSP}}$ near the observed limit in the BP1-type scan, compared with QCD and $t\bar{t}$ background processes and an MSSM-like scenario with a light~LSP.\label{BDP_compare}}
\end{figure}

\begin{figure}
\centering
	\begin{subfigure}[t]{0.3\textwidth}
		\centering
  		\includegraphics[keepaspectratio=true,width=\columnwidth]{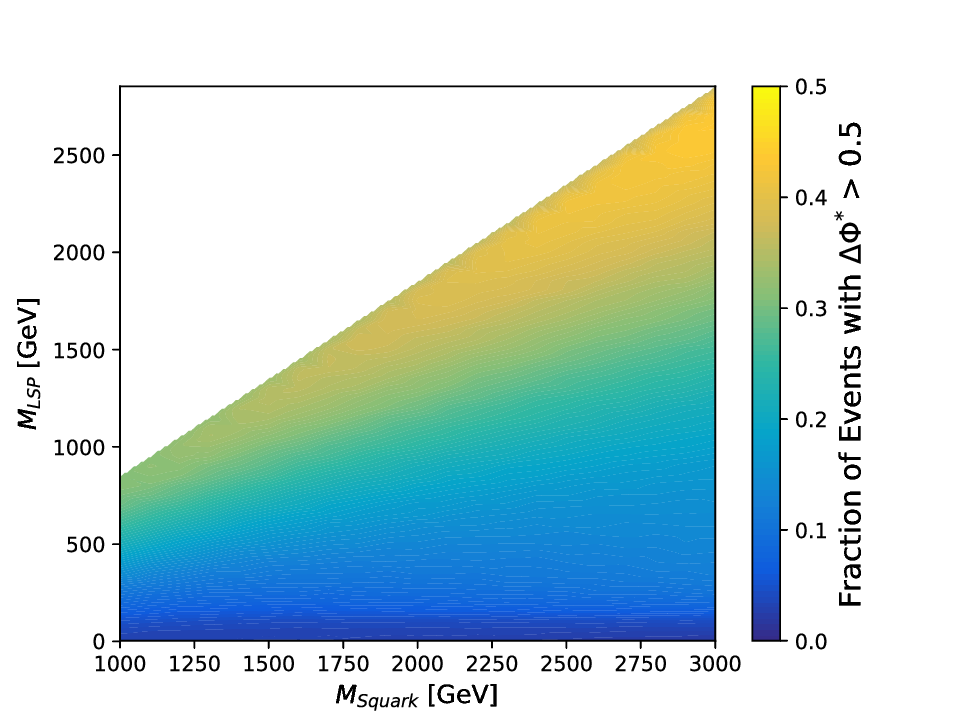}
  		\caption{BP1-type Mass Scan\label{P1_BDPeff}}
	\end{subfigure}\quad%
	\begin{subfigure}[t]{0.3\textwidth}
		\centering
  		\includegraphics[keepaspectratio=true,width=\columnwidth]{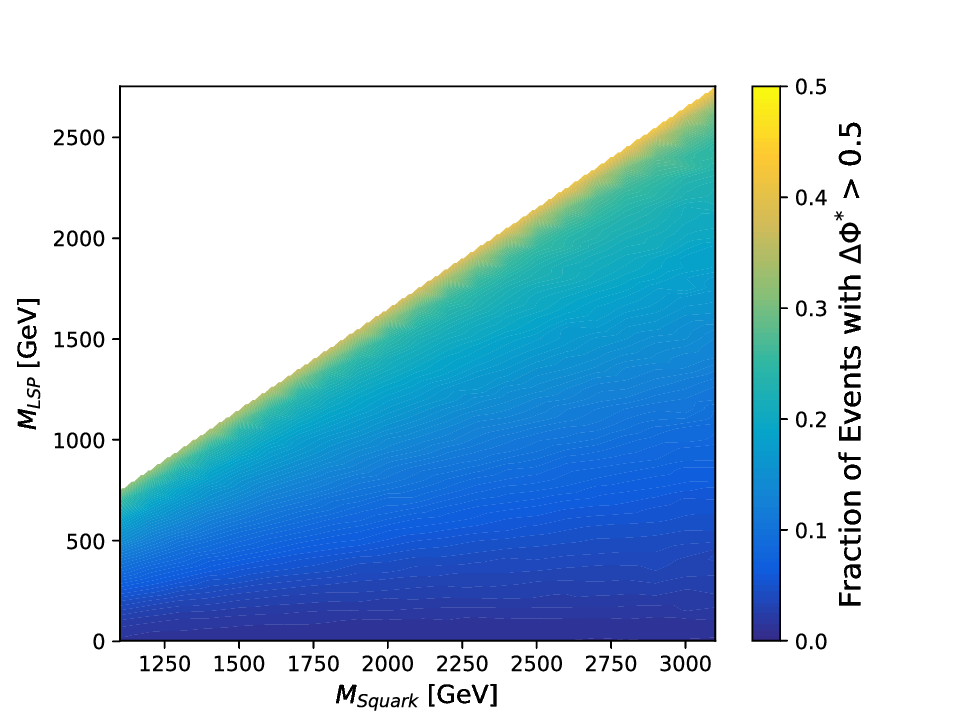}
  		\caption{BP3-type Mass Scan\label{P3_BDPeff}}
	\end{subfigure}\quad%
	\begin{subfigure}[t]{0.3\textwidth}
		\centering
  		\includegraphics[keepaspectratio=true,width=\columnwidth]{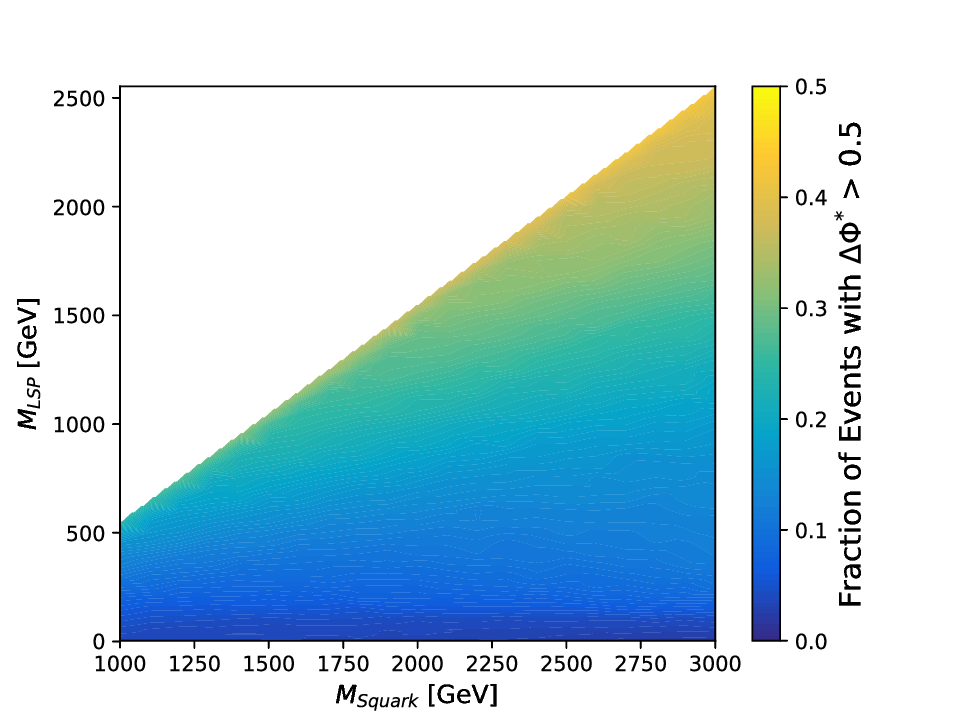}
  		\caption{BP5-type Mass Scan\label{P5_BDPeff}}
	\end{subfigure}\\[5pt]

	\begin{subfigure}[t]{0.3\textwidth}
		\centering
  		\includegraphics[keepaspectratio=true,width=\columnwidth]{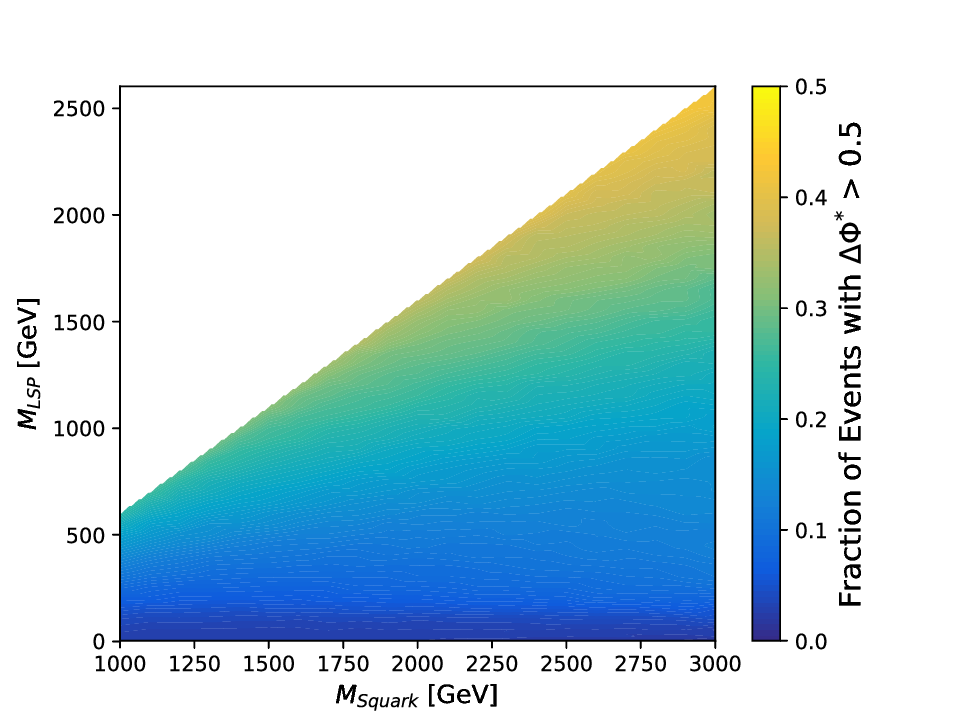}
  		\caption{BP6-type mass scan\label{P6_BDPeff}}
	\end{subfigure}\quad%
	\begin{subfigure}[t]{0.3\textwidth}
		\centering
  		\includegraphics[keepaspectratio=true,width=\columnwidth]{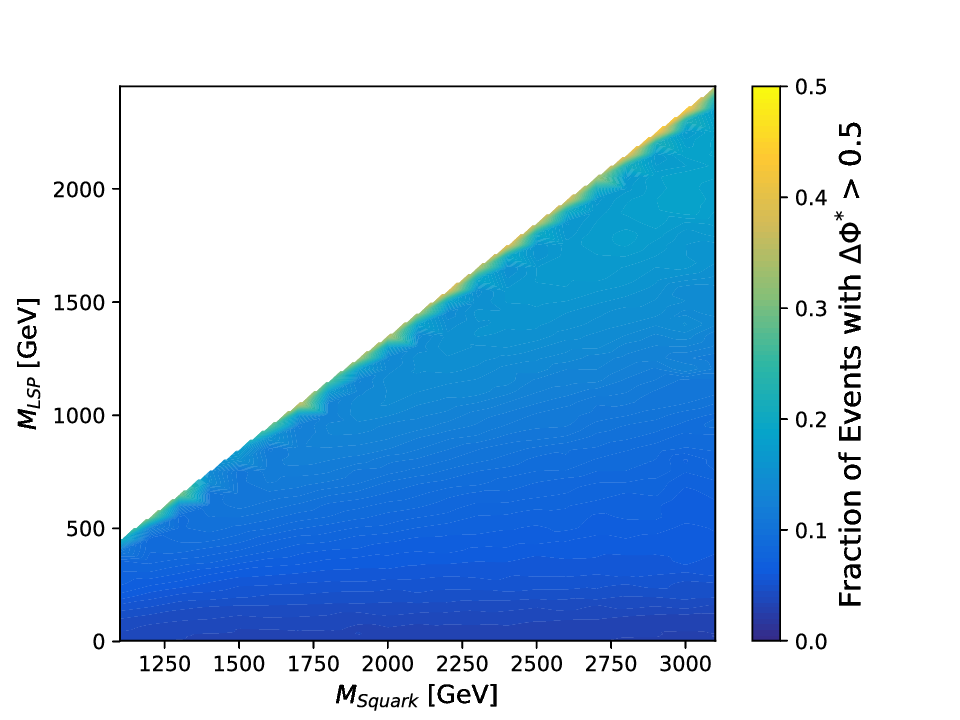}
  		\caption{BP7-type mass scan\label{P7_BDPeff}}
	\end{subfigure}\quad%
	\begin{subfigure}[t]{0.3\textwidth}
		\centering
  		\includegraphics[keepaspectratio=true,width=\columnwidth]{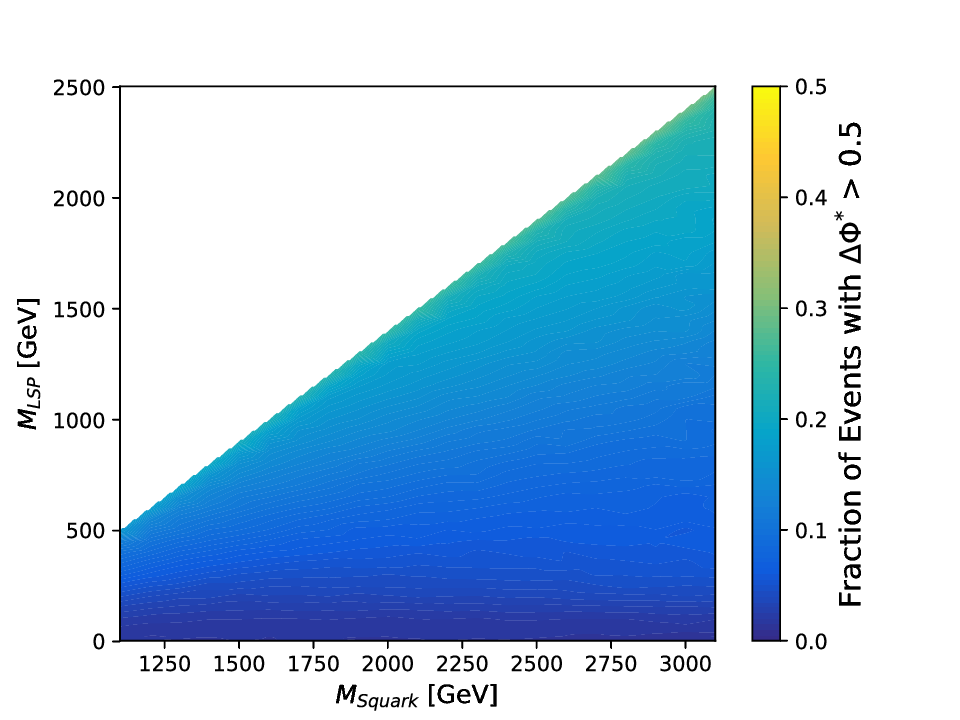}
  		\caption{BP8-type mass scan\label{P8_BDPeff}}
	\end{subfigure}
\caption{Fraction of events with $\Delta\phi^{*} > 0.5$ for the BP1-BP8-type mass scans.\label{BDP_eff}}
\end{figure}

However, the NMSSM scenarios considered in this paper do not generate many events containing large $\Delta\phi^{*}$, as shown in figure~\ref{BDP_compare}. Considering the fraction of events with $\Delta\phi^{*} > 0.5$, shown in figure~\ref{BDP_eff}, it becomes clear that a large fraction of events are rejected by the event selection in~\cite{CMS-SUS-16}, despite this variable being designed to reject QCD background events and to allow events with genuine \met.

The relatively low $\Delta\phi^{*}$ values produced in these signal mass points indicates this particular background-reduction variable is not tuned to the type of scenario under consideration in this paper.

A likely reason for small $\Delta\phi^{*}$ lies in the number of hadronic jets produced. Without of course assuming a uniform distribution in the direction of the hadronic jets, it would still be expected to become increasingly unlikely for the azimuthal angular separation between a hadronic jet and the \mht\ value computed without that jet to exceed the cut of $0.5$ as the number of jets in each event increases.

Accordingly, the regions of parameter space for which the LSP mass is close to that of the squarks and gluino display a larger fraction of events with $\Delta\phi^{*} > 0.5$, since the calculation of $\Delta\phi^{*}$ only involves hadronic jets with \ptt\ $> 40~$GeV/$c$, of which there are fewer in these regions.

Given the very large average number of jets per event in the light-LSP mass points presented in this paper, it would be expected that a $\Delta\phi^{*}$ cut would indeed kill much of the signal events in these regions of parameter space.

\subsection{Angular separation between bottom quark jets from Higgs boson decays}

One such quantity which characterises the BPs in this signal model is the angular separation $\Delta R$ between the bottom quarks from each Higgs boson decay, at MC generation (``truth'') level. The bottom quark jets considered in~\cite{CMS-SUS-16} are standard AK4 jets, formed using the anti-$k_{\text{T}}$ algorithm with a jet cone radius $R=0.4$. If two such jets from a boosted Higgs boson are close enough that their angular separation $\Delta R$ is less than the jet cone radius then it might only be possible to resolve one \emph{fat} bottom quark jet.

In order to examine the angular separation of the bottom quark jets stemming from the Higgs bosons, the MC truth information is examined. This contains the four-momenta of all individual particles in the event, rather than the emulated detector measurement of the hadronised jets, allowing for measurements of angular separation smaller than the jet cone radius of $0.4$.

\begin{figure}
\centering
	\begin{subfigure}[t]{0.3\textwidth}
		\centering
  		\includegraphics[keepaspectratio=true,width=\columnwidth]{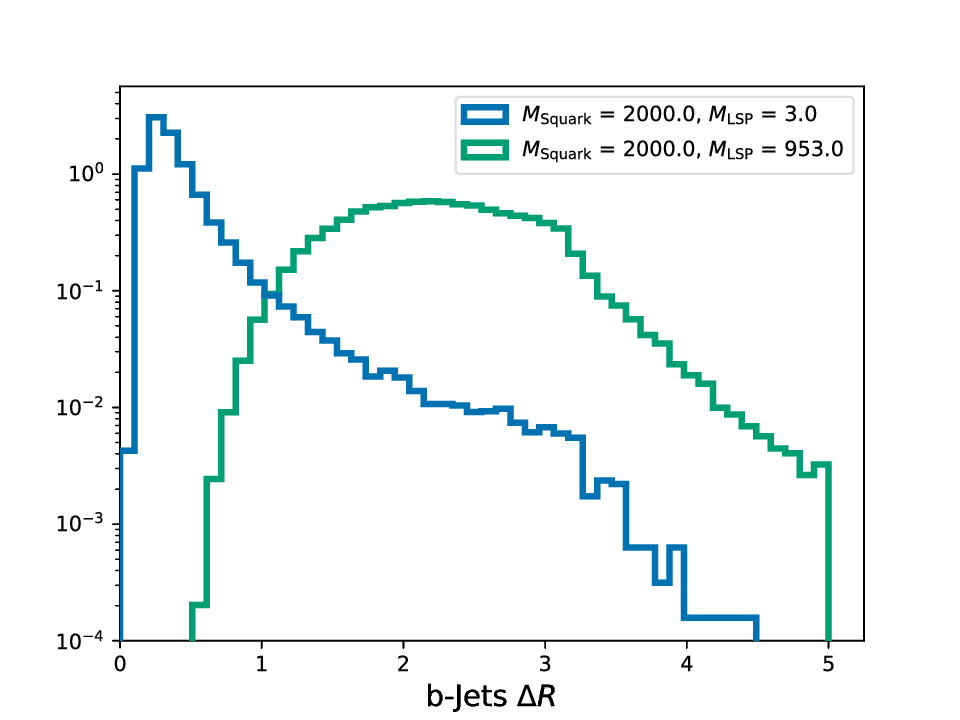}
  		\caption{BP1\label{P1_DeltaR}}
	\end{subfigure}\quad%
	\begin{subfigure}[t]{0.3\textwidth}
		\centering
  		\includegraphics[keepaspectratio=true,width=\columnwidth]{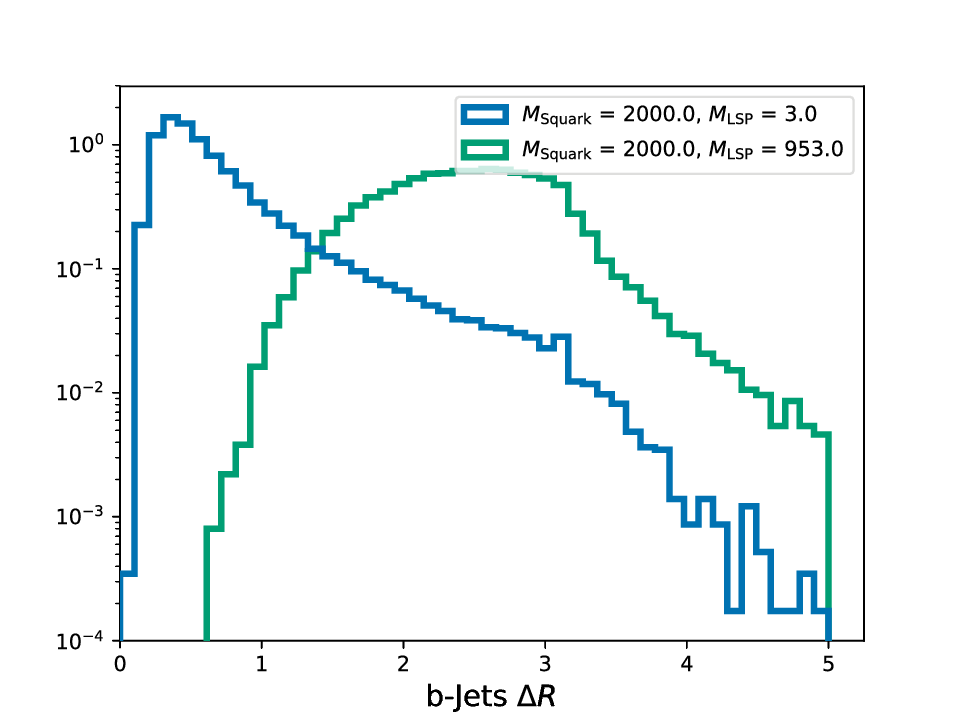}
  		\caption{BP3\label{P3_DeltaR}}
	\end{subfigure}\quad%
	\begin{subfigure}[t]{0.3\textwidth}
		\centering
  		\includegraphics[keepaspectratio=true,width=\columnwidth]{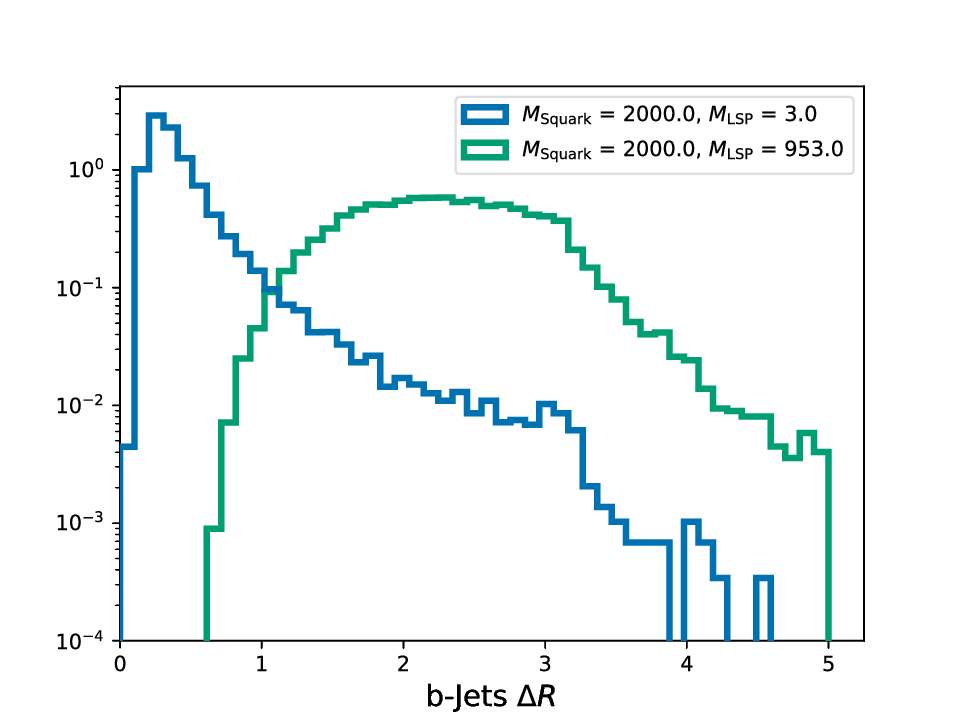}
  		\caption{BP5\label{P5_DeltaR}}
	\end{subfigure}\\[5pt]

	\begin{subfigure}[t]{0.3\textwidth}
		\centering
  		\includegraphics[keepaspectratio=true,width=\columnwidth]{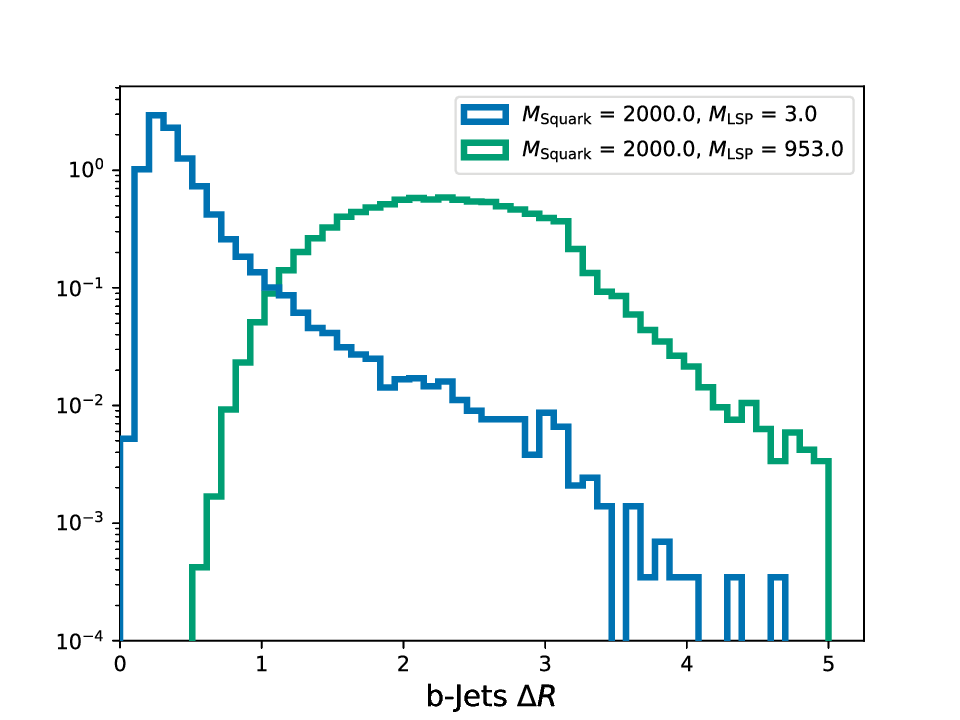}
  		\caption{BP6\label{P6_DeltaR}}
	\end{subfigure}\quad%
	\begin{subfigure}[t]{0.3\textwidth}
		\centering
  		\includegraphics[keepaspectratio=true,width=\columnwidth]{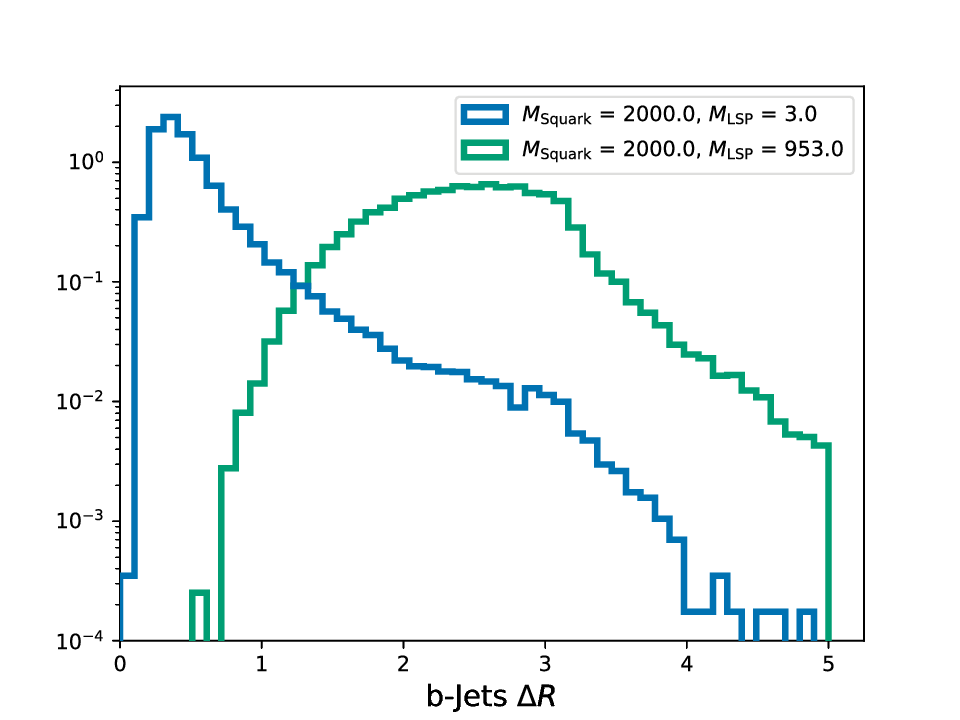}
  		\caption{BP7\label{P7_DeltaR}}
	\end{subfigure}\quad%
	\begin{subfigure}[t]{0.3\textwidth}
		\centering
  		\includegraphics[keepaspectratio=true,width=\columnwidth]{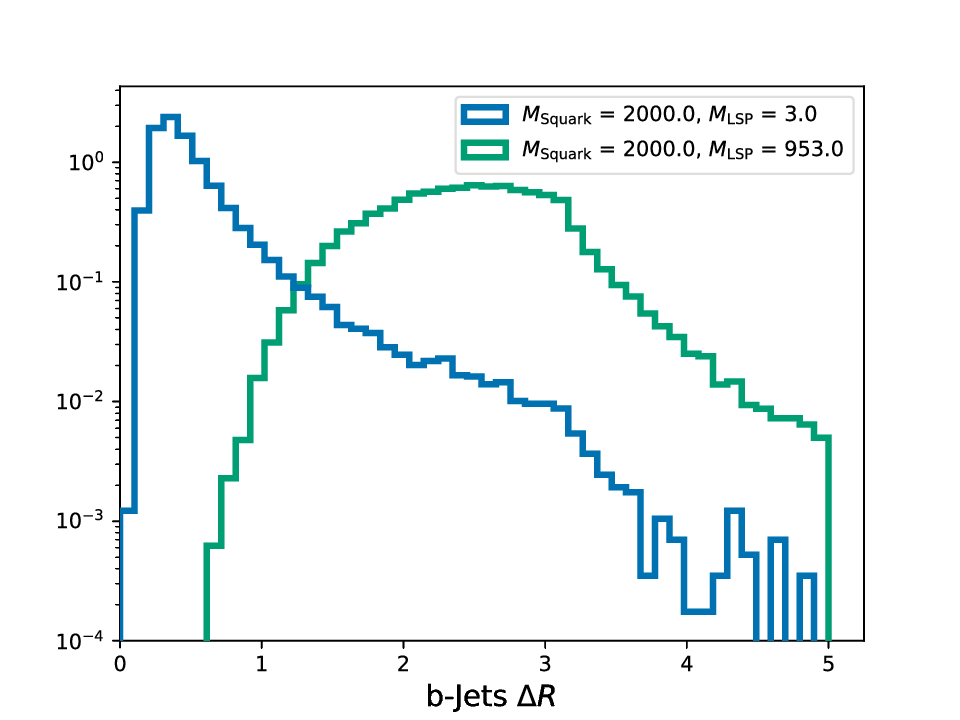}
  		\caption{BP8\label{P8_DeltaR}}
	\end{subfigure}
\caption{$\Delta R$ distributions for low and mid-range $M_{\text{LSP}}$ near the observed limit in the BP1-BP8-type mass scans.\label{Delta-R}}
\end{figure}

\looseness=1
Considering the BP1-BP8-type mass scans it may be seen in figure~\ref{Delta-R} how for example a heavy squark and a light LSP will correspond to a more boosted Higgs boson, thus decreasing the mean $\Delta R$ value. As the LSP mass increases, however, the bottom quark jets become more separated, since the LSP momentum increases relative to that of the associated Higgs boson, resulting in a more boosted LSP and a less boosted Higgs~\mbox{boson}.

Thus, in the extreme light LSP limit, it is expected that a large fraction of the bottom quarks stemming from the decay of each Higgs boson will overlap to the extent that resolving and $b$-tagging both hadronic jets from each Higgs boson will become very difficult.

Here, it may be noted that the behaviour is essentially the same for each of these six mass scans, since in all cases heavy quarks combined with a light LSP leads to a boosted Higgs boson and, as such, small angular separation between the bottom quark-antiquark~pair.


\section{Results}

A fit across the signal, background and data yields in the nine measurement bins described in table \ref{yield_table} is performed in order to determine the strength parameter $\mu$, defined as the upper limit placed on the fraction of the signal cross-section which could not quite be ruled out at $95\%$ Confidence Level (CL). As such, if for a particular point in parameter space $\mu < 1$ this point may be excluded, whereas if $\mu > 1$ the point may not be ruled out.

These limits are calculated given the data and background yields as well as background uncertainties in table \ref{yield_table}, with the systematic uncertainty on the signal yields assumed as $25$\%.

Here we present the contour limit plots for the six scan types performed within the NMSSM\@. The $X$- and $Y$-axis represent the masses of the squark and LSP, respectively, while the colour scale shows the strength parameter, $\mu$. The black contour at $\mu = 1$ then identifies the areas in parameter space inside which all mass points are ruled out, and outside which  may not be excluded, given the data yield and the background and signal estimations.

The expected limit, shown as a red contour, is defined as the upper limit we would observe if the data yields matched the background expectation exactly, and allows us to quantify any excess or deficit in the observation compared to what we would expect given the background-only hypothesis.

\begin{figure}
\centering
	\begin{subfigure}[t]{0.45\textwidth}
		\centering
  		\includegraphics[keepaspectratio=true,width=\columnwidth]{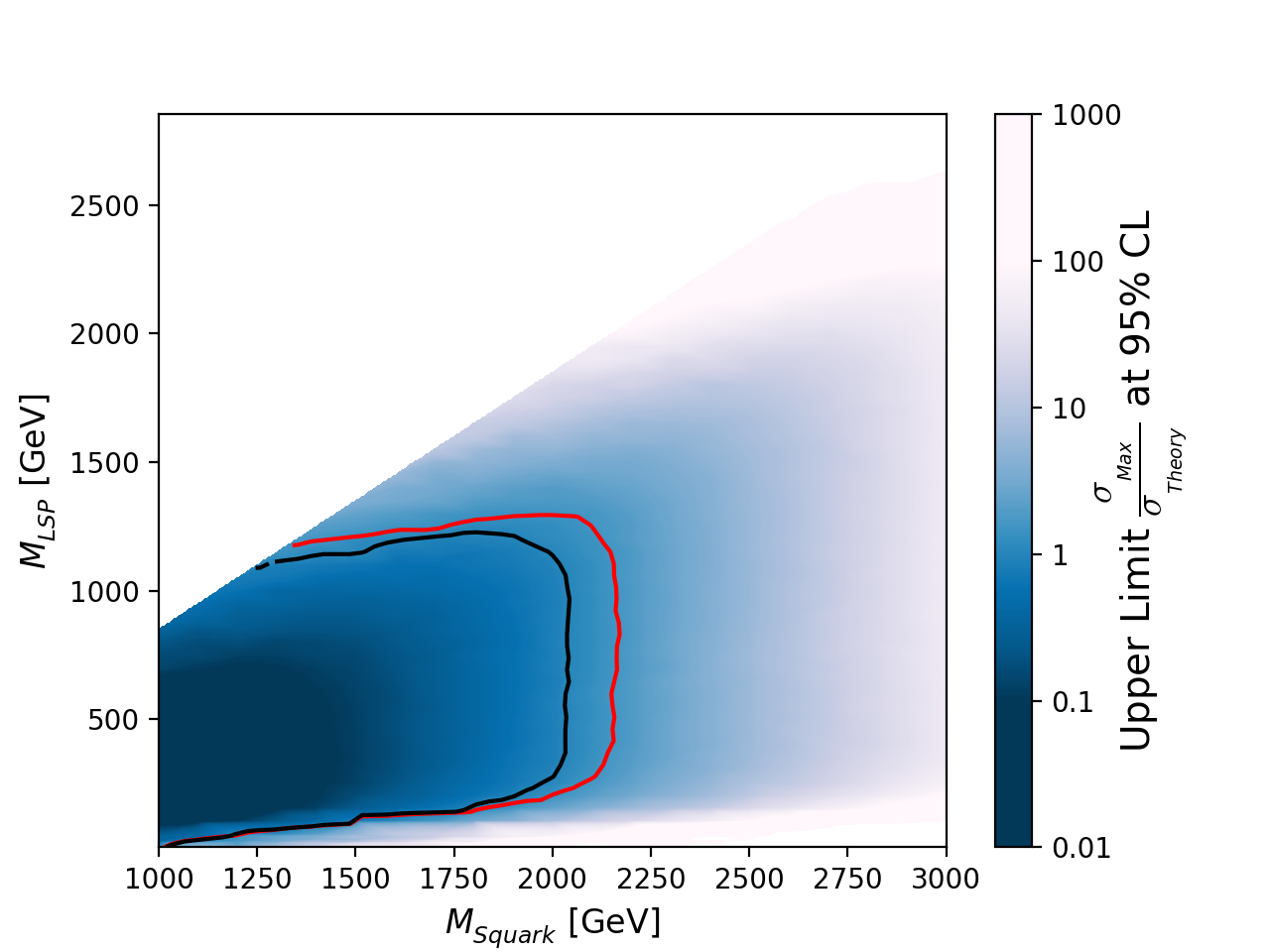}
  		\caption{BP1-type Mass Scan\label{P1_Limit}}
	\end{subfigure}\quad%
	\begin{subfigure}[t]{0.45\textwidth}
		\centering
  		\includegraphics[keepaspectratio=true,width=\columnwidth]{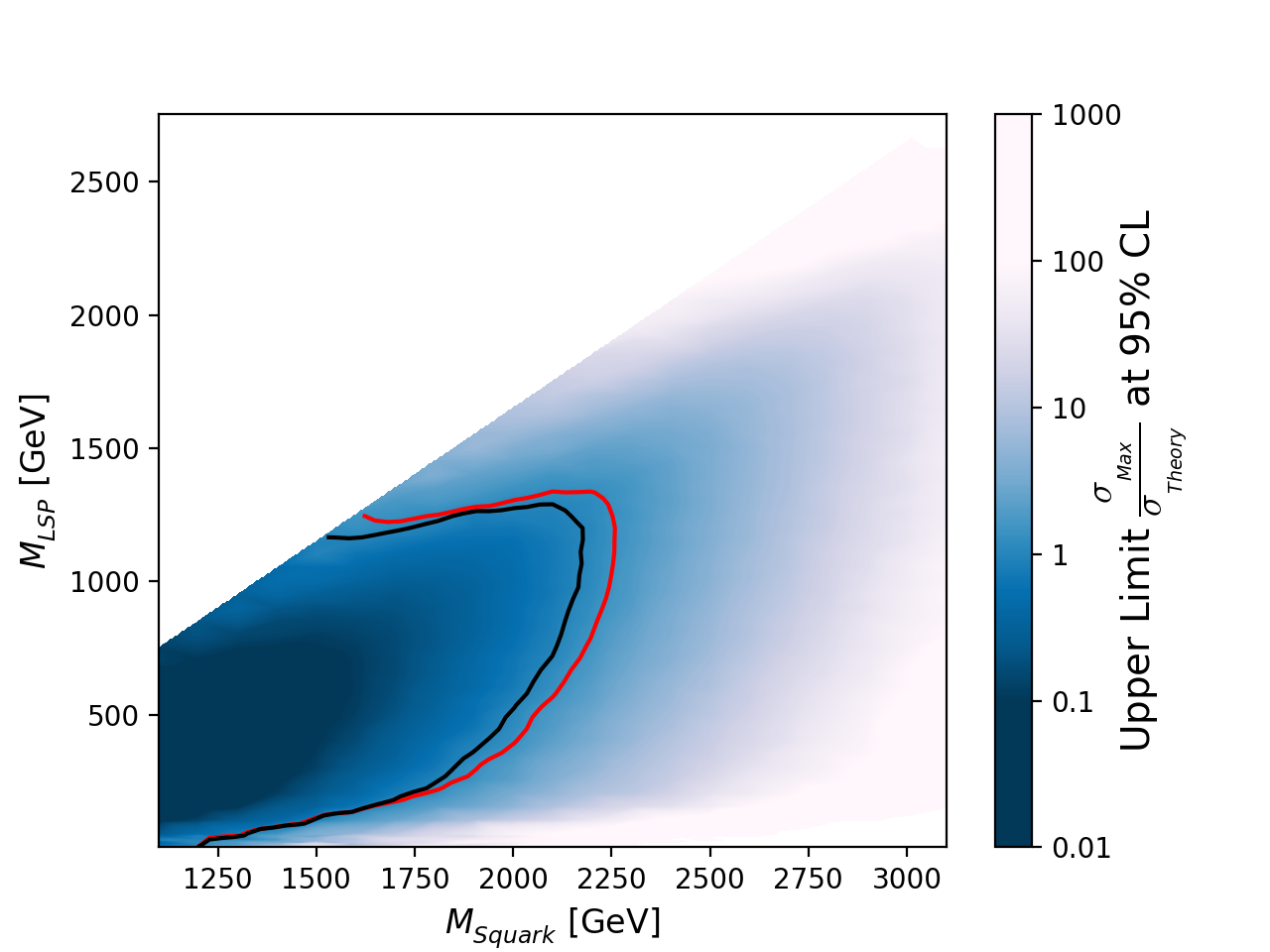}
  		\caption{BP3-type Mass Scan\label{P3_Limit}}
	\end{subfigure}\\[5pt]

	\begin{subfigure}[t]{0.45\textwidth}
		\centering
  		\includegraphics[keepaspectratio=true,width=\columnwidth]{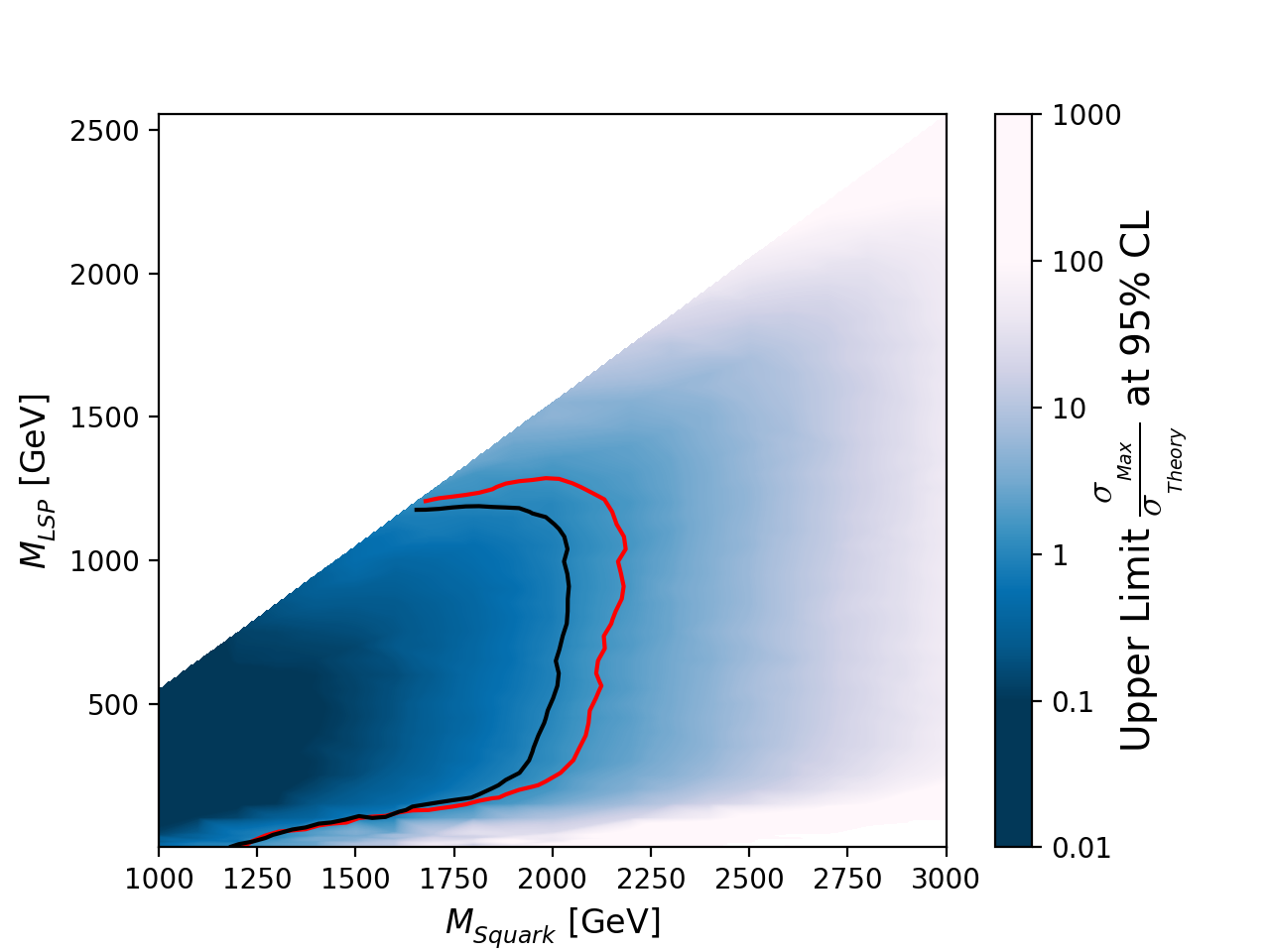}
  		\caption{BP5-type Mass Scan\label{P5_Limit}}
	\end{subfigure}\quad%
	\begin{subfigure}[t]{0.45\textwidth}
		\centering
  		\includegraphics[keepaspectratio=true,width=\columnwidth]{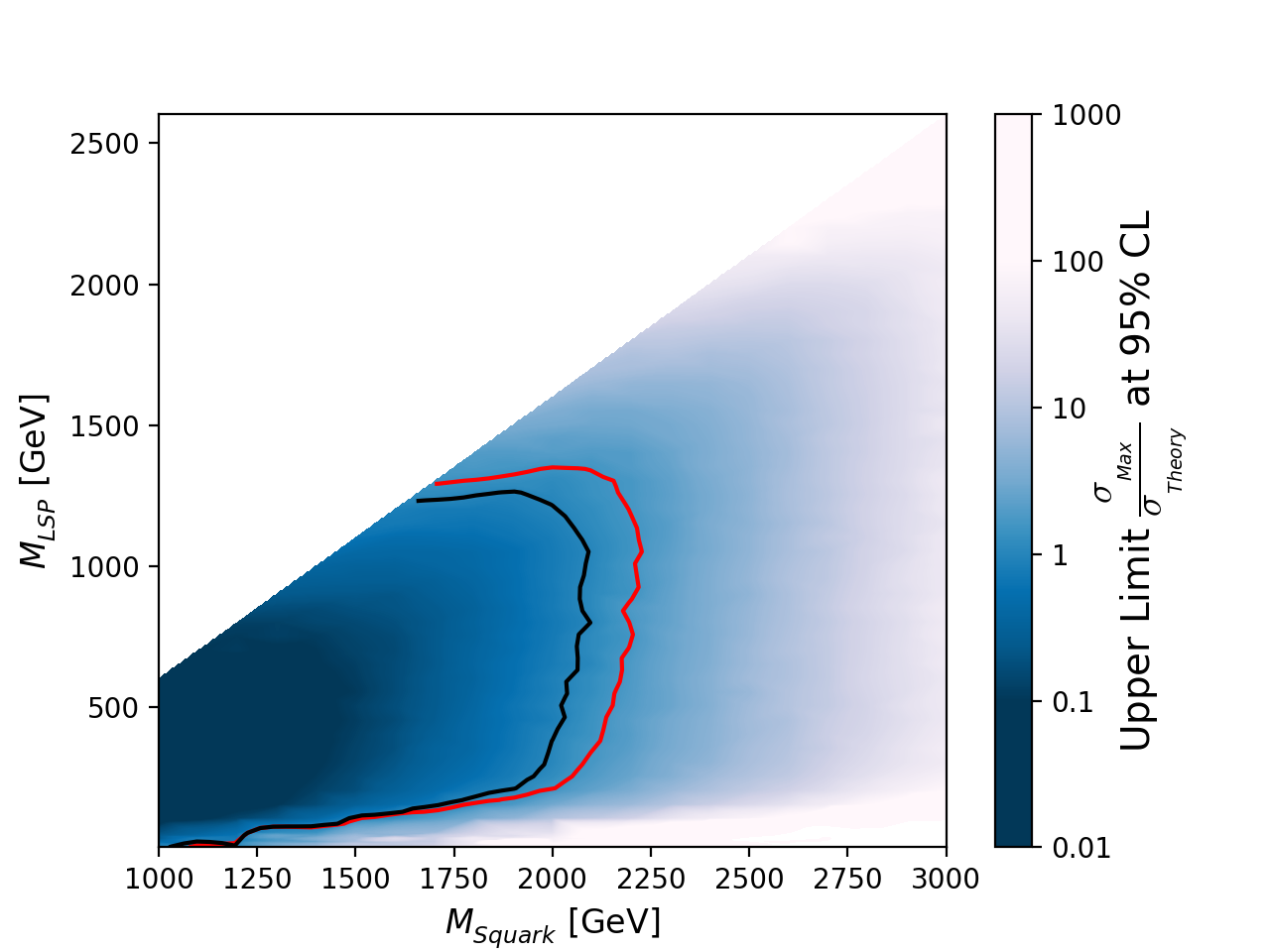}
  		\caption{BP6-type mass scan\label{P6_Limit}}
	\end{subfigure}\\[5pt]

	\begin{subfigure}[t]{0.45\textwidth}
		\centering
  		\includegraphics[keepaspectratio=true,width=\columnwidth]{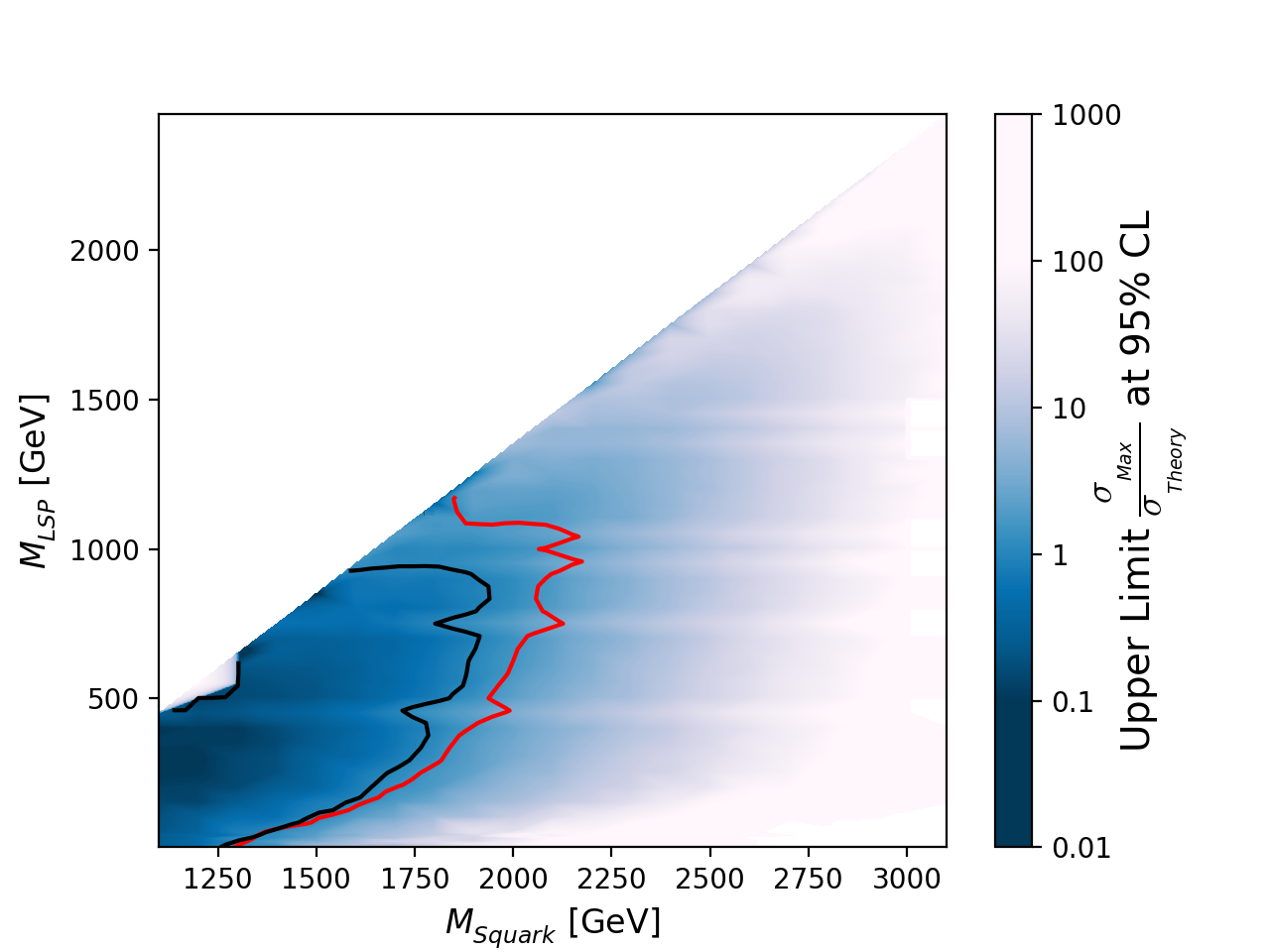}
  		\caption{BP7-type mass scan\label{P7_Limit}}
	\end{subfigure}\quad%
	\begin{subfigure}[t]{0.45\textwidth}
		\centering
  		\includegraphics[keepaspectratio=true,width=\columnwidth]{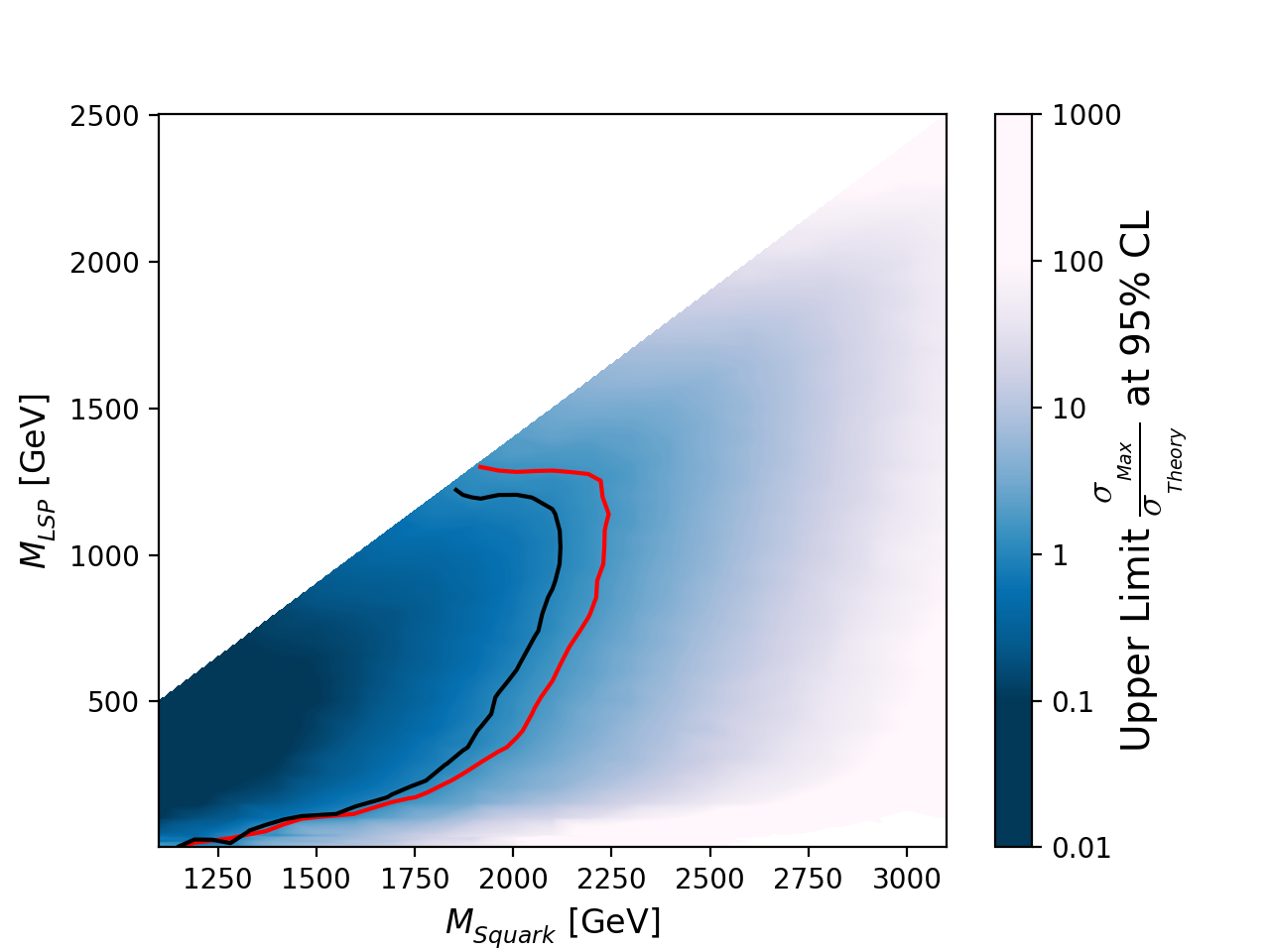}
  		\caption{BP8-type mass scan\label{P8_Limit}}
	\end{subfigure}
\caption{Observed and expected limits for the BP1-BP8-type mass scans. The $X$- and $Y$-axes represent the squark and LSP masses, respectively, whilst the colour scale represents the upper limit on the strength parameter $\mu$.\label{Scans}}
\end{figure}

In figure~\ref{Scans} it can be seen that both the observed and expected limits are far weaker for a very light LSP mass, with both contours bending to the left such that points with much lower squark masses are no longer excluded. As such it would appear the sensitivity of the analysis in~\cite{CMS-SUS-16} decreases dramatically in the limit of a light LSP in these NMSSM scenarios, with the lower bound on the squark mass decreasing from around $2~$TeV/$c^{2}$ to as little as $1~$TeV/$c^{2}$ in some cases, as shown in table~\ref{limit_table}.

Additionally whilst the black observed limit contour is generally further to the left than the red, expected limit contour, indicating a slight excess in some of the data yields compared with the background estimation, the agreement between the two limits is reasonably strong.

These weaker lower bounds on the squark masses for the lightest LSP mass of $3$\,GeV/$c^{2}$ are summarised in table~\ref{limit_table}.
\begin{table}
\center
	\begin{tabular}{ |c|c|c|c|c|c|c| }
	\hline
	Scan & BP1 & BP3 & BP5 & BP6 & BP7 & BP8 \\
 	\hline
 	$M_{\tilde{q}, \text{ min}}$ [GeV/$c^{2}$] & 1000 & 1200 & 1250 & 1000 & 1250 & 1200\\
 	\hline
    $M_{\tilde{g}, \text{ min}}$ [GeV/$c^{2}$] & 1010 & 1000 & 1260 & 1010 & 1050 & 1000\\
 	\hline
 	\end{tabular}
 	\caption{Approximate lower bounds on the squark mass and corresponding gluino mass at 95\% CL for a $3$\,GeV/$c^{2}$ LSP.}
 	\label{limit_table}
\end{table}

It may be noted that the lower bounds on the squark and gluino masses are considerably weaker for these light-LSP, low-\met\ scenarios compared with the simplified models considered in~\cite{CMS-SUS-16}.

As the LSP mass is increased above around $100~$GeV/$c^{2}$ the converse becomes true, with the limits being more harsh for these NMSSM scenarios than for the simplified models. This is expected due to the larger \mht\  in these heavier LSP regions.

However, as the LSP mass is increased closer towards the masses of the squarks and gluinos, the sensitivity appears once again to decrease for heavier neutralinos. Unlike the light LSP region, however, this lack of sensitivity for heavy LSP likely arises from the high \htt\ cut, since few events in this region pass this cut as shown in figure~\ref{P1_HTeff}. Thus, in order to explore this area of mass space a wider \htt\ range would be required than is considered in this paper.

Similar experimental limits can be placed on the other types of scan. We recall here that the BP3-type scan has the gluino mass $200$\,GeV/$c^{2}$ lower than the squark mass, rather than $10$\,GeV/$c^{2}$ higher, and the BP5/BP6- and BP7/BP8-type scans are the same as the BP1- and BP3-types, respectively, but with the appropriate stop or sbottom squark masses $250$\,GeV/$c^{2}$ lighter than the squark/gluino, rather than being decoupled.

The observed and expected limits for these remaining mass scans exhibit a similar behaviour, that is, the cross-section appears to dominate the sensitivity for points with mid-range LSP mass, where the contours are closer to vertical. However, in all cases the sensitivity for regions with lower LSP masses and featuring high \htt\ and low \mht, to the latter of which the analysis in~\cite{CMS-SUS-16} is not optimised, decreases dramatically.

\subsection{MSSM-like scenarios with light LSP}

The main feature of the light LSP and low \met\ scenarios under consideration relies on the LSP being singlino. In this case it is possible for the decay cascades to end exclusively in an NLSP decaying to an LSP and a Higgs boson, which is not the case in the MSSM.

\begin{figure}
\centering
  \includegraphics[keepaspectratio=true,width=90mm]{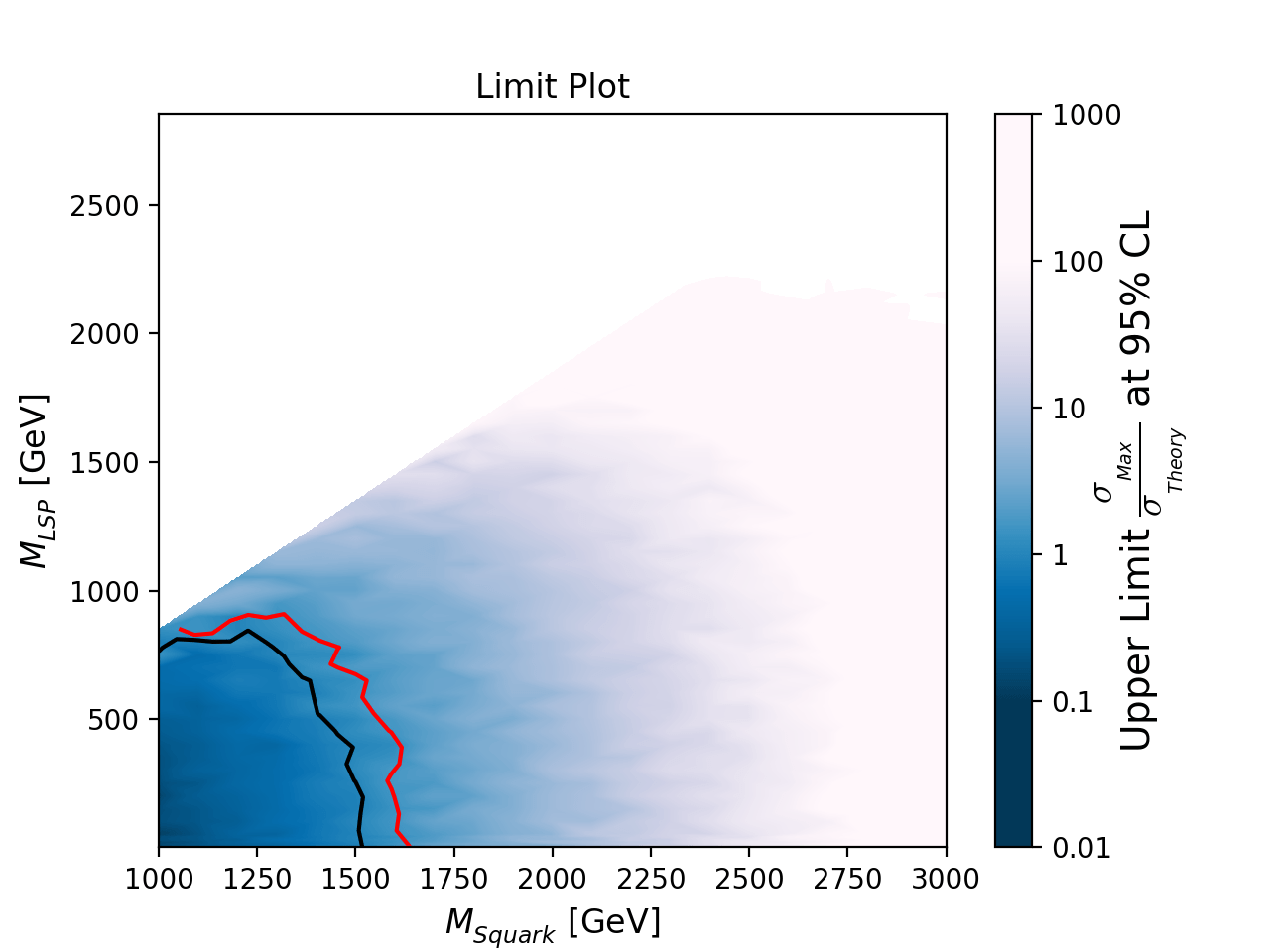}
\captionof{figure}{Observed (black) and expected (red) limits for an MSSM-like scenario demonstrating the higher sensitivity to regions with low LSP mass.\label{MSSM_scan}}
\end{figure}

For comparison, figure~\ref{MSSM_scan} shows the sensitivity to the simplified MSSM-like scenario seen previously. Applying the same event selections as in~\cite{CMS-SUS-16}, the observed and expected limits are calculated at 95\% CL, akin to the limits in figure~\ref{Scans}.

In the limit plots in figures~\ref{Scans} and~\ref{MSSM_scan} it may be observed that the experimental sensitivity to this simplified MSSM-like model is in fact strongest for the lightest LSP mass of $3$\,GeV/$c^{2}$, contrary to the NMSSM-specific low-\met\ mass scenarios.
Conversely,  it is also clear that the overall sensitivity for the  regions of parameter space in which \mlsp\ $> 200$\,GeV/$c^{2}$ is weaker for this model compared with the NMSSM scenarios considered. Since the decay cascade is truncated and there are no Higgs boson decays in this model, the expected numbers of hadronic jets and $b$-tagged jets per event are lower. Therefore, it is unlikely for as many events to contain more than five hadronic jets or as many as two $b$-tagged jets,  so they will not pass the event selections which are imposed in this paper. However, exploiting the full 254 bins in~\cite{CMS-SUS-16} would be expected to increase the sensitivity to this MSSM-like scenario.

\section{Conclusion}

Fairly strong limits of around $2~$TeV/$c^{2}$ have been placed on the squark/gluino masses for a singlino-like LSP in the NMSSM for a LSP mass above $100$\,GeV/$c^{2}$ or so. However, below this mass the limits weaken by a considerable amount in all cases, as shown in table \ref{limit_table}, with limits for some scenarios decreasing by as much as $1~$TeV/$c^{2}$. The eight original BPs in~\cite{UlrichAna}, all featuring a light $3$\,GeV/$c^{2}$ LSP, despite having a large direct production cross-section, are still on or around the limit of sensitivity of the analysis in~\cite{CMS-SUS-16} with $35.9~\text{fb}^{-1}$ data from the CMS detector at the LHC,  thus they cannot be completely excluded at this stage.

It is clear that these light LSP and low \met\ scenarios present further challenges for jets+\met\ based searches akin to~\cite{CMS-SUS-16} at the LHC\@. In order to develop a search for these stealthy scenarios one might wish to access regions of low \mht, however, this would require careful techniques so as to not allow yields from background processes to dominate.

Given the expected high $b$-tag multiplicities in these scenarios (cf.\ figure~\ref{NBJet_compare}) sensitivity may be boosted, though, if one concentrates on signal regions with more $b$-\emph{tagged} jets. One may employ new techniques such as a double $b$-tagger, one of which being developed by the CMS collaboration \cite{CMS_doubleb} and another by the ATLAS collaboration \cite{ATLAS_doubleb} whereby, in the former, AK8 jets are formed and, depending on the substructure, a discriminator output is assigned by a Boosted Decision Tree (BDT), describing how likely the jet is to contain two bottom quarks originating from the same decay vertex. Here AK8 refers to the anti-$k_{\text{T}}$ algorithm with a jet cone radius of $0.8$. This tool thus allows for a way to distinguish between single bottom quark jets and two overlapping bottom quark jets, such as may be found originating from the decay of a boosted Higgs boson, a technique employed in~\cite{CMS-SUS-17-006}.

High $b$-tag multiplicities occur also in scenarios where the r\^ole of the SM Higgs boson is played by a lighter NMSSM specific Higgs boson $H_1$~\cite{Ellwanger}. Using the double-$b$-tagging method, distributions of the double-$b$-tagged jet mass can be studied. In any scenario involving SM or NMSSM specific Higgs bosons, distributions of the double-$b$-tagged jet mass should exhibit excesses over a smoothly decreasing background from QCD and top quark pair production around the corresponding Higgs boson mass~\cite{Ellwanger}. This might lead to a simultaneous discovery of an extra lighter Higgs boson $H_1$.

\section*{Acknowledgments}

SM, CHS-T and AT are supported in part through the NExT Institute, AT is also supported by an STFC studentship. SM acknowledges financial support from the STFC CG ST/L000296/1. H.F acknowledges financial support from STFC CG ST/N000250/1. SM and U.E. acknowledge support from the the H2020-MSCA-RISE-2014 grant no. 645722 (NonMinimalHiggs). U.E. also acknowledges support from the Marie Sklodowska-Curie grant agreement no. 690575 (InvisiblesPlus).

\end{document}